\documentclass{aa}  
\usepackage{graphicx}
\usepackage{txfonts}
\usepackage{color}
\usepackage{multirow}
\usepackage{ulem}
\begin{document}

\title{The GUAPOS project:\\ III. Characterization of the O- and N-bearing complex organic molecules content and search for chemical differentiation}

\author{
C. Mininni,\inst{1}\thanks{E-mail: chiara.mininni@inaf.it},  M. T. Beltr\'an\inst{2}, L. Colzi\inst{3}, V. M. Rivilla\inst{3}, F. Fontani\inst{2}, A. Lorenzani\inst{2},\\ \'A. L\'opez-Gallifa\inst{3}, S. Viti\inst{4,5}, \'A. S\'anchez-Monge\inst{6,7,8,9}, 
 P. Schilke\inst{9},   L. Testi\inst{10}}
\institute{ INAF IAPS, Via del Fosso del Cavaliere 100, 00133 Roma, Italy 
 \and INAF Osservatorio Astrofisico di Arcetri, Largo E. Fermi 5, 50125 Firenze, Italy 
\and Centro de Astrobiolog\'ia (CSIC, INTA), Ctra. de Ajalvir, km. 4, Torrej\'on de Ardoz, E-28850 Madrid, Spain
\and Leiden Observatory, Leiden University, Huygens Laboratory, Niels Bohrweg 2, NL-2333 CA Leiden, The Netherlands
\and Department of Physics and Astronomy, University College London, Gower Street, WC1E 6BT, London, UK
\and Institut de Ci\'encies de l'Espai (ICE, CSIC), Can Magrans s/n, E-08193, Bellaterra, Barcelona, Spain
\and Institut d'Estudis Espacials de Catalunya (IEEC), Barcelona, Spain
\and Observatorio Astron\'omico Nacional (OAN), Alfonso XII, 3 28014, Madrid, Spain
\and I. Physikalisches Institut, Universit\"at zu K\"oln, Z\"ulpicher Str. 77, 50937 K\"oln, Germany
\and Dipartimento di Fisica e Astronomia, Universit\'a di Bologna, Via Gobetti 93/2, 40122 Bologna, Italy
             }
   \date{Received ; accepted }



  \abstract
   {The G31.41+0.31 Unbiased ALMA sPectral Observational Survey (GUAPOS) project targets the hot molecular core (HMC) G31.41+0.31 (G31), to unveil the complex chemistry of one of the most chemically rich high-mass star-forming regions outside the Galactic Center (GC). }
   {In the third paper of the project we present a study of nine O-bearing (CH$_3$OH, $^{13}$CH$_3$OH, CH$_3^{18}$OH,  CH$_3$CHO,  CH$_3$OCH$_3$, CH$_3$COCH$_3$  , C$_2$H$_5$OH, aGg'-(CH$_2$OH)$_2$, and gGg'-(CH$_2$OH)$_2$) and six N-bearing (CH$_3$CN, $^{13}$CH$_3$CN, CH$_3^{13}$CN,  C$_2$H$_3$CN,  C$_2$H$_5$CN, and C$_2$H$_5^{13}$CN) complex organic molecules toward G31. The aim of this work is to characterize the abundances in 
   G31 and to compare them with the values estimated in other sources. Moreover, we search for a possible chemical segregation between O-bearing and N-bearing species in G31, which hosts four compact sources as seen with higher angular resolution data. In the discussion we also include the three isomers of C$_2$H$_4$O$_2$ and the O- and N-bearing molecular species NH$_2$CHO, CH$_3$NCO, CH$_3$C(O)CH$_2$, and  CH$_3$NHCHO, analyzed in previous GUAPOS papers. }
   {Observations were carried out with the interferometer ALMA and cover the entire Band 3 from 84 to 116 GHz ($\sim 32$ GHz bandwidth) with an angular resolution of $1.2\arcsec\times1.2\arcsec$ ($\sim4400\,\mathrm{au}\times4400\,\mathrm{au}$) and a spectral resolution of $\sim0.488\,$ MHz ($\sim1.3-1.7\,\mathrm{km\,s^{-1}}$). The transitions of the fourteen molecular species have been analyzed with the tool SLIM of MADCUBA to determine the physical parameters of the emitting gas. Moreover, we have analyzed the morphology of the emission of the molecular species. }
   {The values of abundances w.r.t H$_2$ in G31 range from 10$^{-6}$ to 10$^{-10}$ for the different species. We have compared the abundances w.r.t methanol of O-bearing, N-bearing, and O- and N-bearing COMs in G31 with other twenty-seven sources, including other hot molecular cores inside and outside the Galactic Center, hot corinos, shocked regions, envelopes around young stellar objects, and quiescent molecular clouds, and with chemical models. } 
   {From the comparison with other sources there is not a unique template for the abundances in hot molecular cores, pointing towards the importance of the thermal history for the chemistry of the various sources. The abundances derived from the chemical models are well in agreement, within a factor ten, with those of G31. From the analysis of the maps we derived the peak positions of all the molecular species toward G31.
   Different species peak at slightly different positions, and this, together with the different central velocities of the lines obtained from the spectral fitting, point to chemical differentiation
   of selected O-bearing species.}

\titlerunning{GUAPOSIII}
  \authorrunning{Mininni et al. }
   \maketitle


\setlength{\parindent}{3ex}
\section{Introduction}
The spectrum of hot molecular cores (HMCs) reveals a very rich chemistry with many rotational transitions of a large number of molecular species, including complex organic molecules (COMs, i.e. molecules containing carbon with 6 or more atoms, \citealt{herbst2009}). These regions are associated with an evolved phase of high-mass star-formation, when a protostellar object(s) has been already formed in the core, leading to an increase of the temperature up to a few hundred K. The high temperatures ($> 100 \,$K) reached in HMCs cause the release (by thermal desorption) of the products of the grain-surface reactions into the gas-phase. Furthermore, also the presence of shocks can lead to the release in gas-phase of COMs (e.g. \citealt{palau2011}).  
The study of these regions is important to unveil the chemistry that occurs during the star-formation process and to constrain the chemical pathways responsible for the abundance variations of several species under the physical conditions present in these cores. \\\indent
The abundances of chemical species in HMCs are influenced by several factors, e.g. the physical conditions of the gas (density, $n$, and temperature, $T$), the cosmic rays (CRs) flux, the age of the embedded source,  external heating, and protostellar outflows that produce shocked regions (e.g. \citealt{garrod2013,sipila2021}).  Inhomogeneities in each of these factors potentially can lead  
 to favor, or inhibit, the chemical processes responsible for the formation of different species in different regions of the source. Ultimately, this could also produce chemical differentiation, with the abundances of some molecular species enhanced in regions where others are, on the contrary, less abundant (or even with such low abundances as not to be detected). \\ \indent
As an example, \citet{Blake1987} found spatial segregation between N-bearing molecules and O-bearing molecules toward Orion-KL, the closest high-mass star-forming region. In their study, the emission of N-bearing molecules was detected toward the hot core, while the O-bearing species were detected toward the compact ridge. This has been confirmed by further single-dish and interferometric studies (e.g. \citealt{Friedel2008, crockett2015}). However, this simple distinction does not provide the full picture. In fact, several studies have shown that some O-bearing molecular species, e.g. acetone (CH$_3$COCH$_3$), are more co-spatial to N-bearing species than to other O-bearing species, and also among O-bearing species the emitting regions vary (e.g. \citealt{WidicusWeaver2012,peng2013, Feng2015, tercero2018}). \citet{tercero2018} concluded that among O-bearing molecules in Orion-KL, spatial segregation is observed between molecules containing a C-O-C bond and those containing a C-O-H bond.
 \begin{table*}
    \centering
     \caption[O-bearing and N-bearing species analyzed]{Molecular species discussed in this work.}
    \begin{tabular}{llllll}
         \hline
         \multicolumn{6}{c}{\textit{O-bearing species}}  \\
         CH$_3$OH & $^{13}$CH$_3$OH& CH$_3^{18}$OH & CH$_3$CHO & CH$_3$OCHO$^{a}$ & CH$_3$COOH$^{a}$  \\ CH$_2$OHCHO$^{a}$ & CH$_3$OCH$_3$ & CH$_3$COCH$_3$ & C$_2$H$_5$OH   & aGg'-(CH$_2$OH)$_2$ & gGg'--(CH$_2$OH)$_2$  \\
         \hline 
         \multicolumn{6}{c}{\textit{N-bearing species}}\\
         CH$_3$CN & $^{13}$CH$_3$CN& CH$_3^{13}$CN&    C$_2$H$_3$CN & C$_2$H$_5$CN & C$_2$H$_5^{13}$CN  \\
          \hline 
         \multicolumn{6}{c}{\textit{O- and N-bearing species}}\\
          CH$_3$NCO$^{b}$& NH$_2$CHO$^{b}$& CH$_3$C(O)NH$_2^{b}$& CH$_3$NHCHO$^{b}$ & & \\
         \hline
    \end{tabular}
    \vspace{0.2cm}
      {\\\small{\textbf{Notes.} $a$) molecular species previously analyzed by \citet{mininni2020}; $b$) molecular species previously analyzed by \citet{colziguapos}. }}
    \label{tab:molecoleNObearing}
\end{table*}
The strong chemical differentiation seen in this source could be a peculiar case, caused by an explosive event in the central region of Orion (e.g. \citealt{bally2005,bally2017}), as also indicated by the elongated emission of several molecular species observed by \citet{pagani2019}. However, the chemical differentiation observed in Orion-KL seems to be not unique. Chemical differentiation has also been observed toward other high-mass star-forming regions. \citet{wyrowski1999}, \citet{remijan2004}, \citet{ciccio2010}, \citet{zernickel2012}, \citet{allen2017},\citet{vanderWalt2021}, and \citet{peng2022} observed chemical differentiation between O-bearing and N-bearing molecules toward W3(OH), W51 e1/e2, W75N, G19.61-0.23, NGC 6334I, G35.20-0.74N, CygX-N30 (W75N B), and G9.62+0.19 (especially between core MM8 and MM4), respectively, while \citet{jimenezserra2012chem} found chemical segregation of selected species in AFGL2591. \citet{jimenez2016} also detected chemical differentiation in the cold prestellar core L1544. \citet{bogelund2019} have shown that in the high-mass star-forming region AFGL4176 the peak of the emission of O-bearing molecules is offset 0.2\arcsec\, from the peak of N-bearing species, and that the mean excitation temperature  is higher for N-bearing than for O-bearing species: $\sim120-160\,\rm{K}$ for O-bearing and $\sim190-240\,\rm{K}$ for N-bearing species. A higher  excitation temperature of N-bearing species with respect to O-bearing species were also reported by \citet{widicusweaver2017} from the analysis of the spectra of 30 star-forming regions, half of which were HMCs. 
Recently, \cite{qin2022} have reported a clear shift of the peak of emission of CH$_3$OCHO and C$_2$H$_5$CN in 29 hot cores, over a sample of 60, while the other 28 sources show no shift in the position of the two molecules.
\\ \indent In these regions the origin of the observed chemical differentiation is still unclear. 
\citet{Caselli1993} proposed that the chemical differentiation between the Orion Hot Core and Compact Ridge could be related to differences in the thermal history of the sources. The use of time-dependent evaporation in chemical models by \citet{viti1999} and \citet{Viti2004}, and later study such as  \citet{suzuki2018} and \citet{garrod2022} have confirmed that such a chemical differentiation can be related to differences in the evolution of temperature within the sources. 
Thus the presence of multiple young stellar objects embedded in high-mass star-forming regions, with different thermal history, could explain the chemical differentiation. Another possible explanation is the presence of accretion shocks at the centrifugal barrier around an accreting protostar, like for example, in the case of the source  G328.2551-0.5321 \citep{Csengeri19shock}. In summary, the suggested chemical differentiation between O- and N-bearing molecules has not found a clear explanation, and further studies are needed to fully understand the underlying processes.\\ 
 \indent The target of the G31.41+0.31 (hereafter, G31) Unbiased ALMA sPectral Observational Survey (GUAPOS, \citealt{mininni2020, colziguapos}) is a well-known and studied HMC located at a distance of 3.75 kpc \citep{immer2019} with a
luminosity of $4.4 \times 10^{4}\, L_{\odot}$ and a gas mass
$M\sim70\, M_{\odot}$ \citep[respectively from][after rescaling the value to the new distance estimate adopted]{oso2009,cesa2019}. The core presents a velocity gradient observed in methyl cyanide and other COMs, such as methyl formate, firstly observed by \citet{cesa1994b}, associated with a rotating toroid \citep{beltran2004,beltran2005,beltran2018, gir2009,cesa2010,cesa2011,cesa2017}. Observations of molecular lines detected the presence of molecular outflows and infall in this source \citep{gir2009, cesa2011, may2014, beltran2018}, which is also associated with free–free sources at 0.7 and 1.3 cm
\citep{cesa2010, beltran2021}. Previous observations of this source have always revealed a single millimeter compact core, until observations by \citet{beltran2021} made with ALMA at 1.4\,mm and 3.5\,mm and with the Very Large Array (VLA) at 7\,mm and 1.3\,cm with a resolution of  $\sim0\farcs15$ , $\sim0\farcs075$,  $\sim0\farcs05$, and  $\sim0\farcs07$, respectively, have revealed the presence of four separated cores, named source A, B, C, and D  (gas masses of 16\,$M_{\odot}$, 15\,$M_{\odot}$, 26\,$M_{\odot}$, and 26\,$M_{\odot}$, respectively) all having part (and in two cases the predominant fraction) of the emission at wavelengths equal or above 7\,mm coming from free-free emission. The total mass of the four fragments is $83\pm19\, M_{\odot}$, consistent within the errors with the mass of $70\, M_{\odot}$ estimated by \citet{cesa2019}. Further high angular resolution analysis (at $\sim0.09\arcsec$) have revealed the presence of infall in all the four cores and the presence of at least six outflows detected in SiO, suggesting that each of the four sources embedded in the Main core drives a molecular outflow \citep{beltran2022outflow}. This reveals that all of the sources are still accreting material. Moreover, complementary large scale observation of N$_2$H$^{+}$ performed with IRAM 30m telescope by \citet{beltran2022cloudcollision} have shown that the environment in which G31 has formed is a typical hub-filament system resulting from a cloud-cloud collision.\\
\indent Previous studies have explored the chemical richness of this HMC (e.g. \citealt{beltran2005,beltran2009,fontani2007,iso2013, Calcutt2014, rivilla2017a, gorai2021, colziguapos, garciaconception2022}), revealing the presence of several O-bearing and N-bearing COMs. \citet{mininni2020} presented the GUAPOS project (G31.41+0.31 Unbiased ALMA sPectral Observational Survey), aimed at studying the full  $\sim$32\,GHz bandwidth spectrum at 3\,mm, namely the whole Atacama Large Millimeter/submillimeter Array (ALMA) Band 3.  The preliminary line identification has revealed an extremely chemically-rich source, with the spectrum showing only a few channels free of molecular line emission. Thus, G31 is an ideal candidate to investigate the complex chemistry in high-mass star-forming regions outside the Galactic Center (GC), because the extreme conditions in the GC, in terms of interstellar radiation field and cosmic-rays flux can have an impact on the chemistry (see e.g. \citealt{bonfand2019} and references therein). Most of the  line emission in G31 arises from  COMs, including both O-bearing and N-bearing species. Therefore, we aim to characterize the emission of these COMs in G31 and to investigate whether spatial segregation between N-bearing and O-bearing species is present also in this source. A first attempt to search for chemical segregation in G31, together with other five HMCs, was done by \citet{fontani2007}, who found a velocity difference of $\sim0.6\,\rm{km\,s^{-1}}$ in the peak velocity of C$_2$H$_5$CN and CH$_3$OCH$_3$, with observations taken with the IRAM 30m telescope. \\\indent
In this paper we analyze the emission of the following COMs: methanol, CH$_3$OH, and its isotopologues $^{13}$CH$_3$OH and CH$_3^{18}$OH, acetaldehyde, CH$_3$CHO, dimethyl ether, CH$_3$OCH$_3$, acetone, CH$_3$COCH$_3$, ethanol, C$_2$H$_5$OH, ethylene glycol, aGg'-(CH$_2$OH)$_2$ and gGg'-(CH$_2$OH)$_2$ conformers,  methyl cyanide, CH$_3$CN, and its isotopologues $^{13}$CH$_3$CN and CH$_3^{13}$CN, vinyl cyanide, C$_2$H$_3$CN, and ethyl cyanide, C$_2$H$_5$CN, and its isotopologue, C$_2$H$_5^{13}$CN. Moreover, in the discussion we will consider the results from the analysis presented in this paper together with the results from the analysis of the three isomers of C$_2$H$_4$O$_2$ by \citet{mininni2020}, and of the O- and N-bearing COMs formamide, NH$_2$CHO, methyl isocyanate, CH$_3$NCO, acetamide, CH$_3$C(O)CH$_2$, and N-methylformamide, CH$_3$NHCHO, by \citet{colziguapos} (see Table \ref{tab:molecoleNObearing}). In this work, molecules including both O and N atoms will be considered as a separate class in the discussion, unlike other works in literature where those molecules are considered among the N-bearing species. This choice would allow us, in case of  observed spatial segregation in G31, to better investigate the behavior of these species and study whether their emission is spatially closer to that of O-bearing species or N-bearing species. \\ \indent In Section 2 we present the observations; in Section 3 we present the spectral analysis (Sect. 3.1) and the analysis of the maps (Sect. 3.2); in Section 4 we discuss the results of both the analysis of the spectrum and of the maps, to reveal the possible presence of chemical differentiation between N-bearing and O-bearing species, or between selected molecular species. We also compare the abundances of COMs in G31 with other high-mass and low-mass star-forming regions, shocked regions and quiescent clouds, and with chemical models. In section 5 we summarize the main conclusions.
 \section{Observations}
 \indent The observations were carried out with ALMA during Cycle 5, (project 2017.1.00501.S, P.I.: M. T. Beltr\'an), and cover the complete spectral range of ALMA band 3, between 84.05 GHz and 115.91 GHz ( $\sim 32\,\mathrm{GHz}$ bandwidth), with a spectral resolution of $\sim0.488\,\mathrm{MHz}$ ($\sim1.3\,-\, 1.7\,\mathrm{km\,s^{-1}}$). The observations were divided in nine correlator configurations, and for each of them four contiguous basebands were observed. The data have been firstly presented by \citet{mininni2020}, where it is possible to find more details about the spectral setups, the flux and phase calibrators, and the configurations of the correlators.
  The uncertainties in the flux calibration are  $\sim5\%$ (from Quality Assesment 2 reports), in good agreement with ALMA Band 3 flux uncertainties reported by \citet{bonatoALMAcalibrator}.\newline\indent 
The data were calibrated and imaged with  CASA\footnote{https://casa.nrao.edu} (the \textit{Common Astronomy Software Applications} package, \citealt{CASAmcmullin}). The maps were created using a robust parameter of \citet{briggs} set equal to 0 and a common restoring synthesized beam of $1\farcs2\times1\farcs2$. The noise of the maps, $rms$, varies between 0.5 mJy/beam and 1.9 mJy/beam.\\ \indent  For all the observed basebands, the spectra have been extracted from an area equal to the size of the beam  and centered toward the peak of the continuum.  \citet{mininni2020} described the steps made to align the spectra of the different correlator configurations and basebands to obtain the final spectrum. The baseline has been analyzed by \citet{colziguapos} with the software STATCONT \citep{alvaro2018}, to statistically determine the value of the continuum level and then subtract it from the spectrum.\\ \indent Both from the alignment of the different spectra and from the statistical analysis performed with STATCONT on the final spectrum, the uncertainty on the determination of the continuum level is $\sim11\%$. 
\section{Analysis}
  
\begin{table*}
    \centering
     \caption{Results of the spectral analysis}
    \begin{tabular}{rccccccc}
    \hline\hline
    species & $T_{\rm{ex}}$ & $N$ &  $X$  &  FWHM &  V--V$_0$ \\ 
     & [K] & [$10^{17}$cm$^{-2}$] &   [$10^{-8}$] & [km\,s$^{-1}$] & [km\,s$^{-1}$]\\
    \hline
     \multicolumn{6}{c}{\textit{O-bearing species}}  \\
        CH$_3$OH\,v$_t=1$ & $208\pm24$ & $100\pm12$& $100\pm30$&$8.3\pm0.4$& $0.94\pm0.15$\\
        $^{13}$CH$_{3}$OH & $152\pm20$ &$9.6\pm1.7$ &$10\pm3$ &$7.2\pm0.3$ &$1.09\pm0.10$\\
        CH$_{3}^{18}$OH & $153\pm20$ & $2.4\pm0.4$ & $2.4\pm0.7$ & $7.0^{a}$ & $1.28\pm0.13$\\
        \hline
        CH$_{3}$OH$^{b}$ &  & $800\pm130$ & $800\pm200$  &  \\
        \hline
        CH$_3$CHO & $82\pm20$ & $0.34\pm0.16$ &$0.34\pm0.18$ &$7.5^{a}$ & $0.45\pm0.11$ \\
        CH$_3$OCH$_3$ &$98\pm11$ &$8.1\pm1.0$ &$8\pm2$ &$7.0^{a}$ &$1.07\pm0.06$ \\
        CH$_3$COCH$_3$ &$170\pm21$ &$5.6\pm1.3$ &$5.6\pm1.4$ &$7.0^{a}$ & $0.84\pm0.06$ \\
        C$_2$H$_5$OH &$119\pm14$ &$4.7\pm0.6^{c}$ &$4.7\pm1.2^{c}$  &$7.02\pm0.09$ &$1.00\pm0.04$ \\
        aGg'-(CH$_2$OH)$_2$ &$120^{a}$ &$1.45\pm0.17$ &$ 1.5\pm0.4$  &$7.2^{a}$ & $0.48\pm0.06$\\
        gGg'-(CH$_2$OH)$_2$ &$120\pm28$ & $0.87\pm0.19$ &$0.9\pm0.3$  &$7.2^{a}$ & $1.0\pm0.2$\\
         \hline
     \multicolumn{6}{c}{\textit{N-bearing species}}  \\
     CH$_3$CN\,v$_8=1$ & $197\pm51$ &$3.2\pm0.9$ &$3.2\pm1.1$  &$8.6\pm0.3$ & $0.68\pm0.15$\\
     $^{13}$CH$_{3}$CN &$111\pm17$  &$0.073\pm0.011$  &$0.073\pm0.019$  &$7.1\pm0.2$ & $0.95\pm0.09$\\
      CH$_{3}^{13}$CN & $54\pm10$ &$0.069\pm0.011$ &$0.069\pm0.018$  &$7.2\pm0.5$ &$0.97\pm0.19$\\
      \hline
      CH$_3$CN$^{d}$ &  & $2.7\pm0.4$ &$2.7\pm0.7$ & &\\
      \hline
      C$_2$H$_3$CN &$104\pm17$ &$0.21\pm0.04$ &$0.21\pm0.06$ &$9.7^{a}$ &$1.05\pm0.15$\\
      C$_2$H$_5$CN &$83\pm10$ &$0.56\pm0.08^{e}$ &$0.56\pm0.14^{e}$ &$7.94\pm0.13$ &$1.37\pm0.06$\\
      C$_2$H$_5^{13}$CN & $126\pm37$ & $0.042\pm0.012^{f}$&$0.042\pm0.017^{f}$ &$6.0^{a}$ &$1.8\pm0.3$\\
      \hline
      C$_2$H$_5$CN$^{g}$ &  & $1.6\pm0.5$ & $1.6\pm0.7$ &  \\
    \hline\hline
    \end{tabular}
    \vspace{0.2cm}
   {\\\small{\textbf{Notes.}  Excitation temperature, $T_{\rm{ex}}$, column density, $N$,  abundance w.r.t H$_2$, $X$, FWHM, and velocity with respect to V$_0=96.5$\,km\,s$^{-1}$,  V--V$_0$. The abundances have been calculated using $N_{\rm{H_2}}=(1.0\pm0.2)\times10^{25}\,\rm{cm^{-2}}$ \citep{mininni2020}; to calculate the corrected values of column densities and abundances of a species from the analysis of its $^{13}$C or $^{18}$O isotopologue, we have used the ratios $^{12}$C/$^{13}$C\,=\,$37\pm12$ ( from \citealt{yan2019}) and $^{16}$O/$^{18}$O\,=\,$333\pm143$ (from \citealt{wilsonrood}) , considering the Galactocentric distance of G31, $D_{\rm{GC}}=5.02$\,kpc; $a$): quantity kept fixed during the fitting; $b$) corrected values for column density and abundance derived from the values of CH$_{3}^{18}$OH; $c$) corrected for the partition function factor at 119\,K of 1.13;  $d$) corrected values for column density and abundance derived from the values of $^{13}$CH$_{3}$CN  (the estimate using the CH$_{3}^{13}$CN would give corrected $N =(2.6\pm0.4)\times10^{17}\,\rm{cm^{-2}}$ and $X = 2.6\pm0.7$); $e$) corrected for the partition function factor at 83\,K of 1.062; $f$) corrected for the partition function factor at 126\,K of 1.24;  $g$) corrected values for column density and abundance derived from the values of C$_2$H$_5^{13}$CN. }}
    \label{tab:resultchapter5}
\end{table*}

 \subsection{Spectral fitting}
 The baseline-subtracted spectrum has been analyzed with the SLIM (Spectral Line Identification and Modeling) tool within the MADCUBA package\footnote{Madrid Data Cube Analysis (MADCUBA) is a software developed in the Center of Astrobiology (Madrid) to visualize and analyze data cubes and single spectra: https://cab.inta-csic.es/madcuba/} \citep{martin2019}. The spectroscopic data were taken from the Cologne Database for Molecular Spectroscopy\footnote{https://cdms.astro.uni-koeln.de.} (CDMS, \citealt{cdms2001,cdms2005}) and from the JPL database of molecular spectroscopy\footnote{https://spec.jpl.nasa.gov/.} \citep{jpl1998}. Precise reference papers for each molecule analyzed in this work are given in Appendix A. For all the molecular species, except C$_2$H$_5$CN and its isotopologues, and C$_2$H$_5$OH, the partition function from the chosen catalog has either been calculated using also the first excited states or their contribution has been estimated to be less than 3\% at temperatures below 300\,K. For C$_2$H$_5$CN and its isotopologues the correction factor to the partition function is given in CDMS\footnote{https://cdms.astro.uni-koeln.de/classic/predictions/catalog/\\archive/EtCN/Qvib.txt} and is above 3\% from 75\,K. Therefore, for C$_2$H$_5$CN and C$_2$H$_5^{13}$CN we will apply the correction factor (interpolated at the T resulting from the fit) to the column density obtained with MADCUBA. In the case of C$_2$H$_5$OH the partition function has been calculated in CDMS (and also in JPL) taking into account the ground vibrational states only, but correction factors for temperatures different than 150\,K (correction factor 1.24 \citealt{Muller2016correctionfactor}) are not available yet. Because the temperature for this specie in G31 is of 119 K (see Sect. 3.1.1), we have not applied any correction. Once the correction factors will be available, authors who want to use the column density value of G31 can rescale it. \\ \indent The analysis has been performed after a preliminary line identification, where we identified a large fraction of the molecular species present in the source (see Appendix E of \citealt{colziguapos}). This line identification has allowed us to check the presence of possible blendings of the transitions of the molecular species analyzed in this work with other species and thus to select the most unblended transitions for each species to be used to constrain the fit.\\ \indent
To obtain the physical parameters of the molecular emission (column density, $N$, excitation temperature, $T_{\rm{ex}}$, full width at half maximum, FWHM, and velocity,  V$_{\rm LSR}$) from the data, we assumed a single temperature component that fills the beam and used the AUTOFIT tool of MADCUBA-SLIM, which finds the best agreement, minimizing the $\chi^{2}$, between the observed spectra and the predicted LTE model, taking into account also the optical depth. The simultaneous fit of multiple transitions for each molecular species allows the determination of V$_{\rm LSR}$ with an error smaller than the resolution in velocity of the spectrum ($\sim1.5\,\rm{km/s}$).
During the fit we have assumed local thermodynamic equilibrium (LTE), justified by the high density of G31 (rough estimate of $n\sim10^8\,\mathrm{cm^{-3}}$\,, \citealt{mininni2020}). The molecular species analyzed in this work are listed in Table \ref{tab:molecoleNObearing}.\\ \indent 
 The emission of methanol, methyl cyanide, and ethyl cyanide is optically thick, especially for the first two species, thus we have analyzed their isotopologues with $^{13}$C and for CH$_3$OH and CH$_3$CN, also the transitions of the torsionally or vibrationally excited state (as already done in \citealt{colziguapos}). To obtain the column density of the species containing only $^{12}$C (i.e. the most abundant isotopologues) we have multiplied the column density derived from the analysis of the $^{13}$C isotopologues for the $^{12}$C/$^{13}$C ratio. We have adopted a value of $^{12}$C/$^{13}$C\,=\,$37\pm12$, derived from \citet{yan2019}. and of $^{16}$O/$^{18}$O\,=\,$333\pm143$, derived from \citet{wilsonrood}, using the galactocentric distance of 5.02\,kpc for G31\footnote{Calculated from the heliocentric distance of 3.75\,kpc \citep{immer2019}}.\\ \indent For all the species we have assumed an error of 11\% on the values of $N$ and $T_{\rm{ex}}$, to reflect the uncertainty on the level of continuum. The best LTE fit results are listed in Table \ref{tab:resultchapter5}. For the species in which the fit of the spectra has been performed in previous GUAPOS papers (see Table 1, \citealt{mininni2020, colziguapos}) the results are listed in Table \ref{tab:previous results}. The abundances have been calculated using the column density of H$_2$ derived in the first paper of the GUAPOS project, $N_{\rm{H_2}}=(1.0\pm0.2)\times 10^{25}\,\rm{cm^{-2}}$ \citep{mininni2020}.\\ \indent From the spectral analysis it is already possible to infer preliminary hints of chemical differentiation. These are given by a clear difference in V$_{\rm LSR}$, FWHM or $T_{\rm{ex}}$ between O-bearing species, N-bearing species, and O- and N-bearing  species or between selected molecular species. A difference in V$_{\rm LSR}$ would be the strongest indication, since different embedded sources - resolved with observations at angular resolution of $\sim0\farcs075$ by \citet{beltran2021} - would likely have sligthly different velocities, while the values of FWHM or $T_{\rm{ex}}$ could be affected not only by a different position of the peak of the emission, but also by a different size of the emission. Figure \ref{fig:meanmap} 
shows the mean maps of the emission of the different molecular species (see Sect. 3.2 for more details), where we can see that different species have different emission sizes. \\ \indent The parameters derived from the best fit are given in Table \ref{tab:resultchapter5}, 
 where instead of the absolute value of V$_{\rm LSR}$, we give the difference V--V$_0$  using as reference velocity V$_0=96.5\,$km\,s$^{-1}$ \citep{beltran2018}.

\subsubsection{O-bearing species}
\paragraph{Methanol: CH$_3$OH, $^{13}$CH$_3$OH, and CH$_3^{18}$OH}
We have  not been able to constrain the fit of CH$_3$OH\,v=0  with a single temperature component. This could be connected both to the presence of a temperature gradient in the source, studied with high angular resolution data by \citet{beltran2018},  and to the high abundance of this molecular species that leads to optically thick transitions. In Fig. \ref{fig:spectrach3oh} we show some of the brightest transitions and we can see that some of them, which correspond to those having the lowest energies of the lower level and the highest Einstein coefficients, show inverse P Cygni profiles. Taking into account that the minimum maximum recoverable scale of the GUAPOS project is $\sim11\arcsec$, it is unlikely that the filtering effect has a major impact. Therefore, we conclude that the inverse P-Cygni profiles observed are due to the high optical depths of the lines and the presence of infall in the core, as reported by \citet{beltran2018, beltran2022outflow}.   
\\\indent We have analyzed the emission of the rotational transitions in the first torsionally excited state, v$_{t}=1$, whose transitions have optical depths estimated by MADCUBA between 0.1 and 0.2, with only two exceptions. We have constrained the fit using the 7 unblended transitions available (see Table \ref{table:ch3ohvib1}). Both the observed spectrum and the synthetic spectrum from the best fit are given in  Fig. \ref{fig:spectrach3ohvib}. The best-fit parameters are given in  Table \ref{tab:resultchapter5}. These parameters are consistent with those found by the best visual agreement of the simulated spectrum with the transitions of CH$_3$OH in the ground state (Fig. \ref{fig:spectrach3oh}), with the exception of the column density which is a factor $\sim2$ larger.\\
We also analyzed the emission of $^{13}$CH$_3$OH and CH$_3^{18}$OH, whose lines are optically thin. The 12 unblended transitions used to constrain the fit of $^{13}$CH$_3$OH are given in Table \ref{table:c13h3oh}, and the result of the fit is given in Table \ref{tab:resultchapter5}. The 12 transitions cover a broad range in $E_{\rm{U}}$ (energy of the upper state), from $\sim 7\,\rm{K}$ to $\sim330\,\rm{K}$. The spectrum of the transitions used to constrain the fit and the synthetic spectrum obtained with the best-fit parameters are given in Fig. \ref{fig:spectrac13h3oh}. The column density for CH$_3$OH derived multiplying the column density from the fit of $^{13}$CH$_3$OH for the ratio $^{12}$C/$^{13}$C=37 is $(3.6\pm0.7)\times10^{19}\rm{cm^{-2}}$. \\
\indent For CH$_3^{18}$OH we selected 15 transitions to constrain the fit (see Table \ref{table:ch3c18oh}). These transitions cover a broad range in $E_{\rm{U}}$ from $\sim 5\,\rm{K}$ to $\sim330\,\rm{K}$. The spectrum of the transitions used to constrain the fit and the synthetic spectrum obtained with the best-fit parameters are given in Fig. \ref{fig:spectrach3o18h}, while the results of the fit are given in Table \ref{tab:resultchapter5}.  The column density for CH$_3$OH derived multiplying the column density from the fit of CH$_3^{18}$OH for the ratio $^{16}$O/$^{18}$O=333 is $(8.0\pm1.3)\times10^{19}\rm{cm^{-2}}$.\indent The estimate of the column density of CH$_3$OH\,vt=1 is a factor 4 and 8 lower than the one derived for CH$_3$OH from $^{13}$CH$_3$OH and CH$_3^{18}$OH, respectively. This could be due to the different energy range covered by the transitions of the species: for CH$_3$OH\,vt=1 the available transitions have all $E_{\rm{U}}/\kappa_{\rm{B}}>300$\,K, while for the two isotopologues the transitions used to constrain the fit are in the range $\sim 5\,\rm{K}-\sim330\,\rm{K}$. From the second-last column of Table \ref{table:c13h3oh}, and \ref{table:ch3c18oh} we can see that CH$_3^{18}$OH is optically thinner than $^{13}$CH$_3$OH, therefore in the discussion we will adopt the value derived from CH$_3^{18}$OH for the column density of CH$_3$OH, since the two isotopologues give estimates of column density that differs by a factor 2.\\

 \paragraph{Acetaldehyde: CH$_3$CHO}
 We have detected 10 unblended lines of acetaldehyde. The spectral parameters of the corresponding transitions are given in Table \ref{table:ch3cho}. The energy range of the upper level covers only energy transitions below 100\,K ($\sim15-80\,\rm{K}$). To help the fitting algorithm to converge, we have fixed the FWHM to 7.5\,\,km\,s$^{-1}$. This value has been selected visually evaluating the best value of FWHM simulating the emission of individual transitions. The plot of the selected transitions with superimposed the synthetic spectrum calculated using the parameters from the best fit is given in Fig. \ref{fig:spectrach3cho}. The estimate of $T_{\rm{ex}}$ is lower than the mean value found for the rest of the molecular species, and could be due to the small range of $E_{\rm{U}}$ of the unblended transitions. To quantify how the column density would change if the temperature were higher, we have performed a fit using the same transitions and fixing the value of $T_{\rm{ex}}$ to 120\,K. The column density value from this fit is $6.0\times10^{16}\,\rm{cm^{-2}}$, changing by a factor of $\sim2$ from the column density derived from the best fit.
 
 \paragraph{Dimethyl ether: CH$_3$OCH$_3$}
 We have identified 33 unblended transitions of dimethyl ether covering energies from $\sim11\,\rm{K}$ to $\sim230\,\rm{K}$,  listed in Table \ref{table:ch3och3}. To help the fitting algorithm to converge, we have fixed the FWHM to 7.0\,\,km\,s$^{-1}$, the best value after a visual inspection. The results of the fit are given in Table \ref{tab:resultchapter5}. We found $T_{\rm{ex}}\sim100\,\rm{K}$, only slightly lower than that of some other molecular species. However, this value of $T_{\rm{ex}}$ is likely real and not biased  by the selected transitions used for the fit, since in the fit we have included several  high-energy transitions. We have plotted the synthetic spectrum derived with the parameters of the best fit in Fig. \ref{fig:spectrach3och3}.
\begin{table*}

    \centering
     \caption{Results from the spectral and map analysis for O-bearing and O- and N-bearing COMs analyzed in previous papers of the GUAPOS project.}
    \scalebox{0.8}{
    \begin{tabular}{rcccccccc}
    \hline\hline
         species & $T_{\rm{ex}}$ & $N$ & $X$ & FWHM & V--V$_{0}$ & R.A.\,(J2000) & Dec.\,(J2000) & reference\\
          & [K] & [$10^{17}\,\rm{cm^{-2}}$] & [$10^{-8}$] & [km\,s$^{-1}$]& [km\,s$^{-1}$] & [18h 47m s] & [-01$^{\circ}$ 12\arcmin \,\,\,\arcsec] & \\
    \hline
    \multicolumn{9}{c}{\textit{O-bearing species}} \\
        CH$_3$OCHO & $221\pm27$& $20\pm4$ & $20\pm6$ & 6.8$^{a}$&1.1$^{a}$ &34.318 &46.054 & \citet{mininni2020} \\
        CH$_3$COOH &$250\pm50^{b}$ &$6.4\pm2.1^{b}$ & $6.2\pm1.9^{b}$ & 7.8$^{a}$&0.0$^{a}$ &34.315 &46.057 & \citet{mininni2020}\\
        CH$_2$OHCHO&  $128\pm17$&$0.5\pm0.09$ &$0.50\pm0.14$ &8.8$^{a}$ &0.0$^{a}$ &34.309 &46.094 & \citet{mininni2020}\\
        \hline
       \multicolumn{9}{c}{\textit{O- and N-bearing species}} \\
        CH$_3$NCO$^{c}$ &$91\pm37$ &$1.2\pm0.3$ &$1.2\pm0.4$ &$7.15^{a}$ &$0.5^{a}$ &34.313 &46.054 & \citet{colziguapos}\\
        NH$_2$CHO$^{d}$ &$150^{a}$ &$1.7\pm0.6$ &$1.7\pm0.7$ &$8.6^{a}$ &$0.5^{a}$ & 34.316& 46.062& \citet{colziguapos}\\
        CH$_3$C(O)NH$_2$ & $285\pm50$ & $0.8\pm0.4$ & $0.8\pm0.4$ & $6.2\pm0.4$ & $0.2^{a}$  & 34.310 & 46.044 & \citet{colziguapos}\\
        CH$_3$NHCHO & $285^{a}$ & $0.37\pm0.16$ & $0.37\pm0.17$ & $7.0^{a}$ & $0.0^{a}$ & 34.307 & 46.058 & \citet{colziguapos}\\
    \hline\hline
    \end{tabular}}\\
    \vspace{0.3cm}
   {\small{\textbf{Notes:} The coordinates of the center of the emission of each species are derived from the task \textit{imfit} applied to the mean maps. $a$): quantity kept fixed during the fitting procedure; $b$): for CH$_3$COOH \citet{mininni2020} found that the values of $T_{\rm{ex}}$, $N$, and $X$ could range from 200$-$299\,K, $4.3-8.4\times10^{17}\,\rm{cm^{-2}}$, and $4.3-8\times10^{-8}$, respectively. The values reported in this table are the mean values of the ranges, while the errors cover the entire ranges.
   c) from CH$_3$NCO $v_{\rm b}=1$; d) from NH$_2^{13}$CHO, where the column density has been multiplied by the factor   $^{12}$C/$^{13}$C = 37.}}
    \label{tab:previous results}
\end{table*}
  \paragraph{Acetone: CH$_3$COCH$_3$}
   We have identified 38 unblended transitions of acetone, covering energies from $\sim15\,\rm{K}$ to $\sim170\,\rm{K}$, listed in Table \ref{table:ch3coch3}. We report the best-fit parameters in Table \ref{tab:resultchapter5}. A small number of transitions of CH$_3$COCH$_3$ are overestimated by the best-fit model.  A better agreement over the whole spectra would require a more accurate description of the temperature gradient of the source G31, as already seen for other molecular species (e.g. methanol, Sect. 3.1.1.1). In Fig. \ref{fig:spectrach3coch3part1} and \ref{fig:spectrach3coc3part2} we plot the most unblended transitions and the synthetic spectra obtained with the parameters of the fitting procedure.
 \paragraph{Ethanol: C$_2$H$_5$OH}
 We have detected 39 unblended transitions of ethanol. They are listed in Table \ref{table:c2h5oh} and plotted in Figs. \ref{fig:spectrac2h5ohp1} and \ref{fig:spectrac2h5ohp2}. The energy of the upper level $E_{\rm{U}}$ covers a broad range, $\sim 17-350\,\rm{K}$. The results of the best fit are given in Table \ref{tab:resultchapter5}, and the synthetic spectrum obtained using the parameters of the best fit is plotted in red over the observed spectrum in Figs. \ref{fig:spectrac2h5ohp1} and \ref{fig:spectrac2h5ohp2}. We note that the partition function from the catalog considers the ground vibrational state only, so we used the values of the energy of the vibrational states by \citet{2011Durig} to calculate the correction factor. For a temperature of 119\,K the correction factor is 1.13, which we applied to correct the value of the column density.
\paragraph{Ethylene glycol: gGg'-(CH$_2$OH)$_2$ and aGg'-(CH$_2$OH)$_2$ }
  We have detected 7 most unblended transitions of gGg'-(CH$_2$OH)$_2$, covering energies from $\sim20\,\rm{K}$ to $\sim120\,\rm{K}$. They are listed in table \ref{table:gGg} and plotted in Fig. \ref{fig:spectra_gGg}. To help the fitting algorithm to converge, we have fixed the FWHM to 7.2 $\mathrm{km\,,s^{-1}}$. This value has been selected visually evaluating the best value of FWHM simulating the emission of individual transitions. The results of the best fit are given in Table \ref{tab:resultchapter5}, and the synthetic spectrum obtained using the parameters of the best fit is plotted in red over the observed spectrum in Fig \ref{fig:spectra_gGg}.\\ \indent
For the conformer aGg'-(CH$_2$OH)$_2$, already detected in G31 by  \citet{rivilla2017a}, we have detected 20 unblended transitions. They are listed in Table \ref{table:aGg} and plotted in Fig. \ref{fig:spectra_aGg}. The energy of the upper level $E_{\rm{U}}$ covers an extremely narrow range between $\sim 15-55\,\rm{K}$, with the exception of two transitions with $E_{\rm{U}}\sim100\,\rm{K}$ (at 85522.2 and 96259.9 MHz), which are blended with other transitions of lower energy of the same specie. Therefore, to help the fitting algorithm to converge, we have fixed the excitation temperature to the value found for gGg'-(CH$_2$OH)$_2$ and the FWHM to 7.2 $\mathrm{km\,,s^{-1}}$.  The results of the best fit are given in Table \ref{tab:resultchapter5}, and the synthetic spectrum obtained using the parameters of the best fit is plotted in red over the observed spectrum in Fig \ref{fig:spectra_aGg}. The ratio of the column density of the confermers aGg'/gGg' is >1.
\paragraph{Isomers of C$_2$H$_4$O$_2$: CH$_3$OCHO, CH$_3$COOH, and CH$_2$OHCHO} The analysis of methyl formate, acetic acid and glycolaldehyde has been presented in \citet{mininni2020}. For CH$_3$OCHO, CH$_3$COOH, and CH$_2$OHCHO \citet{mininni2020} selected 22, 14, and 12 most unblended transitions to perform the fit, respectively. These transitions covered a range in the energy of the upper level from $\sim$20\,K to 220\,K  for  CH$_3$OCHO, $\sim$20\,K to 270\,K for CH$_3$COOH, and $\sim$20\,K to 190\,K for CH$_2$OHCHO. The transitions were optically thin. The physical parameters derived from the spectral fitting are summarized in Table \ref{tab:previous results}.

\subsubsection{N-bearing species}
 \paragraph{Methyl cyanide: CH$_3$CN, $^{13}$CH$_3$CN, and CH$_3^{13}$CN} 
 The transitions of CH$_3$CN\,v$=0$ present within the observed GUAPOS bandwidth are   CH$_3$CN(5$_{\rm{K}}-4_{\rm{K}}$) and   CH$_3$CN(6$_{\rm{K}}-5_{\rm{K}}$). For CH$_3$CN\,v$=0$, it has not been possible to have a good convergence of the fit because the majority of the transitions (K=0, 1, and 2) have high optical depths. The spectrum is plotted in Fig. \ref{fig:spectrach3cn}. Note that the lower K-components show a lower synthesised beam temperature than higher-K components, a clear indication that the lower-K transitions are optically thick.\\ \indent Because of the optical thickness of the ground transitions, we have analyzed the vibrationally excited ones, CH$_3$CN v$_8 =1$. The transitions used to constrain the fit are given in Table \ref{table:ch3cnvib}. The results of the fit are given in Table \ref{tab:resultchapter5}, and  the synthetic spectrum obtained with the best-fit parameters is given in Fig. \ref{fig:spectrach3cnvib}. \\ \indent 
 As for CH$_3$OH, we have also analyzed the emission of the isotopologue $^{13}$CH$_3$CN. We detected only 5 transitions not affected by blending, with  $E_{\rm{U}}$ between $\sim10\,\rm{K}$ and $\sim80\,\rm{K}$. They are reported in Table \ref{table:c13h3cn}. The results of the best fit are given in Table \ref{tab:resultchapter5} and plotted in Fig. \ref{fig:spectrac13h3cn}. Moreover, we have analyzed the isotopologue CH$_3^{13}$CN. We detected only 3 unblended transitions, listed in Table \ref{table:ch3c13n}. The range in $E_{\rm{U}}$ is in between $\sim40$ and $\sim80\,\rm{K}$. The results of the best fit are given in Table \ref{tab:resultchapter5}, and the spectrum is shown in Fig. \ref{fig:spectrach3c13n}. The value of $T_{\rm{ex}}$ is lower than that of CH$_3$CN v$_8 =1$ and $^{13}$CH$_3$CN. This discrepancy in $T_{\rm{ex}}$ could be due to the different (lower) energy range of the transitions used in the fit. Fixing the value of $T_{\rm{ex}}$ to 100\,K the column density from the fitting procedure would be $8.1\times10^{15}\,\rm{cm^{-2}}$, consistent inside the errors with the estimate leaving $T_{\rm{ex}}$ as a free parameter. From the second-last column of Table  \ref{table:c13h3cn} and \ref{table:ch3c13n} we can see that $^{13}$CH$_3$CN is optically thinner than CH$_3^{13}$CN, therefore in the discussion we will adopt the value derived from $^{13}$CH$_3$CN for the column density of CH$_3$CN.\\
 
\paragraph{Vinyl cyanide: C$_2$H$_3$CN}
 We have detected 11 unblended transitions of vinyl cyanide. They are listed in Table \ref{table:c2h3cn} and plotted in Fig. \ref{fig:spectrac2h3cn}. The range in $E_{\rm{U}}$ is limited, between $\sim30$ and $\sim90\,\rm{K}$. To better constrain the fit, we have fixed the value of the FWHM to 9.7\,\,km\,s$^{-1}$. The results of the best fit, given in Table \ref{tab:resultchapter5}, show a low value of $T_{ \rm{ex}}$. Such a low value could be a consequence of the low energy of the unblended transitions used for the fit.  
   
 \begin{figure*}
    \centering
    \includegraphics[width=13cm]{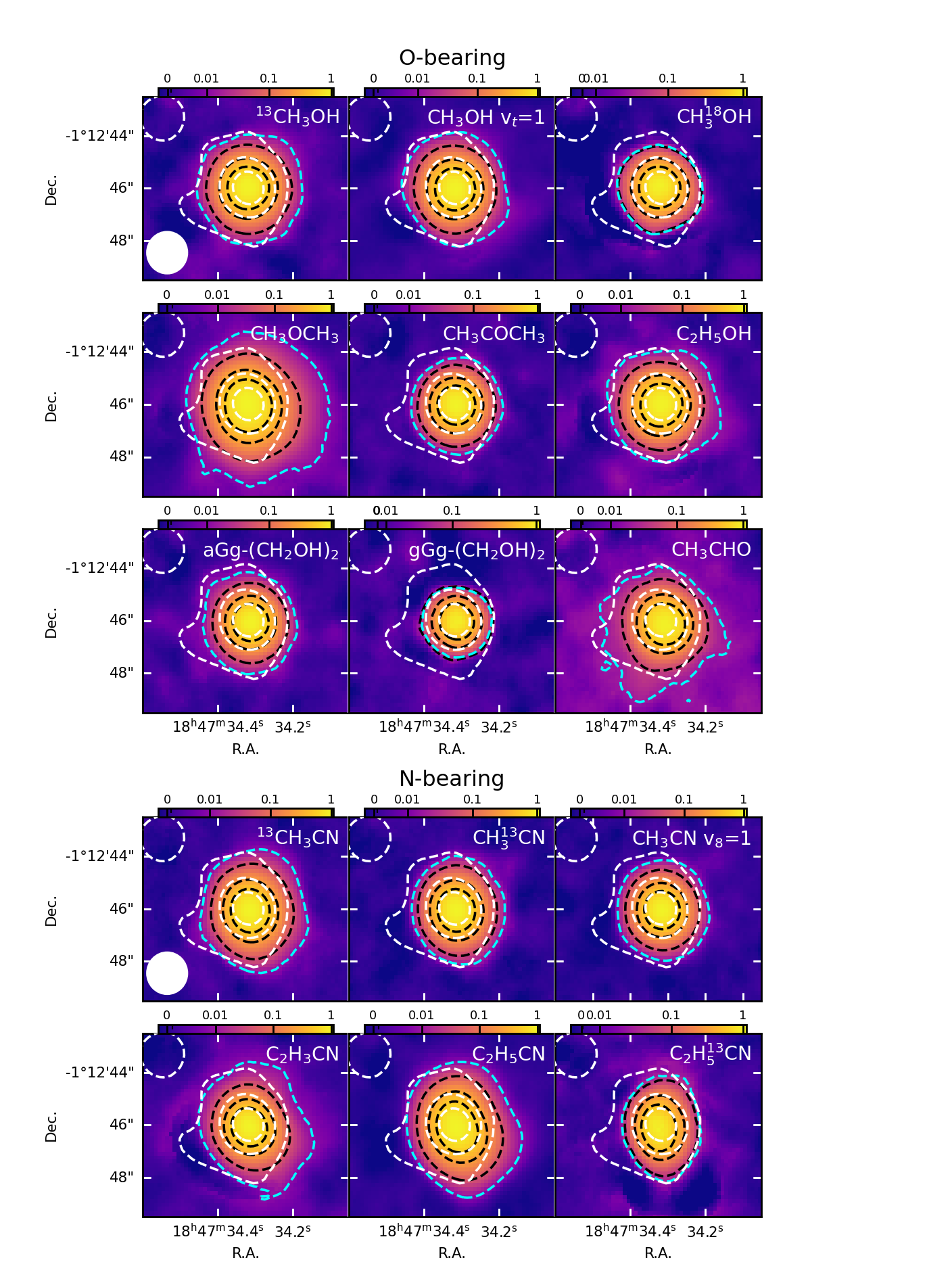}
    \caption{Mean maps of the species for which the analysis has been presented in this work for the first time (see Table 1). The mean maps have been obtained averaging the normalized maps of different transitions of the same species, therefore they are in arbitrary units with peak close to unity. In dashed white line we have plotted the contour levels of the continuum map at 150, 60, and 20 times the value of rms\,=\,0.8\,mJy\,beam$^{-1}$ (from \citealt{mininni2020}). The black dashed lines delimit the contour where the intensity of the mean map drops to 50\%, 25\%, and 5\% of the peak value. The 2D Gaussian fit ellipse to the emission is coincident with the inner black-dashed contour (50\% level) for all the molecular species.} The cyan dashed line delimits the contour where the intensity of the mean map is equal to 10 times the $rms$ of the map. The size of the $1\farcs2$ beam is indicated in the lower-left corner of the first panel on top-left. 
    \label{fig:meanmap}
\end{figure*}
\begin{table*}
    \centering
    \caption{Results of the 2D Gaussian fit to the mean maps. }
    \begin{tabular}{rccccc}
    \hline\hline
         species & $E_{\rm{U}}/k_{\rm{B}}$ & R.A.\,(J2000) & Dec.\,(J2000) & $\theta_{\rm{max}}\times\theta_{\rm{min}}$ & P.A.  \\
          & [K]& [18h 47m s] & [-01$^{\circ}$ 12\arcmin \,\,\,\arcsec] & [$\arcsec\times\arcsec$] & [$\deg$]\\
          \hline
          \multicolumn{6}{c}{\textit{O-bearing species}}\\
        CH$_3$OH\,v$_t=1$ &$\sim300-800$ &34.317 &46.035 &$1.61\times1.58$ & 34  \\
        $^{13}$CH$_{3}$OH &$\sim10-330$ &34.319 &45.990 &$1.66\times1.62$ & 88 \\
        CH$_3^{18}$OH & $\sim10-330$ & 34.323 & 45.972 & $1.62\times1.57$ & 75 \\
        CH$_3$CHO &$\sim10-70$ & 34.308 & 46.132 &$1.64\times1.60$ & 50  \\
        CH$_3$OCH$_3$ &$\sim10-220$ &34.319 &46.052  & $1.98\times1.83$ & 22 \\
        CH$_3$COCH$_3$ &$\sim10-170$ &34.314  &46.026 &$1.51\times1.49$ &80 \\
        C$_2$H$_5$OH &$\sim10-280$ &34.318 &46.018 &$1.67\times1.62$ &113 \\
        aGg'-(CH$_2$OH)$_2$ & $\sim15-55$ & 34.312& 46.068 &$1.46\times1.44$ &57 \\
        gGg'-(CH$_2$OH)$_2$ &$\sim25-115$ &34.313 &46.035 &$1.42\times1.38$ & 66\\
         \hline
          \multicolumn{6}{c}{\textit{N-bearing species}}\\
        CH$_3$CN\,v$_8=1$ & $\sim530-590$&34.316 &46.047 &$1.47\times1.43$ &24  \\
        $^{13}$CH$_{3}$CN &$\sim10-80$ &34.313 &46.069 &$1.71\times1.57$ &16  \\
        CH$_{3}^{13}$CN &$\sim40-80$ &34.314 &46.056 &$1.64\times1.51$ &14  \\
        C$_2$H$_3$CN &$\sim20-80$ &34.316 &46.084 &$1.57\times1.41$ &35  \\
        C$_2$H$_5$CN &$\sim20-120$ &34.311 &46.094 &$1.95\times1.59$ &20  \\
        C$_2$H$_5^{13}$CN &$\sim20-100$ & 34.322&46.105 &$1.77\times1.41$ & 7  \\
          \hline\hline\\
    \end{tabular}
      \label{tab:resultmaps2d}\\
    \vspace{0.2cm}
   {\small{\textbf{Notes:} Results of the 2D Gaussian fit only for the species for which the analysis has been presented in this work for the first time (see Table 1), performed with the task \textit{imfit}. In the second column we give the range in upper state energy $E_{\rm{u}}$ of the transitions used to create the mean maps.}}

\end{table*}
\begin{figure*}
    \centering
    \includegraphics[width=13cm]{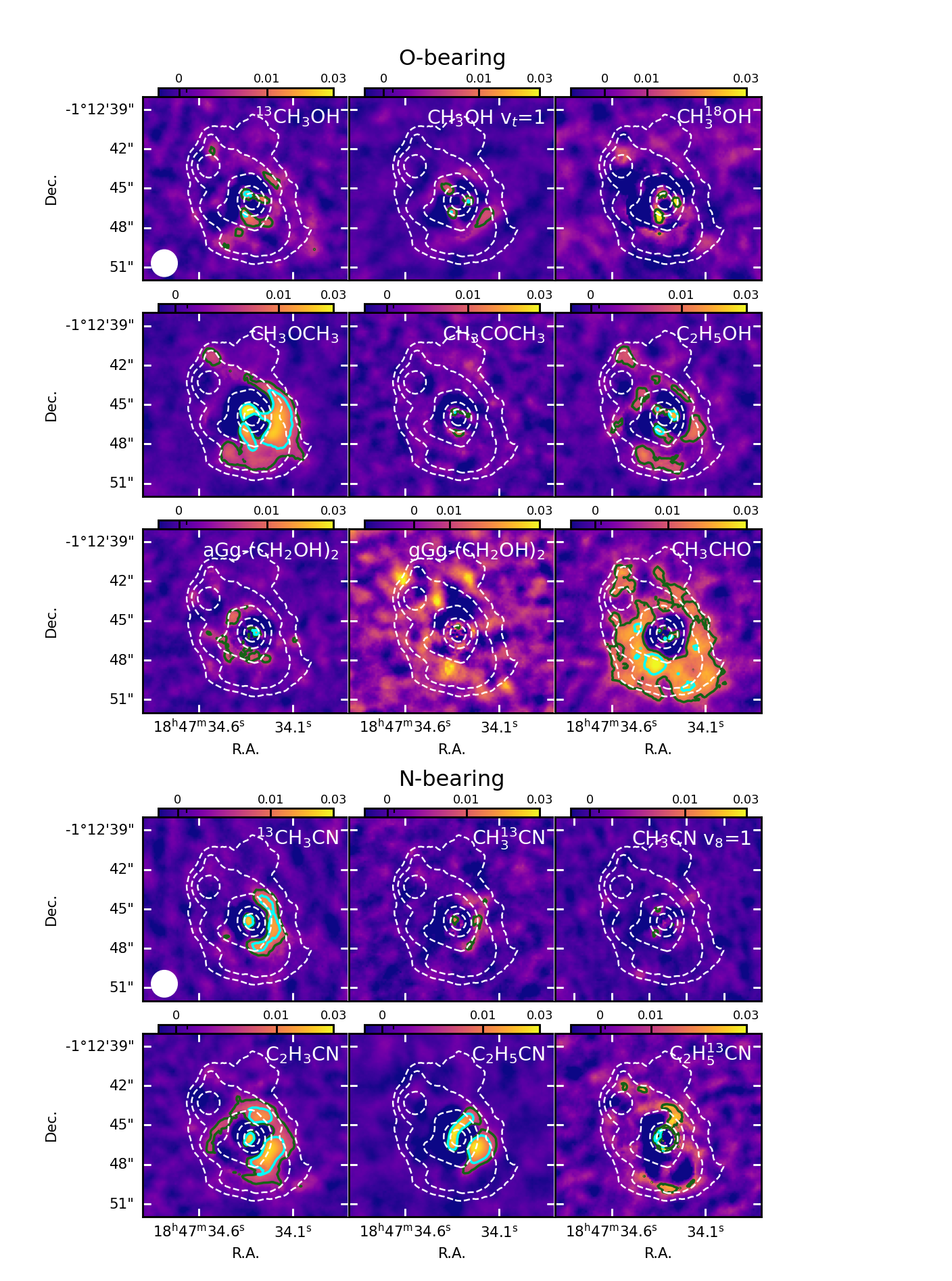}
    \caption{Maps of residuals after the 2D Gaussian fit to the mean maps, for the species for which the analysis has been presented in this work for the first time (see Table 1). The size of the plotted region is larger than that plotted in Fig. \ref{fig:meanmap}. The units are the arbitrary units of the mean maps. In white we have plotted the contours level of the continuum map at 150, 60, 20, 10, and 5 times the value of rms\,=\,0.8\,mJy\,beam$^{-1}$ (from \citealt{mininni2020}). The green and cyan solid lines delimit the contours of the regions where the emission in the residual map is larger than 5 and 10 times the $rms$, respectively. The size of the $1\farcs2$ beam is indicated in the lower-left corner of the top-left panel. }
    \label{fig:meanmapresi}
\end{figure*}
 
  \paragraph{Ethyl cyanide: C$_2$H$_5$CN and C$_2$H$_5^{13}$CN}
  We have detected 16 unblended transitions of C$_2$H$_5$CN, listed in Table \ref{table:c2h5cn}. The $E_{\rm{U}}$ ranges from $\sim25$ to $\sim120\,\rm{K}$. The best-fit parameters are given in  Table \ref{tab:resultchapter5} and the associated synthetic spectrum is shown in Fig. \ref{fig:spectrac2h5cn}. The column density given in  Table \ref{tab:resultchapter5} has been multiplied by the correction factor of the partition function at 83\,K, which is 1.062. For C$_2$H$_5$CN  the low value of $T_{\rm{ex}}$ could be a consequence of the low energy of the unblended transitions used for the fit.
  From Table \ref{table:c2h5cn} we can see that some transitions have optical depths very close to unity, and in two cases above it. Therefore, the emission from this species is not optically thin and we decided to analyze also its isotopologue C$_2$H$_5^{13}$CN to have a better estimate of its column density and abundance.\\ \indent 
 We have detected only 7 most unblended transitions of C$_2$H$_5^{13}$CN, listed in Table \ref{table:c2h5c13n}. The $E_{\rm{U}}$ ranges from $\sim30$ to $\sim110\,\rm{K}$. The intensity of the lines is weak, if compared to the intensity of the other COMs considered. The best-fit parameters are given in  Table \ref{tab:resultchapter5} and the associated synthetic spectrum is shown in Fig. \ref{fig:spectrac2h5cn}. The column density given in  Table \ref{tab:resultchapter5} has been multiplied by the correction factor of the partition function at 126\,K, which is 1.24.

\subsubsection{O- and N-bearing species}
The COMs formamide, NH$_2$CHO, methyl isocyanate, CH$_3$NCO, acetamide, CH$_3$C(O)NH$_2$, and N-methylformamide, CH$_3$NHCHO have been analyzed by \citet{colziguapos}. The physical parameters derived from the spectral fitting are summarized in Table \ref{tab:previous results}. NH$_2$CHO was detected in the ground vibrational state and in the first vibrational state. Moreover, NH$_{2} ^{13}$CHO was detected and analyzed. The values listed in Table \ref{tab:previous results} refer to the analysis of NH$_{2} ^{13}$CHO, whose emission is more optically thin, where the column density had been corrected for a factor $^{12}$C/$^{13}$C\,=\,$37\pm12$. For CH$_3$NCO, the ground vibrational state and the first vibrational state were detected and analyzed, with  $^{13}$CH$_3$NCO tentatively detected because most of the transitions are contaminated or blended with transitions of other species. The values listed in Table  \ref{tab:previous results} are derived from the analysis of the first vibrational state. CH$_3$C(O)NH$_2$ ground state and excited states ($v_{\rm{t}}$ =1,2) had optical depths $\ll 1$, and were fitted together. For its isomer CH$_3$NHCHO, only transitions of the ground state were detected, and to perform the fit the temperature was fixed to the value derived for  CH$_3$C(O)NH$_2$. 
 \subsection{Analysis of the maps}
 \label{sec:analysisofmaps}
 To create the emission maps of the species analyzed in this work we selected for each of them eight of the most unblended lines\footnote{What here is called a \textit{line} can be the result of the emission of multiple transitions of the same species, whose frequencies are closer than FWHM/2. Therefore, eight lines do not necessarily correspond to only eight transitions.}, or the maximum number possible for those species for which we detect less than eight (mostly) unblended transitions to fit. We selected the lines for the maps among the range of the detected unblended transitions, to include both low and high upper state energy transitions $E_{\rm{U}}$. The transitions corresponding to the selected lines are indicated in the last column of Tables \ref{table:ch3ohvib1}$-$\ref{table:c2h5c13n}. To create the maps of each line, we first removed the continuum emission from the cubes using the software STATCONT \citep{alvaro2018}, and then created the integrated intensity (moment 0) map for each line. \\ \indent To obtain a mean map of each species, we normalized each map to its peak intensity 
 and then averaged all the maps of the same species together, using the task \textit{immath} of CASA. Normalizing the different maps before averaging allows us to give the same weight to every map, and to obtain a mean map in which the FWHM of the emitting region is not biased by the brightest lines.\\ \indent
 The mean maps obtained with this procedure are shown in Fig. \ref{fig:meanmap}, and have been fitted with a 2D Gaussian using the task \textit{imfit} inside CASA to obtain the position of the peak of emission and the size of the emission. The results of the 2D Gaussian fit are given in Table \ref{tab:resultmaps2d} for the molecular species analyzed in this paper. The maps of the emission of the three isomers of C$_2$H$_4$O$_2$ and of O- and N-bearing species have been already presented in \citet{mininni2020} and \citet{colziguapos} and are not shown in this paper, while we have reported the position of the center of emission in Table \ref{tab:previous results}. The coordinates of the peak of the emission for CH$_3$NCO, NH$_2$CHO, CH$_3$C(O)NH$_2$, and CH$_3$NHCHO have been derived   from the maps presented in \citet{colziguapos} following the same methodology presented in this paper and in \citet{mininni2020}.\\ \indent The task \textit{imfit} produces also the maps of the residuals (i.e. the difference between the maps in input and the best 2D models computed by the task itself), where it is possible to understand if there is some molecular emission not coming from the main core (modeled with the 2D Gaussian), but from more extended regions or from secondary spots in the map. The maps of the residuals are plotted in Fig. \ref{fig:meanmapresi}.

\section{Discussion}
\subsection{Emission morphology}
From the mean maps of the different species shown in Fig. \ref{fig:meanmap}, we can see that the emission of all the COMs analyzed is centered toward the continuum peak of the HMC, with some species showing a shape  not perfectly Gaussian. Among the O-bearing species, we can see that the contours of the emission of CH$_3$CHO are shifted towards the South-West direction when compared to the contours of the continuum, unlike all the other O-bearing species (aGg'-(CH$_2$OH)$_2$ shows a less significant shift in the same direction). Looking at the residual maps, in Fig. \ref{fig:meanmapresi} (note that Fig. 2 shows a region $\sim2$ times larger around G31 w.r.t Fig.1), we can see that CH$_3$CHO, CH$_3$OCH$_3$, $^{13}$CH$_{3}$CN, C$_2$H$_3$CN, and  C$_2$H$_5$CN show a residual emission over 10 times the \textit{rms} value (cyan contours). In all cases, this emission is concentrated in the West and South-West edge of the main core (excluding residuals in the center of the core), with the exception of CH$_3$CHO that shows  very diffuse emission in all the envelope around the HMC, with spots over the 10\,\textit{rms} level, and a prominent more compact spot to the southern edge of the G31 main core. The origin of the residual emission could be related to the presence of outflows in the East$-$West direction and South West$-$North East direction on the plane of the sky, like those mapped in SiO by \citet{beltran2018, beltran2022outflow}. The more extended morphology of
CH$_3$CHO can not be explained only by the fact that we cover lower energy transition, even if the extended emission is more prominent in the lower energy state moment maps of CH$_3$CHO among those used to do the mean-moment map. In fact, in Appendix D we show the moment-0 maps of a low-energy transition for each molecular species, and none of the other moment-0 maps shows such an extended emission, even in the low-energy transitions. \\ \indent 
 All these regions are located outside the $1\farcs2$ central region from which we extracted the spectrum. It has to be noted that the  brightness peak of this residual emission is two orders of magnitude below the brightness toward the main core (which is 1 in the units of Figs. 1 and 2, from the methodology used to obtain the mean maps  described in Section 3.2), thus the column densities in those regions are expected to be accordingly smaller than those derived toward the center of the G31 HMC.
 \subsection{Positions of molecular species emission peaks}
\begin{figure*}
    \centering
    \includegraphics[width=17cm]{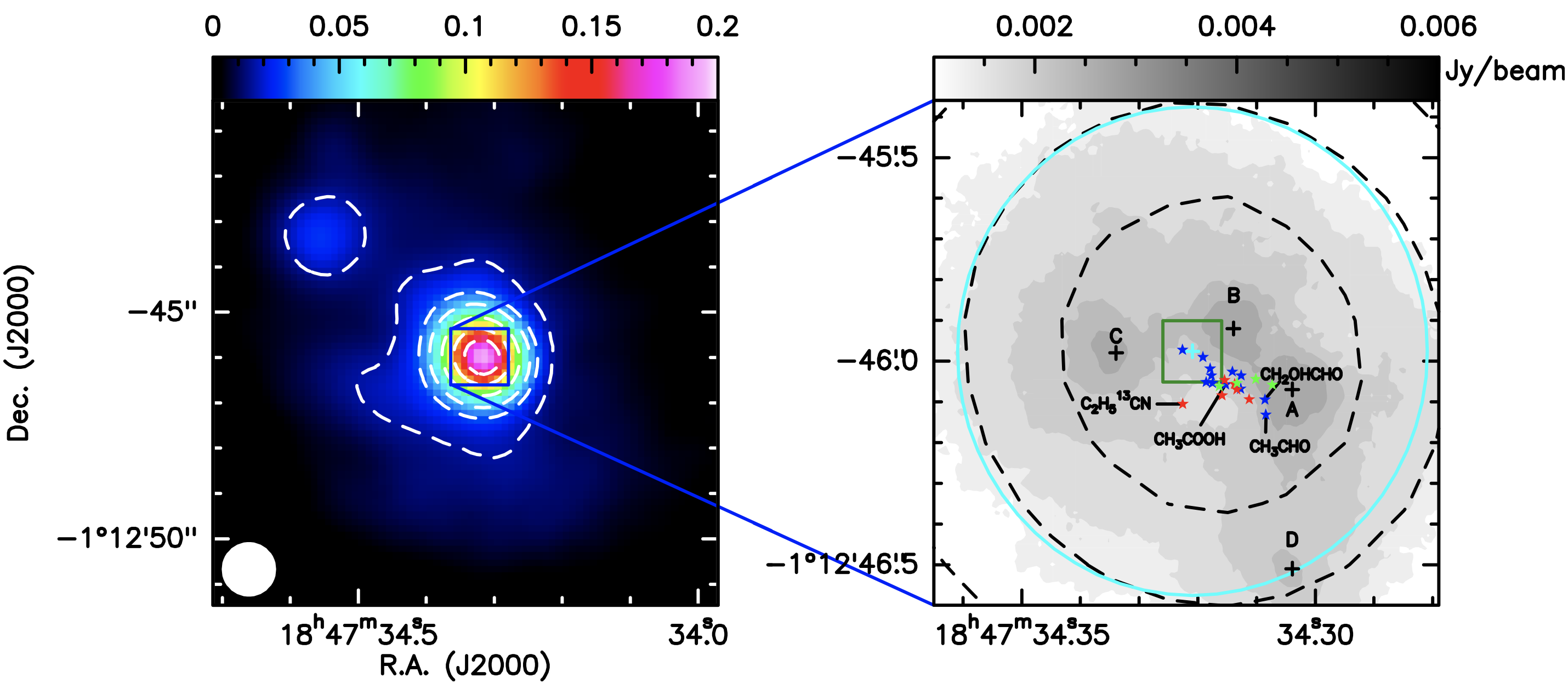}
    \caption{Left panel: continuum map from \citet{mininni2020}. The beam of $1\farcs2$ of the GUAPOS data is given in the lower-left corner. Contour levels are at  20, 40, 60, 100, 150, and 200 times the value of rms = 0.8\,mJy\,beam$^{-1}$. Right panel: zoom in of the left panel, where the black dashed contours are the two inner (150 and 200 times the rms) contours of the continuum image from the GUAPOS data at a resolution of $1\farcs2$, while the gray-levels (0.08, 0.12, 0.2, 0.4 mJy/beam) map is the continuum map at 3.5\,mm from \citet{beltran2021} with an angular resolution of $\sim 0\farcs075$. The dimension of the beam is given in the lower-left corner, while the $1\farcs2$ beam of the GUAPOS data is depicted by the cyan circular contour. The four black crosses marks the position of the four compact sources detected by \citet{beltran2021} and the green square indicates the dimension of the pixel in the GUAPOS maps and cubes, centered around the position of the peak of the GUAPOS continuum, indicated by the cyan cross. The blue stars indicate the positions of the peak of the emission of O-bearing species, red stars indicate the positions of the peak of the emission of N-bearing species, and green stars indicate the positions of the peak of the emission of O- and N-bearing species. The errors of the positions of the peak of molecular species are comparable to or below the dimension of the stars-shaped markers.}
    \label{fig:centerpositions}
\end{figure*}

 To reveal the existence of chemical differentiation inside this source we calculated the position of the peak of the emission for each molecular species, as described in Section \ref{sec:analysisofmaps}. The coordinates are reported in Table \ref{tab:resultmaps2d} for the molecular species analyzed in this work, and in Table \ref{tab:previous results} for the molecular species presented in \citet{mininni2020} and \citet{colziguapos}. \\\indent The distances between the center of emission of the different species are smaller than the beam (1\farcs2), therefore the spatial resolution of the observations is not sufficient to have conclusive results about the possible presence of spatial segregation for some classes of molecular species, or for selected species.\\ \indent Nevertheless, to give a visual reference to be confirmed by higher angular resolution data, in Fig. \ref{fig:centerpositions} we have plotted the continuum emission of the GUAPOS data and the high-resolution continuum data at 3.5\,mm from \citet{beltran2021}, marking with blue stars the position of the peaks of the emission of O-bearing COMs, with red stars the peak positions of N-bearing COMs, and with green stars the peak positions of O- and N-bearing COMs. The error bars on the peak position given in the output of the 2D Gaussian fit performed with the imfit tool of CASA are below the dimension of the maps pixel for all the molecular species. This is due to the fact that the emission is nearly Gaussian, with the residual values contributing at maximum for a factor of 3\% with respect to the peak emission. These error bars were not plotted in the right-side panel of Fig. 3 because they are of the same dimension as the star-shaped markers used for the peak positions. 
 The highest spatial separation is found among the majority of O-bearing molecules and the position of the peak of CH$_3$CHO and CH$_2$OHCHO. In the previous section, we also highlighted a shift of the contour of emission of CH$_3$CHO w.r.t. the emission of the other O-bearing COMs.  
 \\ \indent Only one molecular species, C$_2$H$_5^{13}$CN, is located far away from the position of any of the cores. 

\subsection{Spectral analysis}
\begin{figure*}
    \centering
    \includegraphics[height=15cm, angle=90, trim={2cm, 0, 1cm, 0}, clip]{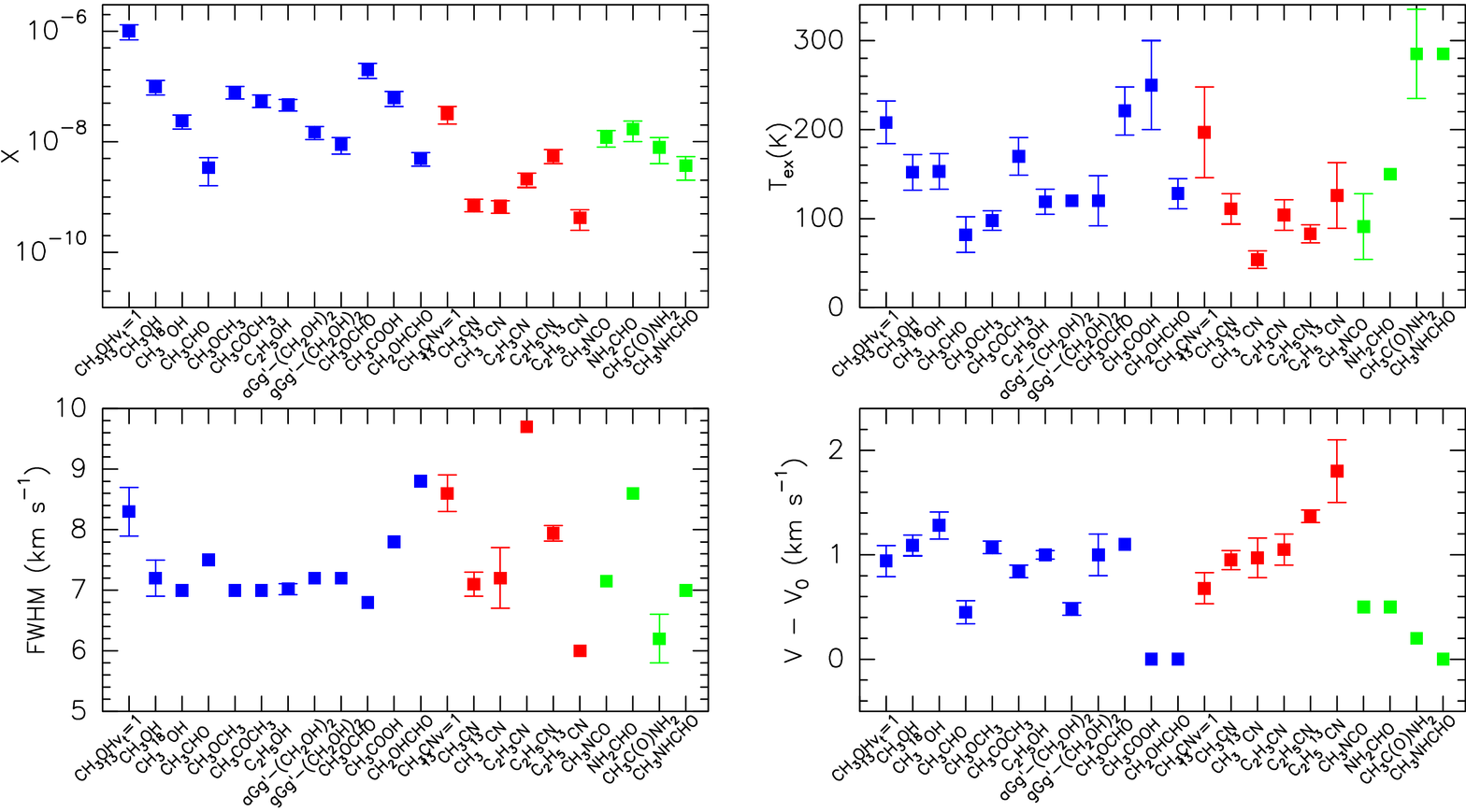}
    \caption{Plots of abundance, $X$, excitation temperature, $T_{\rm{ex}}$, full width at half maximum, FWHM, and line velocity, V--V$_{0}$. In blue the O-bearing species, in red the N-bearing species, while in green the O- and N-bearing species.}
    \label{fig:istogrammi}
\end{figure*}

The abundances w.r.t H$_2$ of the COMs analyzed in G31 range from $\sim10^{-6}$ (CH$_3$OH) to $\sim10^{-10}$ (C$_2$H$_5^{13}$CN). In the case of methanol, the estimate for the column density derived from the v$_t=1$ state is a factor 4 lower than the estimate obtained from the $^{13}$C isotopologue, rescaled with the factor $^{12}$C/$^{13}$C, indicating that the fit of the torsionally excited state might still be affected by optical depths effect. On the contrary, the estimates for the column density and abundance of CH$_3$CN from the v$_8=1$ state and from its two $^{13}$C isotopologues are well consistent within the errors, despite the large differences in $T_{\rm{ex}}$ derived for the three species. The estimate for ethyl cyanide from the $^{13}$C isotopologue is a factor 2 larger than the estimate from the emission of the main isotopologue, whose transitions have optical depths close to 1 (see Table \ref{table:c2h5cn}).
Figure \ref{fig:istogrammi} shows the comparison among the physical parameters of the molecular species presented in this work, together with  CH$_3$OCHO, CH$_3$COOH, CH$_2$OHCHO, CH$_3$NCO, NH$_2$CHO, CH$_3$C(O)NH$_2$, and CH$_3$NHCHO from \citet{mininni2020} and \citet{colziguapos}. We have summarized the data for these molecular species in Table \ref{tab:previous results}.\\\indent As discussed in Section 3.1, the analysis of the physical parameters of the emitting gas derived form the spectral analysis of the COMs can give further hints on the presence of a chemical differentiation. The most robust hints are given by a difference in the peak velocity of different molecular species. Other authors have reported also clear differences in $T_{\rm{ex}}$ among O-bearing and N-bearing species (e.g. \citealt{bogelund2019}), but in G31 part of the discrepancy in $T_{\rm{ex}}$ might be due to the difference in size of the emitting regions of different species, since there are clear evidences of a temperature gradient inside this HMC \citep{beltran2005}. \\ \indent From Fig. \ref{fig:istogrammi}, $T_{\rm{ex}}$ does not show a clear difference between O-bearing species, N-bearing species, and O- and N-bearing species. The values of $T_{\rm{ex}}$ range from $\sim60$\,K to $\sim300$\,K depending on the molecular species. This broad range might be due to emission arising from different regions (see Table \ref{tab:resultmaps2d}), 
or from the fact that 
 some of the molecular species presented in this work do not have unblended transitions with $E_{\rm{U}}>100\,\rm{K}$, leading to possibly underestimated excitation temperatures.\\ \indent  
The FWHM values vary between 6 to 10\,km\,s$^{-1}$, not showing any particular trend among molecular species. On the contrary in V--V$_{0}$ we can see that the majority of the molecular species have V--V$_{0}$ close to 1\,km\,s$^{-1}$ (including C$_2$H$_5^{13}$CN since its value is consistent with $\sim1$\,km\,s$^{-1}$ inside the large uncertainty), with the exception of CH$_3$CHO, CH$_3$COOH, CH$_2$OHCHO, aGg'-(CH$_2$OH)$_2$, and the four O- and N-bearing species  CH$_3$NCO, NH$_2$CHO, CH$_3$C(O)NH$_2$, and CH$_3$NHCHO. In particular,   CH$_3$COOH, CH$_2$OHCHO, CH$_3$C(O)NH$_2$, and CH$_3$NHCHO  have  V--V$_{0}\sim 0.2-0.0$\,km\,s$^{-1}$, while for CH$_3$CHO, CH$_3$NCO, aGg'-(CH$_2$OH)$_2$, and NH$_2$CHO the difference is less pronounced, with V--V$_{0}\sim 0.5$\,km\,s$^{-1}$. A previous study by \citet{fontani2007} found a discrepancy of $\sim0.6\,\rm{km\,s^{-1}}$ between the velocity of C$_2$H$_5$CN and CH$_3$OCH$_3$ in G31 from IRAM 30m telescope data, while in the interferometric data at a resolution of 1\farcs2 presented in this paper the discrepancy is of only $\sim 0.3\,\rm{km\,s^{-1}}$.\\ \indent From these results and the tentative analysis on the positions of the peak of the emission presented in the previous section, there are possible indications of chemical differentiation.  The stronger indications are for CH$_3$CHO and CH$_2$OHCHO, which have the largest separation in the peak position from all the other O-bearing species, and a discrepancy also in  V--V$_{0}$, while for CH$_3$COOH, aGg'-(CH$_2$OH)$_2$ and O- and N-bearing species we have a difference in peak velocity, but not a clear difference in peak position in our 1\farcs2 resolution maps. 

\begin{figure*}
    \centering
    \includegraphics[width=14.3cm]{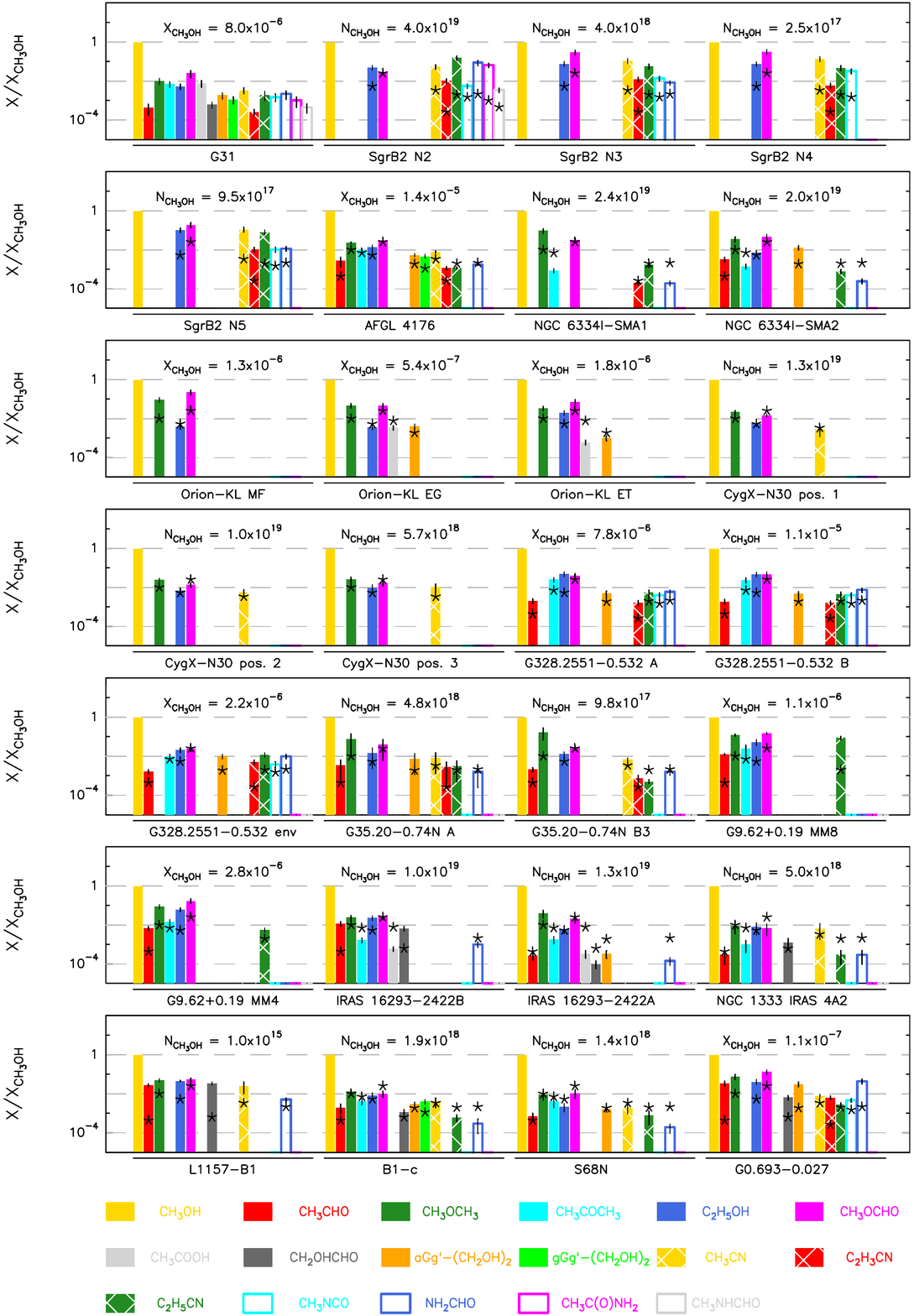}
    \caption{Histograms of O-bearing, N-bearing, and O- and N-bearing COMs abundances with respect to methanol for G31 (this paper, \citealt{mininni2020}, and \citealt{colziguapos}), SgrB2-N2 \citep{belloche2016}, SgrB2-N3/N4/N5 \citep{bonfand2019}, AFGL4176 \citep{bogelund2019}, NGC 6334I-SMA1 and SMA2 (from SMA data analysis, \citealt{zernickel2012}), Orion-KL methyl formate peak (MF), ethylene glycol peak (EG), and ethanol peak (ET) \citep{tercero2018}, CygX-N30 at the position 1, 2, and 3 as described in \citet{vanderWalt2021}, G328.2551-0.532 A, B and inner-envelope position \citep{Csengeri19shock}, G35.20-0.74N A and B3 \citep{allen2017}, IRAS 16293-2422 A and B \citep{jorge2018, coutens2016, Lykke2017, manigand2020}, NGC 1333 IRAS 4A2 \citep{lopezsepulcre2017,taquet2019}, L1157-B1 \citep{lefloch2017}, B1-c \citep{vangelder2020, nazari2021}, S68N \citep{vangelder2020, nazari2021}, and G+0.693-0.027 \citep{requena-torres2006, requena-torres2008, zeng2018, bizzocchi2020, rodriguez-almenida2021, rivilla2022, sanz-novo2022}. For the sources for which the abundances were not available we have plotted the ratio of column densities, since $X/X_{\rm CH_3OH} =N/N_{\rm CH_3OH}$. In the upper part of each panel are reported the absolute values of $X_{\rm CH_3OH}$ (or $N_{\rm CH_3OH}$ in $\rm{cm^{-2}}$, if the abundance is not available) for each source. To better compare the different sources with G31, we marked with asterisks the values of $X/X_{\rm CH_3OH}$ in G31 above the histogram of all the other sources.}
    \label{fig:othersources}
\end{figure*}

\subsection{Comparison with other sources}
In Fig. 5 we plot the abundances w.r.t. methanol of the O-bearing, N-bearing, and O- and N-bearing species detected in G31, together with the abundances of the same COMs found in other twenty-seven sources in literature. These include the HMCs SgrB2 N2/N3/N4/N5 \citep{belloche2016, bonfand2019},  AFGL4176 \citep{bogelund2019}, NGC 6334I- SMA1 and SMA2 \citep{zernickel2012}, Orion-KL observation toward the methyl formate peak (MF), ethylene glycol peak (EG), and ethanol peak (ET) of O-bearing molecules only \citep{tercero2018}, the protostellar source CygX-N30 in the position 1, 2, and 3 analyzed by \citet{vanderWalt2021}, G35.20--0.74N A and B3, the two most-chemically rich sources in G35.20--0.74N analyzed by \citet{allen2017}, the two most chemically rich sources MM8 (HMC) and MM4 (late HMC or hyper compact HII region) in the source G9.62+0.19 \citep{peng2022}, the low-mass hot corinos IRAS 16293-2422 A and B \citep{Jorge2016, manigand2020}, NGC 1333 IRAS 4A2 \citep{lopezsepulcre2017, taquet2019}, B1-c and S68N   \citep{vangelder2020, nazari2021}, the positions of accretion shocks  in high-mass star-forming regions G328.2551-0.532 A and B \citep{Csengeri19shock}, the shocked region L1157-B1 associated with a low-mass young stellar object (YSO, \citealt{lefloch2017, codella2009}), the inner envelope position around a HMC precursor G328.2551-0.532-env \citep{Csengeri19shock}, and the molecular cloud G+0.693-0.027 \citep{requena-torres2006, requena-torres2008, zeng2018, bizzocchi2020, rodriguez-almenida2021, rivilla2022, sanz-novo2022} located toward the North-East of the SgrB2 star-forming complex in the Central Molecular Zone (CMZ). Even though we are interested in discussing the relative abundances of these molecular species in different sources, in Fig. 5 we also report the absolute value of the abundance (or the column density if abundance was not available) of methanol in each source. For G+0.693-0.027 we rescaled all  the abundances to the value of $N(\rm{H_2})$ derived by \citet{martin2008NH2G0.69}, used in all the recent works toward this source. \\ \indent Before comparing the abundances in G31 with the other HMCs, we have checked the presence of a possible beam-dilution effect in the column density and abundance estimates of G31, given that the beam of 1\farcs2 is probing a linear scale of $\sim4400\,\rm{AU}$, larger than the scale resolved in some of the other studies toward HMCs we included in the comparison. \citet{colziguapos} presented the emission maps of O- and N-bearing species at a higher resolution of 0\farcs2 and confirmed that for these species the emission fills the beam of 1\farcs2 of the data analyzed in this paper. Moreover, the estimate of the column density from the high-resolution data is consistent within a factor 2 with the estimate obtained with our resolution. We also compared the column density of CH$_3$OCHO derived from the data at 1\farcs2 with that obtained from the analysis of the data at 0\farcs2 presented by \citet{beltran2018}, and they are well in agreement. \citet{beltran2018} also show the high-resolution map of the integrated emission of CH$_3$CN (12-11) K=2 which covers the region of our 1\farcs2 beam. Therefore, we conclude that possible beam-dilution effects will be minor if present, and will not affect the discussion below.\\ \indent
Comparing the abundances found in G31 and in the other high-mass protostars we found discrepancies: the four sources located toward the GC (SgrB2 N2/N3/N4/N5) show in general higher abundances - of around $\sim1$ order of magnitude or more in most of the cases - of O-bearing species (only CH$_3$OCHO and C$_2$H$_5$OH) and especially of N-bearing species and O- and N-bearing species. This enrichment of COMs with respect to methanol toward the GC may be the result of the peculiar conditions found in those regions, such as the cosmic-ray ionization rate being a factor $\sim50$ larger than the solar neighborhood value \citep{bogelund2019}, that can lead to enhance the efficiency of some chemical pathways. 
 For AFGL4176 we found a good general agreement of the abundances with those in G31, with some smaller differences for CH$_3$CHO, aGg'-(CH$_2$OH)$_2$, gGg'-(CH$_2$OH)$_2$, and C$_2$H$_3$CN. The absolute abundance of methanol w.r.t. H$_2$ in this source is higher by a factor $\sim4$ than the value derived for G31. The abundances of CH$_3$OCH$_3$, CH$_3$COCH$_3$, and NH$_2$CHO toward NGC 6334I-SMA1 show clear discrepancies with the values in G31, while CH$_3$OCHO and C$_2$H$_5$CN abundances are well in agreement. Also in the case of NGC 6334I-SMA2, only half of the COMs selected have abundances comparable to those found in G31. The comparison with the three positions in Orion-KL is limited to only five O-bearing species: CH$_3$OCH$_3$, C$_2$H$_5$OH, CH$_3$OCHO, CH$_3$COOH, and aGg'-(CH$_2$OH)$_2$. The best agreement is found between G31 and the ethylene glycol peak (EG see \citealt{tercero2018}). 
Also in the case of CygX-N30 \citep{vanderWalt2021} the comparison is limited only to four species: CH$_3$OCH$_3$, C$_2$H$_5$OH, CH$_3$OCHO, and CH$_3$CN. The abundances w.r.t methanol  of the three O-bearing species do not vary significantly toward the 3 positions identified by \citet{vanderWalt2021} and are in good agreement with the values measured toward G31, while the abundance of CH$_3$CN presents more variability with position 1 having a value close to G31.\\
\indent The abundances of C$_2$H$_5$OH,  CH$_3$OCHO, CH$_3$CN, C$_2$H$_5$CN, and NH$_2$CHO  in G35.20--0.74N A by \citet{allen2017} are comparable with those in G31, while the other molecular species are in general more abundant than in G31. For G35.20-0.74N B3, the abundance of C$_2$H$_3$CN is similar to that of G31, while the abundance of C$_2$H$_5$CN is clearly lower. The rest of the molecular species have abundances w.r.t methanol close to those found for G35.20--0.74N A. The abundances in cores MM8 and MM4 of G9.62+0.19 are in general higher than those found toward G31, in particular CH$_3$CHO in both sources and C$_2$H$_5$CN in MM8 which show a very high abundance w.r.t. methanol, similar to the values found toward the GC.\\ 
\indent Overall, it seems that there is not a  unique template for the abundances of COMs in HMCs. This could be the result of the peculiar physical properties (and their evolution with time) of each source, together with different environmental conditions that can affect the chemistry. In fact, the thermal history of each source has an impact on the reactions that can occur and their efficiency, likely leading to different values of abundance \citep{Caselli1993,viti1999,Viti2004, suzuki2018}. As example, in the chemical models by \citet{garrod2022} the abundances of N-bearing molecules are more sensitive to the time evolution of the warm-up phase (see Sect. 4.5), while the O-bearing species are more representative of the low-temperature dust-grain chemistry. Thus the different thermal evolution with time of each source can lead to different abundances w.r.t. H$_2$ of N-bearing and O-bearing species, and also of the abundances w.r.t. methanol. Moreover, other environmental factors, such as the cosmic-ray ionization rate and the different N/O elemental abundance in each particular region, have an impact on the chemistry. \\\indent
In the case of hot corinos, the range of abundances w.r.t. methanol of the COMs is similar to the range of abundances found toward G31 and other HMCs. However, among the five sources (IRAS 16293--2422 A and B, NGC 1333 IRAS 4A2, B1-c and S68N) only S68N shows an overall agreement with the majority of the COMs abundances estimated in G31, with the exception of NH$_2$CHO which is one order of magnitude less abundant. In IRAS 16203--22422 B, we have couples of COMs which show inverted abundances compared with G31. Those are CH$_3$CHO and CH$_3$COCH$_3$ together with the two isomers CH$_3$COOH and CH$_2$OHCHO. In IRAS 16203--22422 A, quite all the COMs show lower abundances w.r.t. methanol than in G31, except for CH$_3$OCH$_3$, while for B1-c only half of the abundances are similar to those found towards G31.\\ \indent
The accretion-shock regions G328.2551--0.532 A and B display slightly higher abundances of quite all the COMs presented in Fig. 5 for comparison. This is also true for the shock position driven by a low-mass protostar, L1157-B1, where we have steeper increases of abundance for CH$_3$CHO and CH$_2$OCHO. The abundances in the envelope around a HMC precursor in G328-2551--0.532 also show, in general, higher values. Lastly, we compare the abundances found in G31 with those toward G+0.693--0.027, a giant molecular cloud located inside the CMZ. The abundances of CH$_3$CHO, CH$_2$OHCHO, aGg'-(CH$_2$OH)$_2$, C$_2$H$_3$CN, and NH$_2$CHO w.r.t. CH$_3$OH are a factor $\sim10$ or more higher than the abundances in G31, while  the discrepancies with the other species are less pronounced. However, the value of column densities of CH$_3$OH is 3 orders of magnitude lower than in G31. The presence of these complex molecules in this source is thought to be the result of sputtering of dust grains and their icy mantles (e.g. \citealt{requena-torres2006, requena-torres2008}), together with the action of high cosmic-ray fluxes that enhances the efficiency of ion-neutral reactions increasing the abundances of some molecular species, since this source is located in the GC.\\ \indent
From these comparisons, the chemical content of  G31 in COMs is not comparable to that of sources in the GC due to the peculiar conditions found in those regions, and among other HMCs outside the central region of the galaxy we have a spread in chemical abundances among the sources. The only source with a remarkable similarity in chemical composition with G31, over the large number of molecules presented in this discussion, is AFGL4176. It is likely that the differences seen with respect to the other HMCs are due to different thermal histories of the various sources. Moreover, abundances w.r.t. methanol in hot corinos are mildly lower than those found in G31 as general trend, with the exception of IRAS 16293--2422 B, which however, does not show overall similar ratios. On the other hand, the abundances w.r.t. methanol found toward shocked regions, both in high-mass and low-mass star-forming regions show higher abundances than those of HMCs located outside the GC, like G31.

\begin{figure}
    \centering
    \includegraphics[width=8cm]{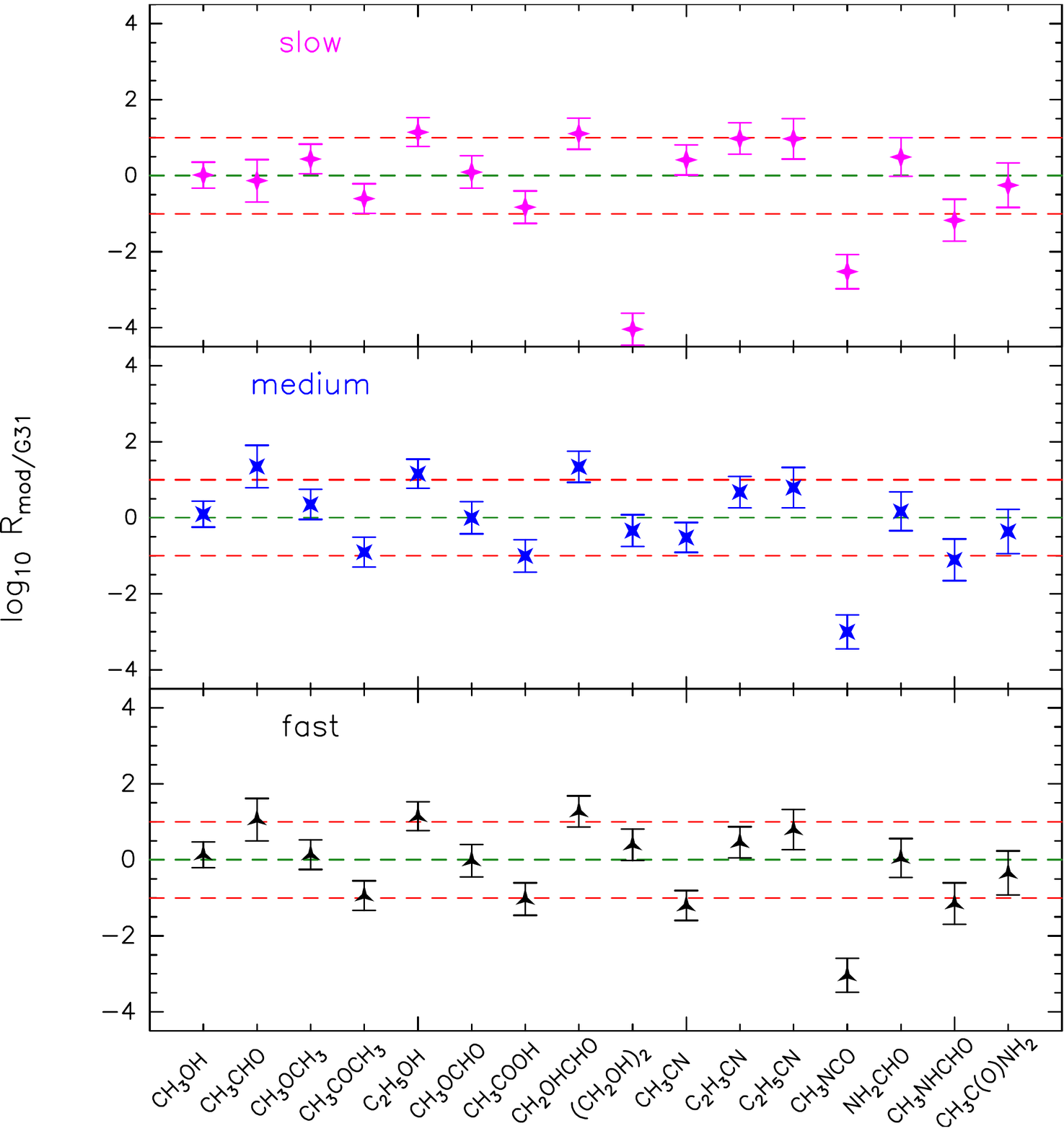}
    \caption{Ratio of abundances from model by \citet{garrod2022} and abundances derived for G31. The three panels use the abundances of the three different warm-up timescale chemical models. We considered an error of 30\% on the values of abundances from chemical models. The green dashed line represent a ratio $R_{\rm{mod/G31}}$ = 1, while the two red dashed lines represent $R_{\rm{mod/G31}}$ = 0.1 and 10.}
    \label{fig:garrod}
\end{figure}
\subsection{Comparison with chemical models}
In this section we compare the abundances of COMs in G31 with the results of the final models presented by \citet{garrod2022}. The work by \citet{garrod2022} is one of the most 
comprehensive models of gas-grain chemistry up to date and predicts the abundances of several molecular species  in HMCs and hot corinos, including a large number of COMs. In particular, nondiffusive reaction mechanism (Eley-Rideal mechanism, three-body reactions, three-body excited-formation reactions, and photodissociation-induced reaction mechanism) are implemented and the chemical network is updated with new routes proposed in recent years for the formation/destruction of dimethyl ether, formaldehyde, glycolaldehyde and ethylene glycol (e.g. \citealt{balucani2015, ayouz2019}).\\\indent The model has two physical phases: the first phase is the cold, free-fall collapse phase in which the gas temperature is fixed to 10\,K, while density increases up to $2\times10^{8}\mathrm{cm^{-3}}$ (consistent with the rough estimate given for G31 of $\sim10^{8}\,\rm{cm^{-3}}$ by \citealt{mininni2018}). The second phase is the warm-up phase, where the density is constant and the temperature increases up to 200\,K in with three different timescales for the three different models (\textit{slow} timescale $5\times 10^{4}$\,yr, \textit{medium} timescale $2\times 10^{5}$\,yr, and \textit{fast} timescale $1\times 10^{6}$\,yr). All the three models then continue until a temperature of 400\, K is reached. \\ \indent
In Fig. \ref{fig:garrod} we plot the ratio between the peak abundances reached in the three final models (\textit{slow} in the upper panel, \textit{medium} in the middle panel, and \textit{fast} in the lower panel) and the abundance derived in G31 for the COMs presented in this work. We assumed an error of 30\% on the values derived from the chemical models. The agreement between the models and the abundances in G31 is remarkable, with only a few molecular species overabundant or underabundant in G31 by more than a factor 10 w.r.t. to the abundances derived by \citet{garrod2022}.
The best overall agreement is with the \textit{slow} model with a few exceptions, but the values from the \textit{medium} model are also very close to those of G31. A large difference between these two models is  seen in (CH$_2$OH)$_2$, for which the value found in G31 is in very good agreement ($R_{\rm{mod/G31}}\sim1$) with the prediction of the \textit{medium} model, while it is 10$^{4}$ times overabundant if compared with the \textit{slow} model. In fact, in the latter model the destruction pathway of (CH$_2$OH)$_2$ on the dust grains via H abstraction has more time to act, leading to the low value presented in Fig. 6. This might indicate that the best match with G31 could be with an intermediate timescale between the \textit{slow} and the \textit{medium} values. The only species that is not well reproduced by any of the models is CH$_3$NCO, which is at least a factor $500$ more abundant in G31. This species is underproduced by the models also when compared with SgrB2(N2) and IRAS 16293B; according to \citet{garrod2022} this could be due to a not accurate estimate of some activation-energy barriers or to the presence of other pathways for its production not present in the chemical network.\\ \indent 
Overall, the agreement of the absolute values of the abundances derived in G31 with those of the model \textit{slow} and \textit{medium} is noteworthy. A good agreement for high-mass star-forming regions was found also for SgrB2(N2) with the \textit{slow} model. However, for some molecular species the agreement was between the trend seen in the model and in the source among different COMs, while a significant fraction  of COMs have  absolute values that do not match well with the model predictions \citep{garrod2022}, unlike the case of G31.
\section{Conclusions}
With the aim of characterizing the emission of the chemically rich HMC G31, and of understanding if chemical differentiation is present in this source, we have analyzed nine O-bearing and six N-bearing COMs using the data of the GUAPOS survey (\citealt{mininni2020, colziguapos}). In fact, \citet{beltran2021} highlighted the presence of four sources (labeled as A, B, C, and D) when observing this region at high angular resolution, which makes G31 an ideal laboratory to test whether different sources embedded in the same core show different chemistry. The analysis has been performed using the SLIM tool within the MADCUBA package. The molecular species analyzed include 
 methanol, CH$_3$OH, and its isotopologues $^{13}$CH$_3$OH and CH$_3^{18}$OH, acetaldehyde, CH$_3$CHO, dimethyl ether, CH$_3$OCH$_3$, acetone, CH$_3$COCH$_3$, ethanol, C$_2$H$_5$OH, methyl cyanide, CH$_3$CN, and its isotopologues $^{13}$CH$_3$CN and CH$_3^{13}$CN, vinyl cyanide, C$_2$H$_3$CN, and ethyl cyanide, C$_2$H$_5$CN, and its isotopologue, C$_2$H$_5^{13}$CN. Moreover, we have included in the analysis the  the three O-bearing isomers of C$_2$H$_4$O$_2$, and the O- and N-bearing COMs CH$_3$NCO, NH$_2$CHO, CH$_2$C(O)NH$_2$, and CH$_3$NHCHO, presented in previous work by \citet{mininni2020} and \citet{colziguapos}.\newline \indent
 The resolution of the maps in not sufficient to give conclusive results on the chemical segregation of molecules in G31. However, from
  the spectral analysis the most reliable parameter that can show hints of possible chemical differentiation is  V--V$_{0}$. In G31 we have found that the values of V--V$_{0}$ (V$_0=96.5$\,km\,s$^{-1}$) are $\sim1.0\,$km\,s$^{-1}$ for the majority of the COMs analyzed in this and previous GUAPOS papers, with the exception of CH$_3$CHO, CH$_3$COOH, CH$_2$OHCHO, and the four O- and N-bearing species  CH$_3$NCO, NH$_2$CHO, CH$_3$C(O)NH$_2$, and CH$_3$NHCHO. In particular,   CH$_3$COOH, CH$_2$OHCHO, CH$_3$C(O)NH$_2$, and CH$_3$NHCHO  have  V--V$_{0}$ of $\sim 0.2-0.0$\,km\,s$^{-1}$,  while for CH$_3$CHO, aGg'-(CH$_2$OH)$_2$, CH$_3$NCO, and NH$_2$CHO the difference is less pronounced, with V--V$_{0}\sim 0.5$\,km\,s$^{-1}$. 
Considering together the hints from spectral analysis and from map analysis, there are multiple indications of a possible chemical differentiation for the two O-bearing species CH$_3$CHO and CH$_2$OHCHO, with respect to the other O-bearing species. Other molecules show less pronounced  differences in V--V$_{0}$ and in shift of the position of the peak. 
\newline\indent  The values of abundances w.r.t. H$_2$ in G31 range from 10$^{-6}$ to 10$^{-10}$ for the different species. We have compared the abundances w.r.t methanol of O-bearing, N-bearing, and O- and N-bearing COMs in G31 with other twenty-seven sources. These include other high-mass HMCs, hot corinos, shocked regions, the position of envelope gas around a HMC precursor, and molecular clouds. No clear trends among all the HMCs have been found. The sources SgrB2-N2/N3/N4/N5 show higher abundances, especially of N-bearing and O- and N-bearing species,  when compared to the rest of the HMCs. The abundances (w.r.t. methanol) found in G31 are in a good agreement with those found toward AFGL4176 \citep{bogelund2019}, and also the abundances of O-bearing species in G31 are in good agreement with those observed toward Orion-KL in the EG peak \citep{tercero2018}. However, there is not a unique template for the abundances in HMCs. From the comparison with other types of sources we can see that as general trend hot corinos show mildly lower abundances w.r.t methanol than those in G31, with the exception of the source IRAS 16293--2422 B, while in shocked regions the abundances w.r.t methanol are enhanced, thanks to the sputtering of dust grains.
\\ \indent The abundances of COMs in G31 have been compared to the results of the three final chemical models of \citet{garrod2022}. The agreement is noteworthy, with most of the species in agreement within a factor ten with the estimate by the models. In particular, the \textit{slow} model shows the best overall agreement, with the exception of (CH$_2$OH)$_2$, which is on the other hand well reproduced by the \textit{medium} model. This might indicate that the best model to reproduce the abundances in G31 could have an intermediate timescale between the \textit{slow} and the \textit{medium} one, when (CH$_2$OH)$_2$ is still not heavily destroyed.
\section*{Acknowledgements}
C. M. acknowledges funding from the European
Research Council (ERC) under the European Union’s Horizon 2020 program,
through the ECOGAL Synergy grant (grant ID 855130).
V.M.R. and L.C. have received funding from the Comunidad de Madrid through the Atracci\'on de Talento Investigador (Doctores con experiencia) Grant (COOL: Cosmic Origins Of Life; 2019-T1/TIC-15379). L.C. has also received partial support from the Spanish State Research Agency (AEI; project number PID2019-105552RB-C41). A.S.M. acknowledges support from the RyC2021-032892-I grant funded by MCIN/AEI/10.13039/501100011033 and by the European Union 'Next GenerationEU'/PRTR, as well as the program Unidad de Excelencia Mar\'ia de Maeztu CEX2020-001058-M.  This paper makes use of the following ALMA data:
ADS/JAO.ALMA\#2017.1.00501.S. ALMA is a partnership of ESO (representing
its member states), NSF (USA) and NINS (Japan), together with NRC (Canada),
MOST and ASIAA (Taiwan), and KASI (Republic of Korea), in cooperation
with the Republic of Chile. The Joint ALMA Observatory is operated by ESO,
AUI/NRAO and NAOJ.




\bibliographystyle{aa}
\bibliography{bibliography.bib}

\begin{thebibliography}{147}
\expandafter\ifx\csname natexlab\endcsname\relax\def\natexlab#1{#1}\fi

\bibitem[{{Allen} {et~al.}(2017){Allen}, {van der Tak}, {S{\'a}nchez-Monge},
  {Cesaroni}, \& {Beltr{\'a}n}}]{allen2017}
{Allen}, V., {van der Tak}, F.~F.~S., {S{\'a}nchez-Monge}, {\'A}., {Cesaroni},
  R., \& {Beltr{\'a}n}, M.~T. 2017, \aap, 603, A133

\bibitem[{{Anderson} {et~al.}(1990){Anderson}, {De Lucia}, \&
  {Herbst}}]{anderson1990}
{Anderson}, T., {De Lucia}, F., \& {Herbst}, E. 1990, \apjs, 72, 797

\bibitem[{{Ayouz} {et~al.}(2019){Ayouz}, {Yuen}, {Balucani}, {Ceccarelli},
  {Schneider}, \& {Kokoouline}}]{ayouz2019}
{Ayouz}, M.~A., {Yuen}, C.~H., {Balucani}, N., {et~al.} 2019, \mnras, 490, 1325

\bibitem[{{Bally} {et~al.}(2017){Bally}, {Ginsburg}, {Arce}, {Eisner},
  {Youngblood}, {Zapata}, \& {Zinnecker}}]{bally2017}
{Bally}, J., {Ginsburg}, A., {Arce}, H., {et~al.} 2017, \apj, 837, 60

\bibitem[{{Bally} \& {Zinnecker}(2005)}]{bally2005}
{Bally}, J. \& {Zinnecker}, H. 2005, \aj, 129, 2281

\bibitem[{{Balucani} {et~al.}(2015){Balucani}, {Ceccarelli}, \&
  {Taquet}}]{balucani2015}
{Balucani}, N., {Ceccarelli}, C., \& {Taquet}, V. 2015, \mnras, 449, L16

\bibitem[{{Baskakov} {et~al.}(1996){Baskakov}, {Dyubko}, {Ilyushin},
  {Efimenko}, {Efremov}, {Podnos}, \& {Alekseev}}]{Baskakov1996}
{Baskakov}, O.~I., {Dyubko}, S.~F., {Ilyushin}, V.~V., {et~al.} 1996, Journal
  of Molecular Spectroscopy, 179, 94

\bibitem[{Bauer \& Godon(1975)}]{bauer1975}
Bauer, A. \& Godon, M. 1975, Canadian Journal of Physics, 53, 1154

\bibitem[{{Belloche} {et~al.}(2016){Belloche}, {M{\"u}ller}, {Garrod}, \&
  {Menten}}]{belloche2016}
{Belloche}, A., {M{\"u}ller}, H.~S.~P., {Garrod}, R.~T., \& {Menten}, K.~M.
  2016, \aap, 587, A91

\bibitem[{{Belov} {et~al.}(1995){Belov}, {Winnewisser}, \&
  {Herbst}}]{belov1995}
{Belov}, S.~P., {Winnewisser}, G., \& {Herbst}, E. 1995, Journal of Molecular
  Spectroscopy, 174, 253

\bibitem[{{Beltr{\'a}n} {et~al.}(2004){Beltr{\'a}n}, {Cesaroni}, {Neri},
  {Codella}, {Furuya}, {Testi}, \& {Olmi}}]{beltran2004}
{Beltr{\'a}n}, M.~T., {Cesaroni}, R., {Neri}, R., {et~al.} 2004, \apjl, 601,
  L187

\bibitem[{{Beltr{\'a}n} {et~al.}(2005){Beltr{\'a}n}, {Cesaroni}, {Neri},
  {Codella}, {Furuya}, {Testi}, \& {Olmi}}]{beltran2005}
{Beltr{\'a}n}, M.~T., {Cesaroni}, R., {Neri}, R., {et~al.} 2005, A\&A, 435, 901

\bibitem[{{Beltr{\'a}n} {et~al.}(2018){Beltr{\'a}n}, {Cesaroni}, {Rivilla},
  {S{\'a}nchez-Monge}, {Moscadelli}, {Ahmadi}, {Allen}, {Beuther}, {Etoka},
  {Galli}, {Galv{\'a}n-Madrid}, {Goddi}, {Johnston}, {Klaassen},
  {K{\"o}lligan}, {Kuiper}, {Kumar}, {Maud}, {Mottram}, {Peters}, {Schilke},
  {Testi}, {van der Tak}, \& {Walmsley}}]{beltran2018}
{Beltr{\'a}n}, M.~T., {Cesaroni}, R., {Rivilla}, V.~M., {et~al.} 2018, \aap,
  615, A141

\bibitem[{{Beltr{\'a}n} {et~al.}(2009){Beltr{\'a}n}, {Codella}, {Viti}, {Neri},
  \& {Cesaroni}}]{beltran2009}
{Beltr{\'a}n}, M.~T., {Codella}, C., {Viti}, S., {Neri}, R., \& {Cesaroni}, R.
  2009, \apjl, 690, L93

\bibitem[{{Beltr{\'a}n} {et~al.}(2022{\natexlab{a}}){Beltr{\'a}n}, {Rivilla},
  {Cesaroni}, {Galli}, {Moscadelli}, {Ahmadi}, {Beuther}, {Etoka}, {Goddi},
  {Klaassen}, {Kuiper}, {Kumar}, {Lorenzani}, {Peters}, {S{\'a}nchez-Monge},
  {Schilke}, {van der Tak}, \& {Vig}}]{beltran2022outflow}
{Beltr{\'a}n}, M.~T., {Rivilla}, V.~M., {Cesaroni}, R., {et~al.}
  2022{\natexlab{a}}, \aap, 659, A81

\bibitem[{{Beltr{\'a}n} {et~al.}(2021){Beltr{\'a}n}, {Rivilla}, {Cesaroni},
  {Maud}, {Galli}, {Moscadelli}, {Lorenzani}, {Ahmadi}, {Beuther}, {Csengeri},
  {Etoka}, {Goddi}, {Klaassen}, {Kuiper}, {Kumar}, {Peters},
  {S{\'a}nchez-Monge}, {Schilke}, {van der Tak}, {Vig}, \&
  {Zinnecker}}]{beltran2021}
{Beltr{\'a}n}, M.~T., {Rivilla}, V.~M., {Cesaroni}, R., {et~al.} 2021, \aap,
  648, A100

\bibitem[{{Beltr{\'a}n} {et~al.}(2022{\natexlab{b}}){Beltr{\'a}n}, {Rivilla},
  {Kumar}, {Cesaroni}, \& {Galli}}]{beltran2022cloudcollision}
{Beltr{\'a}n}, M.~T., {Rivilla}, V.~M., {Kumar}, M.~S.~N., {Cesaroni}, R., \&
  {Galli}, D. 2022{\natexlab{b}}, \aap, 660, L4

\bibitem[{{Bizzocchi} {et~al.}(2020){Bizzocchi}, {Prudenzano}, {Rivilla},
  {Pietropolli-Charmet}, {Giuliano}, {Caselli}, {Mart{\'\i}n-Pintado},
  {Jim{\'e}nez-Serra}, {Mart{\'\i}n}, {Requena-Torres}, {Rico-Villas}, {Zeng},
  \& {Guillemin}}]{bizzocchi2020}
{Bizzocchi}, L., {Prudenzano}, D., {Rivilla}, V.~M., {et~al.} 2020, \aap, 640,
  A98

\bibitem[{{Blake} {et~al.}(1987){Blake}, {Sutton}, {Masson}, \&
  {Phillips}}]{Blake1987}
{Blake}, G.~A., {Sutton}, E.~C., {Masson}, C.~R., \& {Phillips}, T.~G. 1987,
  \apj, 315, 621

\bibitem[{{B{\o}gelund} {et~al.}(2019){B{\o}gelund}, {Barr}, {Taquet},
  {Ligterink}, {Persson}, {Hogerheijde}, \& {van Dishoeck}}]{bogelund2019}
{B{\o}gelund}, E.~G., {Barr}, A.~G., {Taquet}, V., {et~al.} 2019, \aap, 628, A2

\bibitem[{{Bonato} {et~al.}(2018){Bonato}, {Liuzzo}, {Giannetti}, {Massardi},
  {De Zotti}, {Burkutean}, {Galluzzi}, {Negrello}, {Baronchelli}, {Brand},
  {Zwaan}, {Rygl}, {Marchili}, {Klitsch}, \& {Oteo}}]{bonatoALMAcalibrator}
{Bonato}, M., {Liuzzo}, E., {Giannetti}, A., {et~al.} 2018, \mnras, 478, 1512

\bibitem[{{Bonfand} {et~al.}(2019){Bonfand}, {Belloche}, {Garrod}, {Menten},
  {Willis}, {St{\'e}phan}, \& {M{\"u}ller}}]{bonfand2019}
{Bonfand}, M., {Belloche}, A., {Garrod}, R.~T., {et~al.} 2019, \aap, 628, A27

\bibitem[{{Boucher} {et~al.}(1977){Boucher}, {Burie}, {Demaison}, {Dubrulle},
  {Legrand}, \& {Segard}}]{boucher1977}
{Boucher}, D., {Burie}, J., {Demaison}, J., {et~al.} 1977, Journal of Molecular
  Spectroscopy, 64, 290

\bibitem[{{Boucher} {et~al.}(1980){Boucher}, {Dubrulle}, {Demaison}, \&
  {Dreizler}}]{boucher1980}
{Boucher}, D., {Dubrulle}, A., {Demaison}, J., \& {Dreizler}, H. 1980,
  Zeitschrift Naturforschung Teil A, 35, 1136

\bibitem[{{Briggs}(1995)}]{briggs}
{Briggs}, D.~S. 1995, in American Astronomical Society Meeting Abstracts, Vol.
  187, American Astronomical Society Meeting Abstracts, 112.02

\bibitem[{{Calcutt} {et~al.}(2014){Calcutt}, {Viti}, {Codella}, {Beltr{\'a}n},
  {Fontani}, \& {Woods}}]{Calcutt2014}
{Calcutt}, H., {Viti}, S., {Codella}, C., {et~al.} 2014, \mnras, 443, 3157

\bibitem[{{Caselli} {et~al.}(1993){Caselli}, {Hasegawa}, \&
  {Herbst}}]{Caselli1993}
{Caselli}, P., {Hasegawa}, T.~I., \& {Herbst}, E. 1993, \apj, 408, 548

\bibitem[{{Cazzoli} \& {Kisiel}(1988)}]{Cazzoli1988}
{Cazzoli}, G. \& {Kisiel}, Z. 1988, Journal of Molecular Spectroscopy, 130, 303

\bibitem[{{Cazzoli} \& {Puzzarini}(2006)}]{cazzoli2006}
{Cazzoli}, G. \& {Puzzarini}, C. 2006, Journal of Molecular Spectroscopy, 240,
  153

\bibitem[{{Cesaroni}(2019)}]{cesa2019}
{Cesaroni}, R. 2019, \aap, 631, A65

\bibitem[{{Cesaroni} {et~al.}(2011){Cesaroni}, {Beltr{\'a}n}, {Zhang},
  {Beuther}, \& {Fallscheer}}]{cesa2011}
{Cesaroni}, R., {Beltr{\'a}n}, M.~T., {Zhang}, Q., {Beuther}, H., \&
  {Fallscheer}, C. 2011, \aap, 533, A73

\bibitem[{{Cesaroni} {et~al.}(2010){Cesaroni}, {Hofner}, {Araya}, \&
  {Kurtz}}]{cesa2010}
{Cesaroni}, R., {Hofner}, P., {Araya}, E., \& {Kurtz}, S. 2010, \aap, 509, A50

\bibitem[{{Cesaroni} {et~al.}(1994){Cesaroni}, {Olmi}, {Walmsley},
  {Churchwell}, \& {Hofner}}]{cesa1994b}
{Cesaroni}, R., {Olmi}, L., {Walmsley}, C.~M., {Churchwell}, E., \& {Hofner},
  P. 1994, \apjl, 435, L137

\bibitem[{{Cesaroni} {et~al.}(2017){Cesaroni}, {S{\'a}nchez-Monge},
  {Beltr{\'a}n}, {Johnston}, {Maud}, {Moscadelli}, {Mottram}, {Ahmadi},
  {Allen}, {Beuther}, {Csengeri}, {Etoka}, {Fuller}, {Galli},
  {Galv{\'a}n-Madrid}, {Goddi}, {Henning}, {Hoare}, {Klaassen}, {Kuiper},
  {Kumar}, {Lumsden}, {Peters}, {Rivilla}, {Schilke}, {Testi}, {van der Tak},
  {Vig}, {Walmsley}, \& {Zinnecker}}]{cesa2017}
{Cesaroni}, R., {S{\'a}nchez-Monge}, {\'A}., {Beltr{\'a}n}, M.~T., {et~al.}
  2017, \aap, 602, A59

\bibitem[{{Christen} {et~al.}(2001){Christen}, {Coudert}, {Larsson}, \&
  {Cremer}}]{christen2001}
{Christen}, D., {Coudert}, L.~H., {Larsson}, J.~A., \& {Cremer}, D. 2001,
  Journal of Molecular Spectroscopy, 205, 185

\bibitem[{{Christen} {et~al.}(1995){Christen}, {Coudert}, {Suenram}, \&
  {Lovas}}]{christen1995}
{Christen}, D., {Coudert}, L.~H., {Suenram}, R.~D., \& {Lovas}, F.~J. 1995,
  Journal of Molecular Spectroscopy, 172, 57

\bibitem[{{Christen} \& {M{\"u}ller}(2003)}]{christen2003}
{Christen}, D. \& {M{\"u}ller}, H.~S.~P. 2003, Physical Chemistry Chemical
  Physics (Incorporating Faraday Transactions), 5, 3600

\bibitem[{{Codella} {et~al.}(2009){Codella}, {Benedettini}, {Beltr{\'a}n},
  {Gueth}, {Viti}, {Bachiller}, {Tafalla}, {Cabrit}, {Fuente}, \&
  {Lefloch}}]{codella2009}
{Codella}, C., {Benedettini}, M., {Beltr{\'a}n}, M.~T., {et~al.} 2009, \aap,
  507, L25

\bibitem[{{Colmont} {et~al.}(1997){Colmont}, {Wlodarczak}, {Priem},
  {M{\"u}ller}, {Tien}, {Richards}, \& {Gerry}}]{colmont1997}
{Colmont}, J.~M., {Wlodarczak}, G., {Priem}, D., {et~al.} 1997, Journal of
  Molecular Spectroscopy, 181, 330

\bibitem[{{Colzi} {et~al.}(2021){Colzi}, {Rivilla}, {Beltr{\'a}n},
  {Jim{\'e}nez-Serra}, {Mininni}, {Melosso}, {Cesaroni}, {Fontani},
  {Lorenzani}, {S{\'a}nchez-Monge}, {Viti}, {Schilke}, {Testi}, {Alonso}, \&
  {Kolesnikov{\'a}}}]{colziguapos}
{Colzi}, L., {Rivilla}, V.~M., {Beltr{\'a}n}, M.~T., {et~al.} 2021, \aap, 653,
  A129

\bibitem[{{Coutens} {et~al.}(2016){Coutens}, {J{\o}rgensen}, {van der Wiel},
  {M{\"u}ller}, {Lykke}, {Bjerkeli}, {Bourke}, {Calcutt}, {Drozdovskaya},
  {Favre}, {Fayolle}, {Garrod}, {Jacobsen}, {Ligterink}, {{\"O}berg},
  {Persson}, {van Dishoeck}, \& {Wampfler}}]{coutens2016}
{Coutens}, A., {J{\o}rgensen}, J.~K., {van der Wiel}, M.~H.~D., {et~al.} 2016,
  \aap, 590, L6

\bibitem[{{Crockett} {et~al.}(2015){Crockett}, {Bergin}, {Neill}, {Favre},
  {Blake}, {Herbst}, {Anderson}, \& {Hassel}}]{crockett2015}
{Crockett}, N.~R., {Bergin}, E.~A., {Neill}, J.~L., {et~al.} 2015, \apj, 806,
  239

\bibitem[{{Csengeri} {et~al.}(2019){Csengeri}, {Belloche}, {Bontemps},
  {Wyrowski}, {Menten}, \& {Bouscasse}}]{Csengeri19shock}
{Csengeri}, T., {Belloche}, A., {Bontemps}, S., {et~al.} 2019, \aap, 632, A57

\bibitem[{{Demaison} {et~al.}(1994){Demaison}, {Cosleou}, {Bocquet}, \&
  {Lesarri}}]{Demaison1994}
{Demaison}, J., {Cosleou}, J., {Bocquet}, R., \& {Lesarri}, A.~G. 1994, Journal
  of Molecular Spectroscopy, 167, 400

\bibitem[{{Demaison} {et~al.}(1979){Demaison}, {Dubrulle}, {Boucher}, {Burie},
  \& {Typke}}]{demaison1979}
{Demaison}, J., {Dubrulle}, A., {Boucher}, D., {Burie}, J., \& {Typke}, V.
  1979, Journal of Molecular Spectroscopy, 76, 1

\bibitem[{{Demyk} {et~al.}(2007){Demyk}, {M{\"a}der}, {Tercero}, {Cernicharo},
  {Demaison}, {Margul{\`e}s}, {Wegner}, {Keipert}, \& {Sheng}}]{demyk2007}
{Demyk}, K., {M{\"a}der}, H., {Tercero}, B., {et~al.} 2007, \aap, 466, 255

\bibitem[{{Durig} {et~al.}(2011){Durig}, {Deeb}, {Darkhalil}, {Klaassen},
  {Gounev}, \& {Ganguly}}]{2011Durig}
{Durig}, J.~R., {Deeb}, H., {Darkhalil}, I.~D., {et~al.} 2011, Journal of
  Molecular Structure, 985, 202

\bibitem[{{Endres} {et~al.}(2009){Endres}, {Drouin}, {Pearson}, {M{\"u}ller},
  {Lewen}, {Schlemmer}, \& {Giesen}}]{endres2009}
{Endres}, C.~P., {Drouin}, B.~J., {Pearson}, J.~C., {et~al.} 2009, \aap, 504,
  635

\bibitem[{{Feng} {et~al.}(2015){Feng}, {Beuther}, {Henning}, {Semenov},
  {Palau}, \& {Mills}}]{Feng2015}
{Feng}, S., {Beuther}, H., {Henning}, T., {et~al.} 2015, \aap, 581, A71

\bibitem[{{Fisher} {et~al.}(2007){Fisher}, {Paciga}, {Xu}, {Zhao}, {Moruzzi},
  \& {Lees}}]{fisher2007}
{Fisher}, J., {Paciga}, G., {Xu}, L.-H., {et~al.} 2007, Journal of Molecular
  Spectroscopy, 245, 7

\bibitem[{{Fontani} {et~al.}(2007){Fontani}, {Pascucci}, {Caselli}, {Wyrowski},
  {Cesaroni}, \& {Walmsley}}]{fontani2007}
{Fontani}, F., {Pascucci}, I., {Caselli}, P., {et~al.} 2007, \aap, 470, 639

\bibitem[{{Friedel} \& {Snyder}(2008)}]{Friedel2008}
{Friedel}, D.~N. \& {Snyder}, L.~E. 2008, \apj, 672, 962

\bibitem[{{Fukuyama} {et~al.}(1996){Fukuyama}, {Odashima}, {Takagi}, \&
  {Tsunekawa}}]{fukuyama1996}
{Fukuyama}, Y., {Odashima}, H., {Takagi}, K., \& {Tsunekawa}, S. 1996, \apjs,
  104, 329

\bibitem[{{Garc{\'\i}a de la Concepci{\'o}n} {et~al.}(2022){Garc{\'\i}a de la
  Concepci{\'o}n}, {Colzi}, {Jim{\'e}nez-Serra}, {Molpeceres}, {Corchado},
  {Rivilla}, {Mart{\'\i}n-Pintado}, {Beltr{\'a}n}, \&
  {Mininni}}]{garciaconception2022}
{Garc{\'\i}a de la Concepci{\'o}n}, J., {Colzi}, L., {Jim{\'e}nez-Serra}, I.,
  {et~al.} 2022, \aap, 658, A150

\bibitem[{{Garrod}(2013)}]{garrod2013}
{Garrod}, R.~T. 2013, \apj, 765, 60

\bibitem[{{Garrod} {et~al.}(2022){Garrod}, {Jin}, {Matis}, {Jones}, {Willis},
  \& {Herbst}}]{garrod2022}
{Garrod}, R.~T., {Jin}, M., {Matis}, K.~A., {et~al.} 2022, \apjs, 259, 1

\bibitem[{{Gerry} {et~al.}(1976){Gerry}, {Lees}, \& {Winnewisser}}]{gerry1976}
{Gerry}, M.~C.~L., {Lees}, R.~M., \& {Winnewisser}, G. 1976, Journal of
  Molecular Spectroscopy, 61, 231

\bibitem[{{Gerry} \& {Winnewisser}(1973)}]{gerry1973}
{Gerry}, M.~C.~L. \& {Winnewisser}, G. 1973, Journal of Molecular Spectroscopy,
  48, 1

\bibitem[{{Girart} {et~al.}(2009){Girart}, {Beltr{\'a}n}, {Zhang}, {Rao}, \&
  {Estalella}}]{gir2009}
{Girart}, J.~M., {Beltr{\'a}n}, M.~T., {Zhang}, Q., {Rao}, R., \& {Estalella},
  R. 2009, Science, 324, 1408

\bibitem[{{Gorai} {et~al.}(2021){Gorai}, {Das}, {Shimonishi}, {Sahu}, {Mondal},
  {Bhat}, \& {Chakrabarti}}]{gorai2021}
{Gorai}, P., {Das}, A., {Shimonishi}, T., {et~al.} 2021, \apj, 907, 108

\bibitem[{{Groner} {et~al.}(1998){Groner}, {Albert}, {Herbst}, \& {De
  Lucia}}]{Groner1998}
{Groner}, P., {Albert}, S., {Herbst}, E., \& {De Lucia}, F.~C. 1998, \apj, 500,
  1059

\bibitem[{{Groner} {et~al.}(2002){Groner}, {Albert}, {Herbst}, {De Lucia},
  {Lovas}, {Drouin}, \& {Pearson}}]{groner2002}
{Groner}, P., {Albert}, S., {Herbst}, E., {et~al.} 2002, \apjs, 142, 145

\bibitem[{{Herbst} {et~al.}(1984){Herbst}, {Messer}, {De Lucia}, \&
  {Helminger}}]{herbst1984}
{Herbst}, E., {Messer}, J.~K., {De Lucia}, F.~C., \& {Helminger}, P. 1984,
  Journal of Molecular Spectroscopy, 108, 42

\bibitem[{{Herbst} \& {van Dishoeck}(2009)}]{herbst2009}
{Herbst}, E. \& {van Dishoeck}, E.~F. 2009, \araa, 47, 427

\bibitem[{{Hoshino} {et~al.}(1996){Hoshino}, {Ohishi}, {Akabane}, {Ukai},
  {Tsunekawa}, \& {Takagi}}]{hoshino1996}
{Hoshino}, Y., {Ohishi}, M., {Akabane}, K., {et~al.} 1996, \apjs, 104, 317

\bibitem[{{Hughes} {et~al.}(1951){Hughes}, {Good}, \& {Coles}}]{hughes1951}
{Hughes}, R.~H., {Good}, W.~E., \& {Coles}, D.~K. 1951, Physical Review, 84,
  418

\bibitem[{{Ikeda} {et~al.}(1998){Ikeda}, {Duan}, {Tsunekawa}, \&
  {Takagi}}]{Ikeda1998}
{Ikeda}, M., {Duan}, Y.-B., {Tsunekawa}, S., \& {Takagi}, K. 1998, \apjs, 117,
  249

\bibitem[{{Immer} {et~al.}(2019){Immer}, {Li}, {Quiroga-Nu{\~n}ez}, {Reid},
  {Zhang}, {Moscadelli}, \& {Rygl}}]{immer2019}
{Immer}, K., {Li}, J., {Quiroga-Nu{\~n}ez}, L.~H., {et~al.} 2019, \aap, 632,
  A123

\bibitem[{{Isokoski} {et~al.}(2013){Isokoski}, {Bottinelli}, \& {van
  Dishoeck}}]{iso2013}
{Isokoski}, K., {Bottinelli}, S., \& {van Dishoeck}, E.~F. 2013, \aap, 554,
  A100

\bibitem[{{Jim{\'e}nez-Serra} {et~al.}(2016){Jim{\'e}nez-Serra}, {Vasyunin},
  {Caselli}, {Marcelino}, {Billot}, {Viti}, {Testi}, {Vastel}, {Lefloch}, \&
  {Bachiller}}]{jimenez2016}
{Jim{\'e}nez-Serra}, I., {Vasyunin}, A.~I., {Caselli}, P., {et~al.} 2016,
  \apjl, 830, L6

\bibitem[{{Jim{\'e}nez-Serra} {et~al.}(2012){Jim{\'e}nez-Serra}, {Zhang},
  {Viti}, {Mart{\'\i}n-Pintado}, \& {de Wit}}]{jimenezserra2012chem}
{Jim{\'e}nez-Serra}, I., {Zhang}, Q., {Viti}, S., {Mart{\'\i}n-Pintado}, J., \&
  {de Wit}, W.~J. 2012, \apj, 753, 34

\bibitem[{{Johnson} {et~al.}(1977){Johnson}, {Lovas}, {Gottlieb}, {Gottlieb},
  {Litvak}, {Guelin}, \& {Thaddeus}}]{johnson1977}
{Johnson}, D.~R., {Lovas}, F.~J., {Gottlieb}, C.~A., {et~al.} 1977, \apj, 218,
  370

\bibitem[{{J{\o}rgensen} {et~al.}(2018){J{\o}rgensen}, {M{\"u}ller}, {Calcutt},
  {Coutens}, {Drozdovskaya}, {{\"O}berg}, {Persson}, {Taquet}, {van Dishoeck},
  \& {Wampfler}}]{jorge2018}
{J{\o}rgensen}, J.~K., {M{\"u}ller}, H.~S.~P., {Calcutt}, H., {et~al.} 2018,
  \aap, 620, A170

\bibitem[{{J{\o}rgensen} {et~al.}(2016){J{\o}rgensen}, {van der Wiel},
  {Coutens}, {Lykke}, {M{\"u}ller}, {van Dishoeck}, {Calcutt}, {Bjerkeli},
  {Bourke}, {Drozdovskaya}, {Favre}, {Fayolle}, {Garrod}, {Jacobsen},
  {{\"O}berg}, {Persson}, \& {Wampfler}}]{Jorge2016}
{J{\o}rgensen}, J.~K., {van der Wiel}, M.~H.~D., {Coutens}, A., {et~al.} 2016,
  \aap, 595, A117

\bibitem[{{Kalenskii} \& {Johansson}(2010)}]{ciccio2010}
{Kalenskii}, S.~V. \& {Johansson}, L.~E.~B. 2010, Astronomy Reports, 54, 1084

\bibitem[{{Kleiner} {et~al.}(1996){Kleiner}, {Lovas}, \&
  {Godefroid}}]{Kleiner1996}
{Kleiner}, I., {Lovas}, F.~J., \& {Godefroid}, M. 1996, Journal of Physical and
  Chemical Reference Data, 25, 1113

\bibitem[{{Kukolich}(1982)}]{kukolich1982}
{Kukolich}, S.~G. 1982, \jcp, 76, 97

\bibitem[{{Kukolich} {et~al.}(1973){Kukolich}, {Ruben}, {Wang}, \&
  {Williams}}]{kukolich1973}
{Kukolich}, S.~G., {Ruben}, D.~J., {Wang}, J.~H.~S., \& {Williams}, J.~R. 1973,
  \jcp, 58, 3155

\bibitem[{{Lees} \& {Baker}(1968)}]{lees1968}
{Lees}, R.~M. \& {Baker}, J.~G. 1968, \jcp, 48, 5299

\bibitem[{{Lefloch} {et~al.}(2017){Lefloch}, {Ceccarelli}, {Codella}, {Favre},
  {Podio}, {Vastel}, {Viti}, \& {Bachiller}}]{lefloch2017}
{Lefloch}, B., {Ceccarelli}, C., {Codella}, C., {et~al.} 2017, \mnras, 469, L73

\bibitem[{{L{\'o}pez-Sepulcre} {et~al.}(2017){L{\'o}pez-Sepulcre}, {Sakai},
  {Neri}, {Imai}, {Oya}, {Ceccarelli}, {Higuchi}, {Aikawa}, {Bottinelli},
  {Caux}, {Hirota}, {Kahane}, {Lefloch}, {Vastel}, {Watanabe}, \&
  {Yamamoto}}]{lopezsepulcre2017}
{L{\'o}pez-Sepulcre}, A., {Sakai}, N., {Neri}, R., {et~al.} 2017, \aap, 606,
  A121

\bibitem[{{Lovas} {et~al.}(1979){Lovas}, {Lutz}, \& {Dreizler}}]{lovaslutz1979}
{Lovas}, F.~J., {Lutz}, H., \& {Dreizler}, H. 1979, Journal of Physical and
  Chemical Reference Data, 8, 1051

\bibitem[{{Lykke} {et~al.}(2017){Lykke}, {Coutens}, {J{\o}rgensen}, {van der
  Wiel}, {Garrod}, {M{\"u}ller}, {Bjerkeli}, {Bourke}, {Calcutt},
  {Drozdovskaya}, {Favre}, {Fayolle}, {Jacobsen}, {{\"O}berg}, {Persson}, {van
  Dishoeck}, \& {Wampfler}}]{Lykke2017}
{Lykke}, J.~M., {Coutens}, A., {J{\o}rgensen}, J.~K., {et~al.} 2017, \aap, 597,
  A53

\bibitem[{{M{\"a}der} {et~al.}(1974){M{\"a}der}, {Heise}, \&
  {Dreizler}}]{maeder1974}
{M{\"a}der}, H., {Heise}, H.~M., \& {Dreizler}, H. 1974, Zeitschrift
  Naturforschung Teil A, 29, 164

\bibitem[{{Manigand} {et~al.}(2020){Manigand}, {J{\o}rgensen}, {Calcutt},
  {M{\"u}ller}, {Ligterink}, {Coutens}, {Drozdovskaya}, {van Dishoeck}, \&
  {Wampfler}}]{manigand2020}
{Manigand}, S., {J{\o}rgensen}, J.~K., {Calcutt}, H., {et~al.} 2020, \aap, 635,
  A48

\bibitem[{{Mart{\'\i}n} {et~al.}(2019){Mart{\'\i}n}, {Mart{\'\i}n-Pintado},
  {Blanco-S{\'a}nchez}, {Rivilla}, {Rodr{\'\i}guez-Franco}, \&
  {Rico-Villas}}]{martin2019}
{Mart{\'\i}n}, S., {Mart{\'\i}n-Pintado}, J., {Blanco-S{\'a}nchez}, C.,
  {et~al.} 2019, \aap, 631, A159

\bibitem[{{Mart{\'\i}n} {et~al.}(2008){Mart{\'\i}n}, {Requena-Torres},
  {Mart{\'\i}n-Pintado}, \& {Mauersberger}}]{martin2008NH2G0.69}
{Mart{\'\i}n}, S., {Requena-Torres}, M.~A., {Mart{\'\i}n-Pintado}, J., \&
  {Mauersberger}, R. 2008, \apj, 678, 245

\bibitem[{{Matsushima} {et~al.}(1994){Matsushima}, {Evenson}, \&
  {Zink}}]{Matsushima1994}
{Matsushima}, F., {Evenson}, K.~M., \& {Zink}, L.~R. 1994, Journal of Molecular
  Spectroscopy, 164, 517

\bibitem[{{Mayen-Gijon} {et~al.}(2014){Mayen-Gijon}, {Anglada}, {Osorio},
  {Rodr{\'\i}guez}, {Lizano}, {G{\'o}mez}, \&
  {Carrasco-Gonz{\'a}lez}}]{may2014}
{Mayen-Gijon}, J.~M., {Anglada}, G., {Osorio}, M., {et~al.} 2014, \mnras, 437,
  3766

\bibitem[{{McMullin} {et~al.}(2007){McMullin}, {Waters}, {Schiebel}, {Young},
  \& {Golap}}]{CASAmcmullin}
{McMullin}, J.~P., {Waters}, B., {Schiebel}, D., {Young}, W., \& {Golap}, K.
  2007, Astronomical Society of the Pacific Conference Series, Vol. 376, {CASA
  Architecture and Applications}, ed. R.~A. {Shaw}, F.~{Hill}, \& D.~J. {Bell},
  127

\bibitem[{{Mininni} {et~al.}(2020){Mininni}, {Beltr{\'a}n}, {Rivilla},
  {S{\'a}nchez-Monge}, {Fontani}, {M{\"o}ller}, {Cesaroni}, {Schilke}, {Viti},
  {Jim{\'e}nez-Serra}, {Colzi}, {Lorenzani}, \& {Testi}}]{mininni2020}
{Mininni}, C., {Beltr{\'a}n}, M.~T., {Rivilla}, V.~M., {et~al.} 2020, \aap,
  644, A84

\bibitem[{{Mininni} {et~al.}(2018){Mininni}, {Fontani}, {Rivilla},
  {Beltr{\'a}n}, {Caselli}, \& {Vasyunin}}]{mininni2018}
{Mininni}, C., {Fontani}, F., {Rivilla}, V.~M., {et~al.} 2018, \mnras, 476, L39

\bibitem[{{M{\"u}ller} {et~al.}(2008){M{\"u}ller}, {Belloche}, {Menten},
  {Comito}, \& {Schilke}}]{Muellerbelloche2008}
{M{\"u}ller}, H. S.~P., {Belloche}, A., {Menten}, K.~M., {Comito}, C., \&
  {Schilke}, P. 2008, Journal of Molecular Spectroscopy, 251, 319

\bibitem[{{M{\"u}ller} {et~al.}(2016){M{\"u}ller}, {Belloche}, {Xu}, {Lees},
  {Garrod}, {Walters}, {van Wijngaarden}, {Lewen}, {Schlemmer}, \&
  {Menten}}]{Muller2016correctionfactor}
{M{\"u}ller}, H. S.~P., {Belloche}, A., {Xu}, L.-H., {et~al.} 2016, \aap, 587,
  A92

\bibitem[{{M{\"u}ller} {et~al.}(2015){M{\"u}ller}, {Brown}, {Drouin},
  {Pearson}, {Kleiner}, {Sams}, {Sung}, {Ordu}, \& {Lewen}}]{muller2015}
{M{\"u}ller}, H. S.~P., {Brown}, L.~R., {Drouin}, B.~J., {et~al.} 2015, Journal
  of Molecular Spectroscopy, 312, 22

\bibitem[{{M{\"u}ller} \& {Christen}(2004)}]{muller2004gGg}
{M{\"u}ller}, H. S.~P. \& {Christen}, D. 2004, Journal of Molecular
  Spectroscopy, 228, 298

\bibitem[{{M{\"u}ller} {et~al.}(2009){M{\"u}ller}, {Drouin}, \&
  {Pearson}}]{muellerdrouin2009}
{M{\"u}ller}, H.~S.~P., {Drouin}, B.~J., \& {Pearson}, J.~C. 2009, \aap, 506,
  1487

\bibitem[{{M{\"u}ller} {et~al.}(2004){M{\"u}ller}, {Menten}, \&
  {M{\"a}der}}]{mueller2004}
{M{\"u}ller}, H.~S.~P., {Menten}, K.~M., \& {M{\"a}der}, H. 2004, \aap, 428,
  1019

\bibitem[{{M{\"u}ller} {et~al.}(2005){M{\"u}ller}, {Schl{\"o}der}, {Stutzki},
  \& {Winnewisser}}]{cdms2005}
{M{\"u}ller}, H. S.~P., {Schl{\"o}der}, F., {Stutzki}, J., \& {Winnewisser}, G.
  2005, Journal of Molecular Structure, 742, 215

\bibitem[{{M{\"u}ller} {et~al.}(2001){M{\"u}ller}, {Thorwirth}, {Roth}, \&
  {Winnewisser}}]{cdms2001}
{M{\"u}ller}, H.~S.~P., {Thorwirth}, S., {Roth}, D.~A., \& {Winnewisser}, G.
  2001, \aap, 370, L49

\bibitem[{{Nazari} {et~al.}(2021){Nazari}, {van Gelder}, {van Dishoeck},
  {Tabone}, {van't Hoff}, {Ligterink}, {Beuther}, {Boogert}, {Caratti o
  Garatti}, {Klaassen}, {Linnartz}, {Taquet}, \& {Tychoniec}}]{nazari2021}
{Nazari}, P., {van Gelder}, M.~L., {van Dishoeck}, E.~F., {et~al.} 2021, \aap,
  650, A150

\bibitem[{{Neustock} {et~al.}(1990){Neustock}, {Guarnieri}, {Demaison}, \&
  {Wlodarczak}}]{Neustock1990}
{Neustock}, W., {Guarnieri}, A., {Demaison}, J., \& {Wlodarczak}, G. 1990,
  Zeitschrift Naturforschung Teil A, 45, 702

\bibitem[{{Odashima} {et~al.}(1995){Odashima}, {Matsushima}, {Nagai},
  {Tsunekawa}, \& {Takagi}}]{odashima1995}
{Odashima}, H., {Matsushima}, F., {Nagai}, K., {Tsunekawa}, S., \& {Takagi}, K.
  1995, Journal of Molecular Spectroscopy, 173, 404

\bibitem[{{Oldag} \& {Sutter}(1992)}]{Oldag1992}
{Oldag}, F. \& {Sutter}, D.~H. 1992, Zeitschrift Naturforschung Teil A, 47, 527

\bibitem[{{Osorio} {et~al.}(2009){Osorio}, {Anglada}, {Lizano}, \&
  {D'Alessio}}]{oso2009}
{Osorio}, M., {Anglada}, G., {Lizano}, S., \& {D'Alessio}, P. 2009, \apj, 694,
  29

\bibitem[{{Pagani} {et~al.}(2019){Pagani}, {Bergin}, {Goldsmith}, {Melnick},
  {Snell}, \& {Favre}}]{pagani2019}
{Pagani}, L., {Bergin}, E., {Goldsmith}, P.~F., {et~al.} 2019, \aap, 624, L5

\bibitem[{{Palau} {et~al.}(2011){Palau}, {Fuente}, {Girart}, {Fontani},
  {Boissier}, {Pi{\'e}tu}, {S{\'a}nchez-Monge}, {Busquet}, {Estalella},
  {Zapata}, {Zhang}, {Neri}, {Ho}, {Alonso-Albi}, \& {Audard}}]{palau2011}
{Palau}, A., {Fuente}, A., {Girart}, J.~M., {et~al.} 2011, \apjl, 743, L32

\bibitem[{{Pearson} {et~al.}(2008){Pearson}, {Brauer}, \&
  {Drouin}}]{pearson2008}
{Pearson}, J.~C., {Brauer}, C.~S., \& {Drouin}, B.~J. 2008, Journal of
  Molecular Spectroscopy, 251, 394

\bibitem[{{Pearson} \& {Mueller}(1996)}]{pearson1996}
{Pearson}, J.~C. \& {Mueller}, H.~S.~P. 1996, \apj, 471, 1067

\bibitem[{{Pearson} {et~al.}(1994){Pearson}, {Sastry}, {Herbst}, \& {De
  Lucia}}]{pearson1994}
{Pearson}, J.~C., {Sastry}, K.~V.~L.~N., {Herbst}, E., \& {De Lucia}, F.~C.
  1994, \apjs, 93, 589

\bibitem[{{Pearson} {et~al.}(1995){Pearson}, {Sastry}, {Winnewisser}, {Herbst},
  \& {De Lucia}}]{pearson1995}
{Pearson}, J.~C., {Sastry}, K.~V.~L.~N., {Winnewisser}, M., {Herbst}, E., \&
  {De Lucia}, F.~C. 1995, Journal of Physical and Chemical Reference Data, 24,
  1

\bibitem[{{Peng} {et~al.}(2013){Peng}, {Despois}, {Brouillet}, {Baudry},
  {Favre}, {Remijan}, {Wootten}, {Wilson}, {Combes}, \&
  {Wlodarczak}}]{peng2013}
{Peng}, T.~C., {Despois}, D., {Brouillet}, N., {et~al.} 2013, \aap, 554, A78

\bibitem[{{Peng} {et~al.}(2022){Peng}, {Liu}, {Qin}, {Baug}, {Liu}, {Wang},
  {Garay}, {Zhang}, {Chen}, {Lee}, {Juvela}, {Li}, {Tatematsu}, {Liu}, {Lee},
  {Luo}, {Dewangan}, {Wu}, {Zhang}, {Bronfman}, {Ge}, {Tang}, {Zhang}, {Xu},
  {Wang}, \& {Zhou}}]{peng2022}
{Peng}, Y., {Liu}, T., {Qin}, S.-L., {et~al.} 2022, \mnras, 512, 4419

\bibitem[{{Peter} \& {Dreizler}(1965)}]{peter1965}
{Peter}, R. \& {Dreizler}, H. 1965, Zeitschrift Naturforschung Teil A, 20, 301

\bibitem[{{Pickett} {et~al.}(1981){Pickett}, {Cohen}, {Brinza}, \&
  {Schaefer}}]{pickett1981}
{Pickett}, H.~M., {Cohen}, E.~A., {Brinza}, D.~E., \& {Schaefer}, M.~M. 1981,
  Journal of Molecular Spectroscopy, 89, 542

\bibitem[{{Pickett} {et~al.}(1998){Pickett}, {Poynter}, {Cohen}, {Delitsky},
  {Pearson}, \& {M{\"u}ller}}]{jpl1998}
{Pickett}, H.~M., {Poynter}, R.~L., {Cohen}, E.~A., {et~al.} 1998, \jqsrt, 60,
  883

\bibitem[{{Predoi-Cross} {et~al.}(1997){Predoi-Cross}, {Lees}, {Lichau},
  {Winnewisser}, \& {Drummond}}]{predoicross1997}
{Predoi-Cross}, A., {Lees}, R.~M., {Lichau}, H., {Winnewisser}, M., \&
  {Drummond}, J.~R. 1997, International Journal of Infrared and Millimeter
  Waves, 18, 2047

\bibitem[{{Qin} {et~al.}(2022){Qin}, {Liu}, {Liu}, {Goldsmith}, {Li}, {Zhang},
  {Liu}, {Wu}, {Bronfman}, {Juvela}, {Lee}, {Garay}, {Zhang}, {He}, {Hsu},
  {Shen}, {Lee}, {Wang}, {Tang}, {Tang}, {Zhang}, {Yue}, {Xue}, {Li}, {Peng},
  {Dutta}, {Ge}, {Xu}, {Chen}, {Baug}, {Dewangan}, \& {Tej}}]{qin2022}
{Qin}, S.-L., {Liu}, T., {Liu}, X., {et~al.} 2022, \mnras, 511, 3463

\bibitem[{{Remijan} {et~al.}(2004){Remijan}, {Shiao}, {Friedel}, {Meier}, \&
  {Snyder}}]{remijan2004}
{Remijan}, A., {Shiao}, Y.~S., {Friedel}, D.~N., {Meier}, D.~S., \& {Snyder},
  L.~E. 2004, \apj, 617, 384

\bibitem[{{Requena-Torres} {et~al.}(2008){Requena-Torres},
  {Mart{\'\i}n-Pintado}, {Mart{\'\i}n}, \& {Morris}}]{requena-torres2008}
{Requena-Torres}, M.~A., {Mart{\'\i}n-Pintado}, J., {Mart{\'\i}n}, S., \&
  {Morris}, M.~R. 2008, \apj, 672, 352

\bibitem[{{Requena-Torres} {et~al.}(2006){Requena-Torres},
  {Mart{\'\i}n-Pintado}, {Rodr{\'\i}guez-Franco}, {Mart{\'\i}n},
  {Rodr{\'\i}guez-Fern{\'a}ndez}, \& {de Vicente}}]{requena-torres2006}
{Requena-Torres}, M.~A., {Mart{\'\i}n-Pintado}, J., {Rodr{\'\i}guez-Franco},
  A., {et~al.} 2006, \aap, 455, 971

\bibitem[{{Richard} {et~al.}(2012){Richard}, {Margul{\`e}s}, {Motiyenko}, \&
  {Guillemin}}]{richard2012}
{Richard}, C., {Margul{\`e}s}, L., {Motiyenko}, R.~A., \& {Guillemin}, J.~C.
  2012, \aap, 543, A135

\bibitem[{{Rivilla} {et~al.}(2017){Rivilla}, {Beltr{\'a}n}, {Cesaroni},
  {Fontani}, {Codella}, \& {Zhang}}]{rivilla2017a}
{Rivilla}, V.~M., {Beltr{\'a}n}, M.~T., {Cesaroni}, R., {et~al.} 2017, \aap,
  598, A59

\bibitem[{{Rivilla} {et~al.}(2022){Rivilla}, {Colzi}, {Jim{\'e}nez-Serra},
  {Mart{\'\i}n-Pintado}, {Meg{\'\i}as}, {Melosso}, {Bizzocchi},
  {L{\'o}pez-Gallifa}, {Mart{\'\i}nez-Henares}, {Massalkhi}, {Tercero}, {de
  Vicente}, {Guillemin}, {Garc{\'\i}a de la Concepci{\'o}n}, {Rico-Villas},
  {Zeng}, {Mart{\'\i}n}, {Requena-Torres}, {Tonolo}, {Alessandrini}, {Dore},
  {Barone}, \& {Puzzarini}}]{rivilla2022}
{Rivilla}, V.~M., {Colzi}, L., {Jim{\'e}nez-Serra}, I., {et~al.} 2022, \apjl,
  929, L11

\bibitem[{{Rodr{\'\i}guez-Almeida} {et~al.}(2021){Rodr{\'\i}guez-Almeida},
  {Jim{\'e}nez-Serra}, {Rivilla}, {Mart{\'\i}n-Pintado}, {Zeng}, {Tercero}, {de
  Vicente}, {Colzi}, {Rico-Villas}, {Mart{\'\i}n}, \&
  {Requena-Torres}}]{rodriguez-almenida2021}
{Rodr{\'\i}guez-Almeida}, L.~F., {Jim{\'e}nez-Serra}, I., {Rivilla}, V.~M.,
  {et~al.} 2021, \apjl, 912, L11

\bibitem[{{S{\'a}nchez-Monge} {et~al.}(2018){S{\'a}nchez-Monge}, {Schilke},
  {Ginsburg}, {Cesaroni}, \& {Schmiedeke}}]{alvaro2018}
{S{\'a}nchez-Monge}, {\'A}., {Schilke}, P., {Ginsburg}, A., {Cesaroni}, R., \&
  {Schmiedeke}, A. 2018, \aap, 609, A101

\bibitem[{{Sanz-Novo} {et~al.}(2022){Sanz-Novo}, {Belloche}, {Rivilla},
  {Garrod}, {Alonso}, {Redondo}, {Barrientos}, {Kolesnikov{\'a}}, {Valle},
  {Rodr{\'\i}guez-Almeida}, {Jimenez-Serra}, {Mart{\'\i}n-Pintado},
  {M{\"u}ller}, \& {Menten}}]{sanz-novo2022}
{Sanz-Novo}, M., {Belloche}, A., {Rivilla}, V.~M., {et~al.} 2022, \aap, 666,
  A114

\bibitem[{{Sastry} {et~al.}(1984){Sastry}, {Lees}, \& {De Lucia}}]{sastry1984}
{Sastry}, K.~V.~L.~N., {Lees}, R.~M., \& {De Lucia}, F.~C. 1984, Journal of
  Molecular Spectroscopy, 103, 486

\bibitem[{{Sipil{\"a}} {et~al.}(2021){Sipil{\"a}}, {Silsbee}, \&
  {Caselli}}]{sipila2021}
{Sipil{\"a}}, O., {Silsbee}, K., \& {Caselli}, P. 2021, \apj, 922, 126

\bibitem[{{Stolze} \& {Sutter}(1985)}]{stolze1985}
{Stolze}, M. \& {Sutter}, D.~H. 1985, Zeitschrift Naturforschung Teil A, 40,
  998

\bibitem[{{Suzuki} {et~al.}(2018){Suzuki}, {Majumdar}, {Ohishi}, {Saito},
  {Hirota}, \& {Wakelam}}]{suzuki2018}
{Suzuki}, T., {Majumdar}, L., {Ohishi}, M., {et~al.} 2018, \apj, 863, 51

\bibitem[{{Taquet} {et~al.}(2019){Taquet}, {Bianchi}, {Codella}, {Persson},
  {Ceccarelli}, {Cabrit}, {J{\o}rgensen}, {Kahane}, {L{\'o}pez-Sepulcre}, \&
  {Neri}}]{taquet2019}
{Taquet}, V., {Bianchi}, E., {Codella}, C., {et~al.} 2019, \aap, 632, A19

\bibitem[{{Tercero} {et~al.}(2018){Tercero}, {Cuadrado}, {L{\'o}pez},
  {Brouillet}, {Despois}, \& {Cernicharo}}]{tercero2018}
{Tercero}, B., {Cuadrado}, S., {L{\'o}pez}, A., {et~al.} 2018, \aap, 620, L6

\bibitem[{{Vacherand} {et~al.}(1986){Vacherand}, {Van Eijck}, {Burie}, \&
  {Demaison}}]{vacherand1986}
{Vacherand}, J.~M., {Van Eijck}, B.~P., {Burie}, J., \& {Demaison}, J. 1986,
  Journal of Molecular Spectroscopy, 118, 355

\bibitem[{{van der Walt} {et~al.}(2021){van der Walt}, {Kristensen},
  {J{\o}rgensen}, {Calcutt}, {Manigand}, {el Akel}, {Garrod}, \&
  {Qiu}}]{vanderWalt2021}
{van der Walt}, S.~J., {Kristensen}, L.~E., {J{\o}rgensen}, J.~K., {et~al.}
  2021, \aap, 655, A86

\bibitem[{{van Gelder} {et~al.}(2020){van Gelder}, {Tabone}, {Tychoniec}, {van
  Dishoeck}, {Beuther}, {Boogert}, {Caratti o Garatti}, {Klaassen}, {Linnartz},
  {M{\"u}ller}, \& {Taquet}}]{vangelder2020}
{van Gelder}, M.~L., {Tabone}, B., {Tychoniec}, {\L}., {et~al.} 2020, \aap,
  639, A87

\bibitem[{{Viti} {et~al.}(2004){Viti}, {Collings}, {Dever}, {McCoustra}, \&
  {Williams}}]{Viti2004}
{Viti}, S., {Collings}, M.~P., {Dever}, J.~W., {McCoustra}, M. R.~S., \&
  {Williams}, D.~A. 2004, \mnras, 354, 1141

\bibitem[{{Viti} \& {Williams}(1999)}]{viti1999}
{Viti}, S. \& {Williams}, D.~A. 1999, \mnras, 305, 755

\bibitem[{{Widicus Weaver} \& {Friedel}(2012)}]{WidicusWeaver2012}
{Widicus Weaver}, S.~L. \& {Friedel}, D.~N. 2012, \apjs, 201, 16

\bibitem[{{Widicus Weaver} {et~al.}(2017){Widicus Weaver}, {Laas}, {Zou},
  {Kroll}, {Rad}, {Hays}, {Sanders}, {Lis}, {Cross}, {Wehres}, {McGuire}, \&
  {Sumner}}]{widicusweaver2017}
{Widicus Weaver}, S.~L., {Laas}, J.~C., {Zou}, L., {et~al.} 2017, \apjs, 232, 3

\bibitem[{{Wilson} \& {Rood}(1994)}]{wilsonrood}
{Wilson}, T.~L. \& {Rood}, R. 1994, \araa, 32, 191

\bibitem[{{Wyrowski} {et~al.}(1999){Wyrowski}, {Schilke}, {Walmsley}, \&
  {Menten}}]{wyrowski1999}
{Wyrowski}, F., {Schilke}, P., {Walmsley}, C.~M., \& {Menten}, K.~M. 1999,
  \apjl, 514, L43

\bibitem[{{Xu} {et~al.}(2008){Xu}, {Fisher}, {Lees}, {Shi}, {Hougen},
  {Pearson}, {Drouin}, {Blake}, \& {Braakman}}]{xufisher2008}
{Xu}, L.-H., {Fisher}, J., {Lees}, R.~M., {et~al.} 2008, Journal of Molecular
  Spectroscopy, 251, 305

\bibitem[{{Xu} \& {Lovas}(1997)}]{xulovas1997}
{Xu}, L.-H. \& {Lovas}, F.~J. 1997, Journal of Physical and Chemical Reference
  Data, 26, 17

\bibitem[{{Yan} {et~al.}(2019){Yan}, {Zhang}, {Henkel}, {Mufakharov}, {Jia},
  {Tang}, {Wu}, {Li}, {Zeng}, {Wang}, {Li}, {Huang}, \& {Jian}}]{yan2019}
{Yan}, Y.~T., {Zhang}, J.~S., {Henkel}, C., {et~al.} 2019, \apj, 877, 154

\bibitem[{{Zeng} {et~al.}(2018){Zeng}, {Jim{\'e}nez-Serra}, {Rivilla},
  {Mart{\'\i}n}, {Mart{\'\i}n-Pintado}, {Requena-Torres},
  {Armijos-Abenda{\~n}o}, {Riquelme}, \& {Aladro}}]{zeng2018}
{Zeng}, S., {Jim{\'e}nez-Serra}, I., {Rivilla}, V.~M., {et~al.} 2018, \mnras,
  478, 2962

\bibitem[{{Zernickel} {et~al.}(2012){Zernickel}, {Schilke}, {Schmiedeke},
  {Lis}, {Brogan}, {Ceccarelli}, {Comito}, {Emprechtinger}, {Hunter}, \&
  {M{\"o}ller}}]{zernickel2012}
{Zernickel}, A., {Schilke}, P., {Schmiedeke}, A., {et~al.} 2012, \aap, 546, A87

\end{thebibliography}




\begin{appendix}
\section{Catalog entries documentation for the molecular species analyzed}

\subsection{Methanol CH$_3$OH, $^{13}$CH$_3$OH, and CH$_3^{18}$OH}
We have used the spectroscopy from the CDMS catalog. The entry of CH$_3$OH is based mainly on the work of   \citet{xufisher2008}, with additional data from \citet{lees1968}, \citet{pickett1981}, \citet{sastry1984}, \citet{herbst1984}, \citet{anderson1990}, \citet{Matsushima1994}, \citet{odashima1995}, \citet{belov1995}, and \citet{mueller2004}. More information is available at:\\
\textbf{https://cdms.astro.uni-koeln.de/cgi-bin/cdmsinfo?file=e032504.cat}\\
\indent The entry of $^{13}$CH$_3$OH is based on the review work by \citet{xulovas1997}. More information available at:\\
\textbf{https://cdms.astro.uni-koeln.de/cgi-bin/cdmsinfo?file=e033502.cat}\\
\indent The entry of CH$_3^{18}$OH is based on the work by \citet{fisher2007} with the inclusion of other transitions from \citet{hughes1951}, \citet{gerry1976}, \citet{hoshino1996}, \citet{predoicross1997}, and \citet{Ikeda1998}. More information available at:\\ \textbf{https://cdms.astro.uni-koeln.de/cgi-bin/cdmsinfo?file=e034504.cat}

\subsection{Acetaldehyde CH$_3$CHO}
We have used the spectroscopy from the JPL catalog. The entry is based on the work of \citet{Kleiner1996} and reference therein. More information is available at:\\
\textbf{https://spec.jpl.nasa.gov/ftp/pub/catalog/doc/d044003.pdf}

\subsection{Dimethyl ether CH$_3$OCH$_3$}
We have used the spectroscopy from the CDMS catalog. The entry of CH$_3$OCH$_3$ is based mainly on the work of \citet{endres2009}, \citet{lovaslutz1979}, \citet{Neustock1990}, and \citet{Groner1998}. More information is available at:\\
\textbf{https://cdms.astro.uni-koeln.de/cgi-bin/cdmsinfo?file=e046514.cat}

\subsection{Acetone CH$_3$COCH$_3$}
We have used the spectroscopy from the JPL catalog. The entry is based on the works of \citet{peter1965}, \citet{vacherand1986}, \citet{Oldag1992}, and \citet{groner2002}. More information is available at:\\
\textbf{https://spec.jpl.nasa.gov/ftp/pub/catalog/doc/d058003.pdf}
\subsection{Ethanol C$_2$H$_5$OH}
We have used the spectroscopy from the CDMS catalog. The entry of C$_2$H$_5$OH is based mainly on the work of \citet{pearson1995,pearson1996,pearson2008}. More information is available at:\\
\textbf{https://cdms.astro.uni-koeln.de/cgi-bin/cdmsinfo?file=e046524.cat}

\subsection{Ethylene glycol aGg'-(CH$_2$OH)$_2$ and gGg'-(CH$_2$OH)$_2$}
We have used the spectroscopy from the CDMS catalog. The entries of  aGg'-(CH$_2$OH)$_2$ is based on the works by \citet{christen1995} and \citet{christen2003}. More information is available at:\\
\textbf{https://cdms.astro.uni-koeln.de/cgi-bin/cdmsinfo?file=e062503.cat}\\
\indent The entries of  gGg'-(CH$_2$OH)$_2$ is based on the works by \citet{christen2001} and \citet{muller2004gGg}. More information is available at:\\
\textbf{https://cdms.astro.uni-koeln.de/cgi-bin/cdmsinfo?file=e062504.cat}

\subsection{Methyl cyanide CH$_3$CN, $^{13}$CH$_3$CN, and CH$_3^{13}$CN}
We have used the spectroscopy from the CDMS catalog. The entry of CH$_3$CN  is mainly based on the works by \citet{muller2015}, \citet{kukolich1973,kukolich1982}, \citet{boucher1977}, \citet{cazzoli2006}, \citet{bauer1975}. More information is available at:\\
\textbf{https://cdms.astro.uni-koeln.de/cgi-bin/cdmsinfo?file=e041505.cat}\\
\indent The entry of $^{13}$CH$_3$CN is based on the works by \citet{muellerdrouin2009}, \citet{pearson1996}, and    \citet{demaison1979}. More information is available at:\\
\textbf{https://cdms.astro.uni-koeln.de/cgi-bin/cdmsinfo?file=e042508.cat}\\ 
\indent The entry of CH$_3^{13}$CN is a combined CDMS and JPL entry and is based on the works by \citet{muellerdrouin2009},  \citet{pearson1996},    \citet{demaison1979}, and \citet{kukolich1982}.
More information is available at:\\
\textbf{https://cdms.astro.uni-koeln.de/cgi-bin/cdmsinfo?file=e042509.cat}

\subsection{Vinyl cyanide C$_2$H$_3$CN}
We have used the spectroscopy from the CDMS catalog. The entry of C$_2$H$_3$CN is based mainly on the works of \citet{Muellerbelloche2008}, \citet{gerry1973}, \citet{stolze1985}, \citet{Cazzoli1988}, \citet{Demaison1994}, \citet{Baskakov1996}, and \citet{colmont1997}. More information is available at:\\
\textbf{https://cdms.astro.uni-koeln.de/cgi-bin/cdmsinfo?file=e053515.cat}

\subsection{Ethyl cyanide C$_2$H$_5$CN and C$_2$H$_5^{13}$CN }
We have used the spectroscopy from the CDMS catalog. The entry of C$_2$H$_5$CN is based mainly on the works of \citet{maeder1974}, \citet{johnson1977}, \citet{boucher1980}, \citet{pearson1994}, and \citet{fukuyama1996}. More information is available at:\\
\textbf{https://cdms.astro.uni-koeln.de/cgi-bin/cdmsinfo?file=e055502.cat}\\ \indent 
The entry of C$_2$H$_5^{13}$CN is based mainly on the works of \citet{demyk2007} and \citet{richard2012}. More information is available at:\\
\textbf{https://cdms.astro.uni-koeln.de/cgi-bin/cdmsinfo?file=e056504.cat}

\section{Spectra}
\begin{figure*}
 \centering
\includegraphics[width=0.8\textwidth]{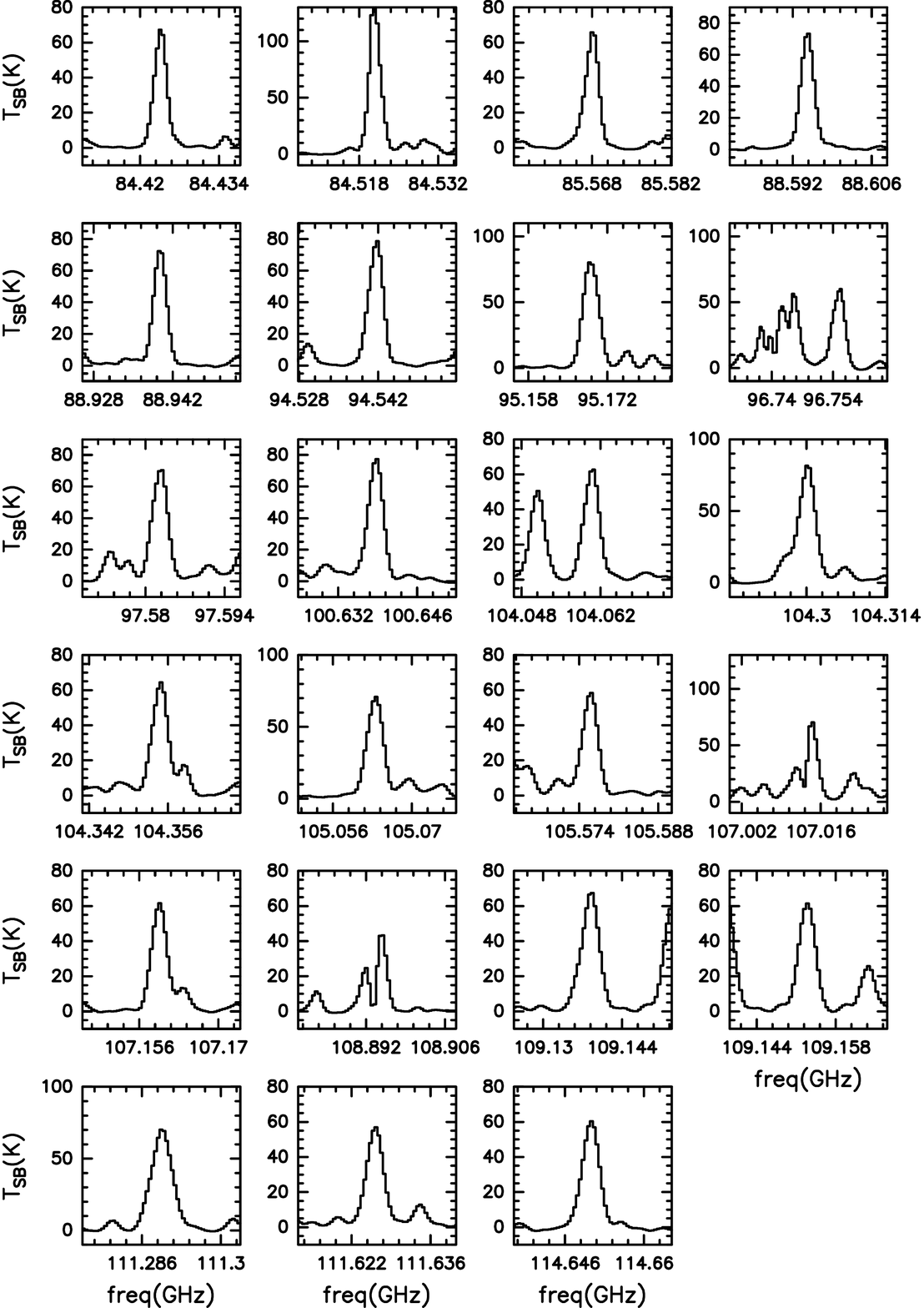}
\caption{Observed spectrum of some of the brightest transitions of
CH$_3$OH\,v$_t = 0$. $T_{\rm{SB}}$ stands for synthesized beam temperature.
}
    \label{fig:spectrach3oh}
\end{figure*}
\begin{figure*}
    \centering
    \includegraphics[width=0.8\textwidth , trim={0, 9cm, 0, 9cm}, clip]{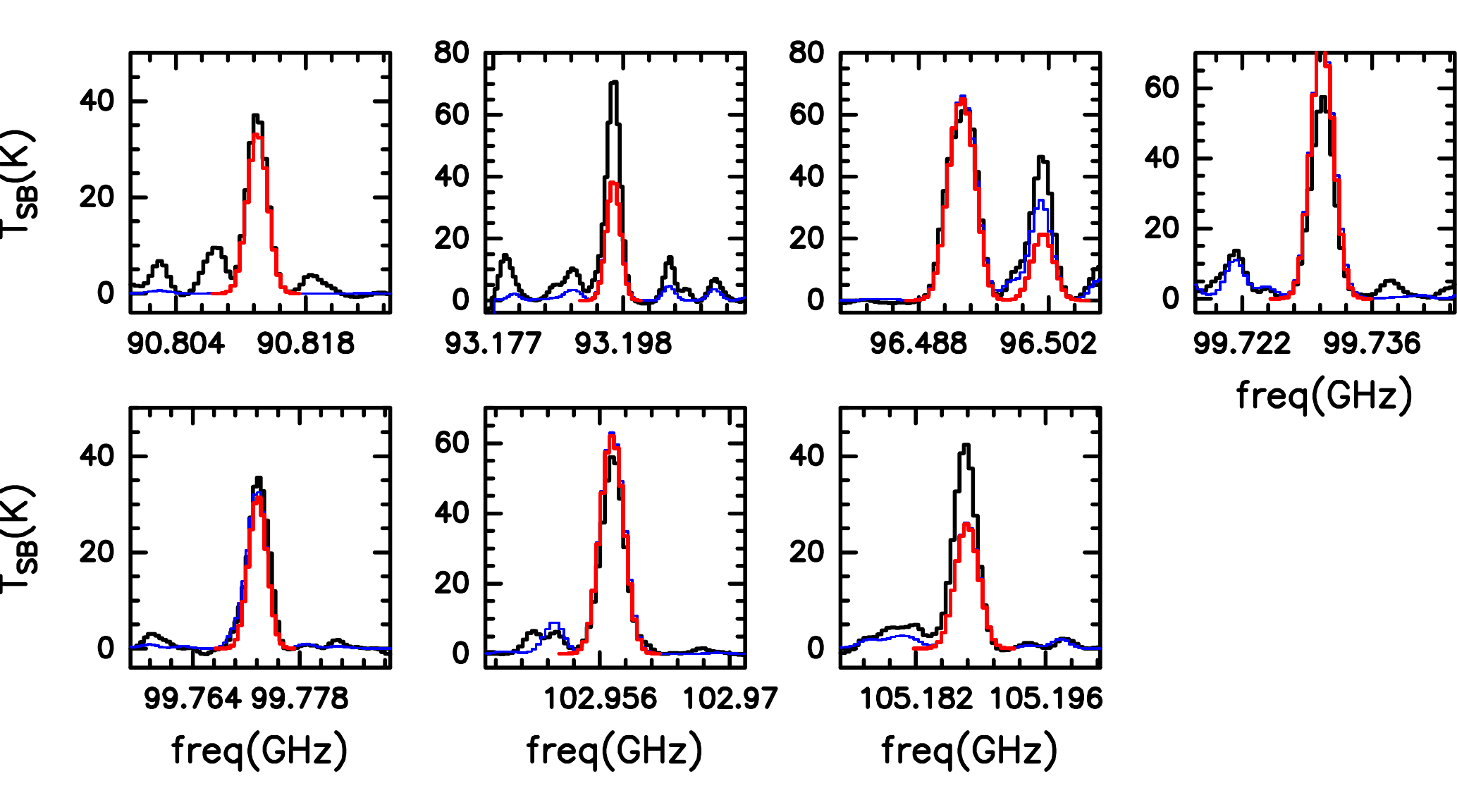}
    \caption{Transitions used to constrain the fit of CH$_3$OH\,v$_t = 1$. In black the
observed spectrum, in red the synthetic spectrum of the best fit for CH$_3$OH\,v$_t = 1$ only,
while in blue the spectrum that takes into account all the species identified in the spectrum, including those published in \citet{mininni2020,colziguapos}. $T_{\rm{SB}}$ stands for synthesized beam temperature.
}
    \label{fig:spectrach3ohvib}
\end{figure*}
\begin{figure*}
    \centering
    \includegraphics[width=0.8\textwidth, trim={0, 7cm, 0, 7cm}, clip]{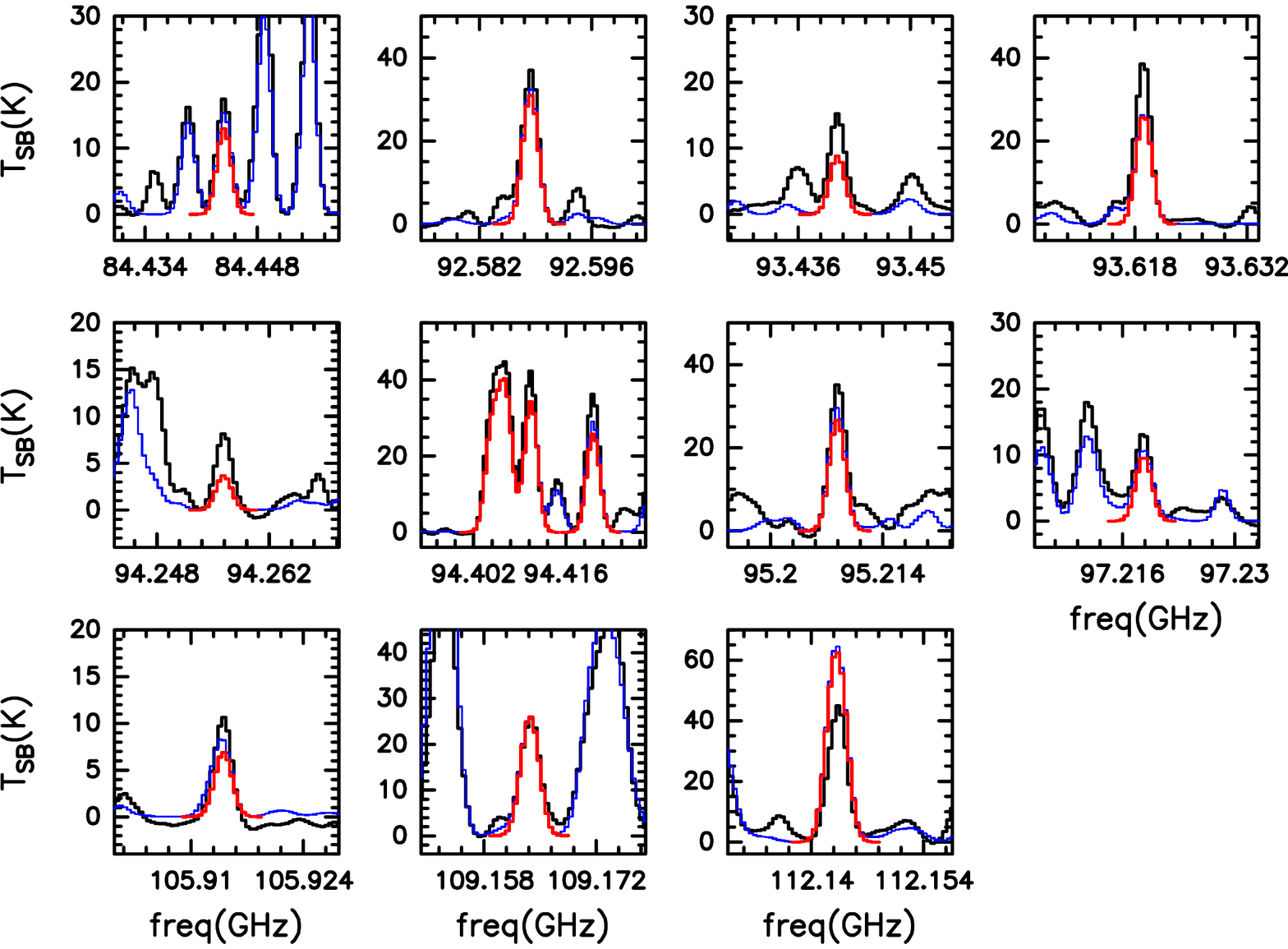}
    \caption{Transitions used to constrain the fit of $^{13}$CH$_3$OH. In black the
observed spectrum, in red the synthetic spectrum of the best fit for $^{13}$CH$_3$OH only,
while in blue the spectrum that takes into account all the species identified in the spectrum, including those published in \citet{mininni2020,colziguapos}. $T_{\rm{SB}}$ stands for synthesized beam temperature.
}
    \label{fig:spectrac13h3oh}
\end{figure*}
\begin{figure*}
    \centering
    \includegraphics[width=0.75\textwidth, trim={0, 7cm, 0, 7cm}, clip]{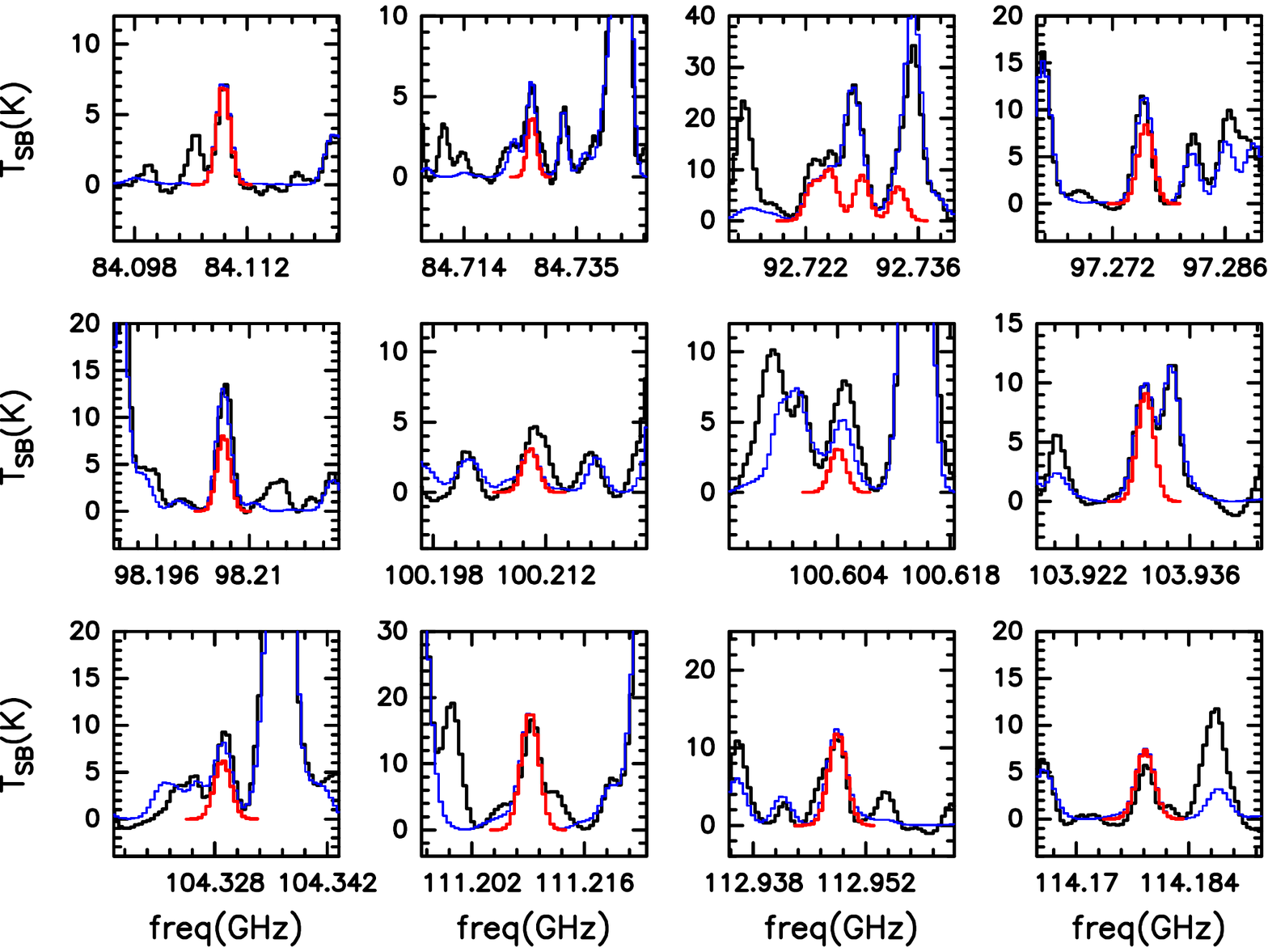}
    \caption{Transitions used to constrain the fit of CH$_3^{18}$OH. In black the
observed spectrum, in red the synthetic spectrum of the best fit for CH$_3^{18}$OH only, while in blue the spectrum that takes into account all the species identified in the spectrum, including those published in \citet{mininni2020,colziguapos}. $T_{\rm{SB}}$ stands for synthesized beam temperature.
}
    \label{fig:spectrach3o18h}
\end{figure*}
\begin{figure*}
    \centering
    \includegraphics[width=0.75\textwidth, trim={0, 7cm, 0, 7cm}, clip]{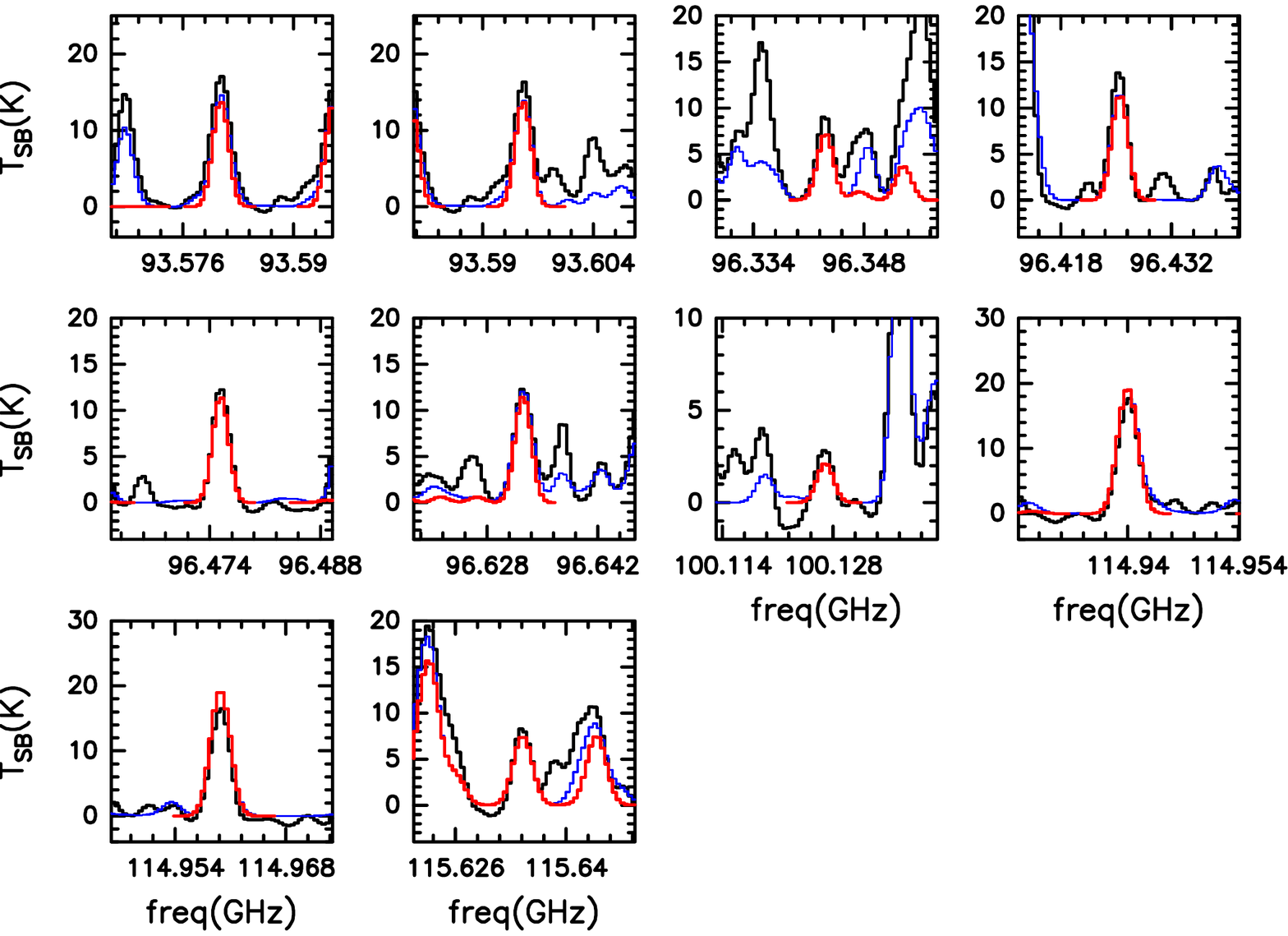}
    \caption{Transitions used to constrain the fit of CH$_3$CHO. In black the
observed spectrum, in red the synthetic spectrum of the best fit for CH$_3$CHO only,
while in blue the spectrum that takes into account all the species identified in the spectrum, including those published in \citet{mininni2020,colziguapos}. $T_{\rm{SB}}$ stands for synthesized beam temperature.
}
    \label{fig:spectrach3cho}
\end{figure*}
\begin{figure*}
    \centering
    \includegraphics[width=0.8\textwidth, trim={0, 1.7cm, 0, 1.7cm}, clip]{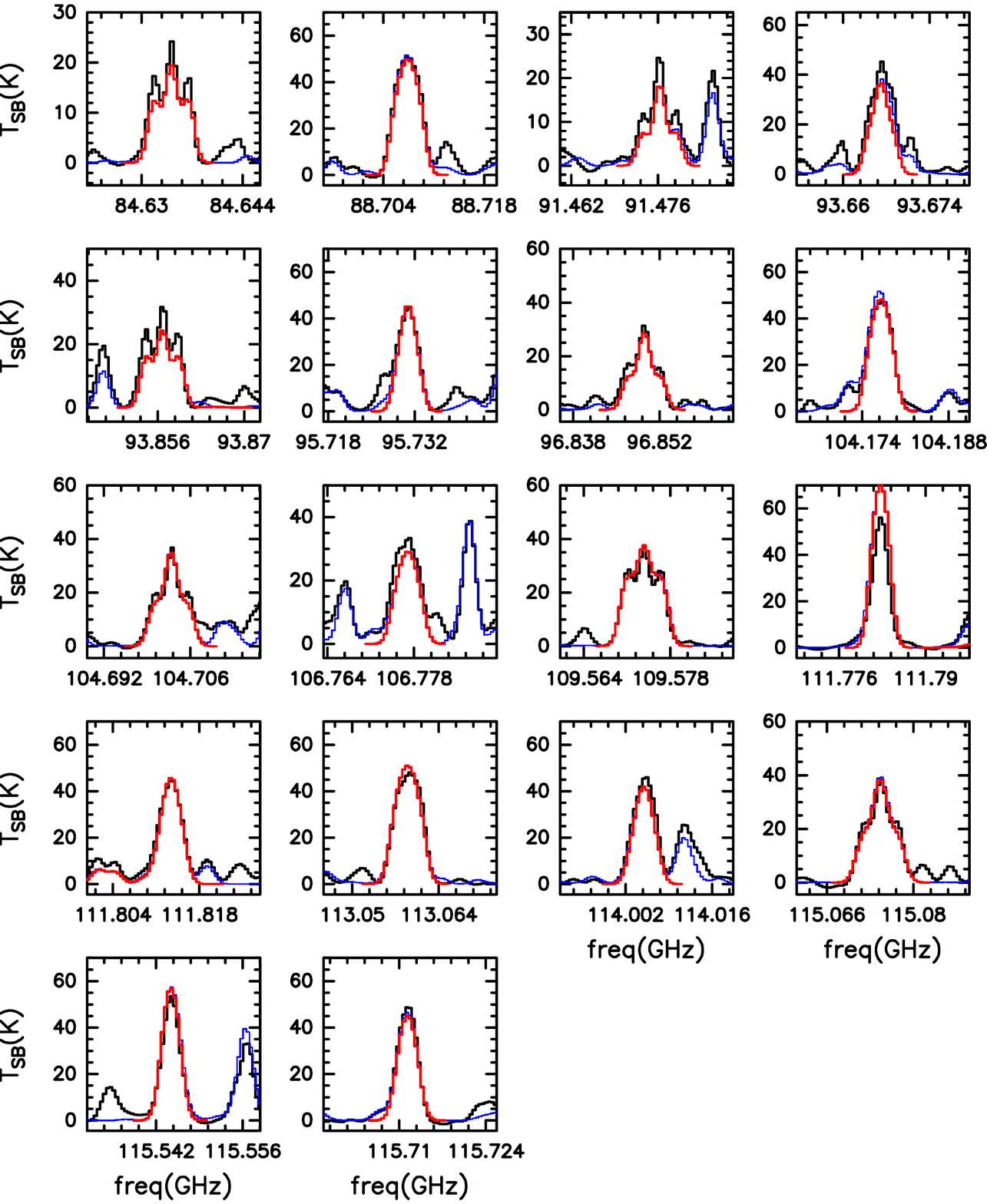}
    \caption{Transitions used to constrain the fit of CH$_3$OCH$_3$. In black the
observed spectrum, in red the synthetic spectrum of the best fit for CH$_3$OCH$_3$ only,
while in blue the spectrum that takes into account all the species identified in the spectrum, including those published in \citet{mininni2020,colziguapos}. $T_{\rm{SB}}$ stands for synthesized beam temperature.
}
    \label{fig:spectrach3och3}
\end{figure*}
\begin{figure*}
    \centering
    \includegraphics[width=0.8\textwidth]{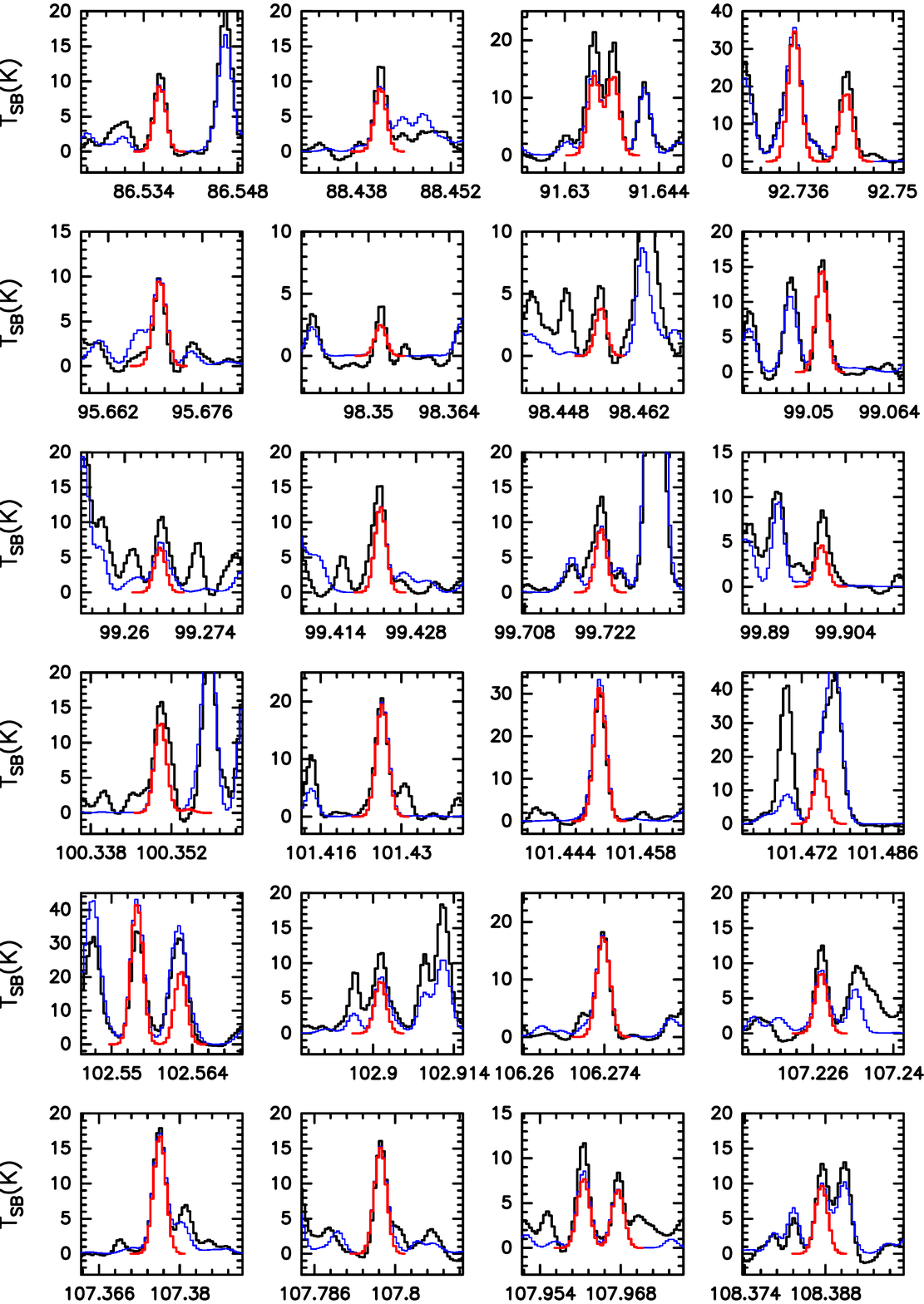}
    \caption{Transitions used to constrain the fit of CH$_3$COCH$_3$. In black the
observed spectrum, in red the synthetic spectrum of the best fit for CH$_3$COCH$_3$ only,
while in blue the spectrum that takes into account all the species identified in the spectrum, including those published in \citet{mininni2020,colziguapos}. $T_{\rm{SB}}$ stands for synthesized beam temperature.
}
    \label{fig:spectrach3coch3part1}
\end{figure*}
\begin{figure*}
    \centering
    \includegraphics[width=0.8\textwidth, trim={0, 7cm, 0, 7cm}, clip]{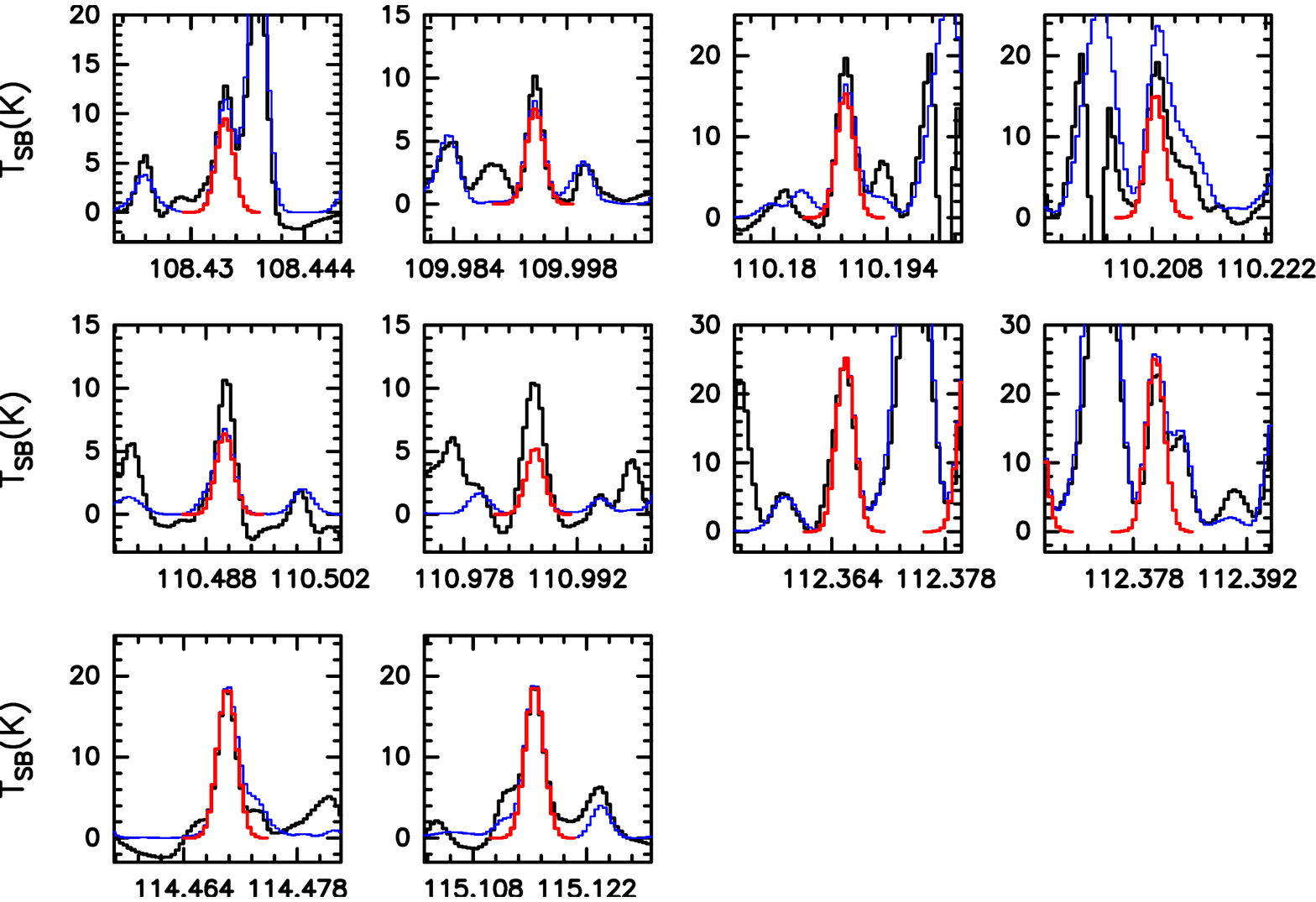}
    \caption{Continue}
    \label{fig:spectrach3coc3part2}
\end{figure*}

\begin{figure*}
    \centering
    \includegraphics[width=0.8\textwidth]{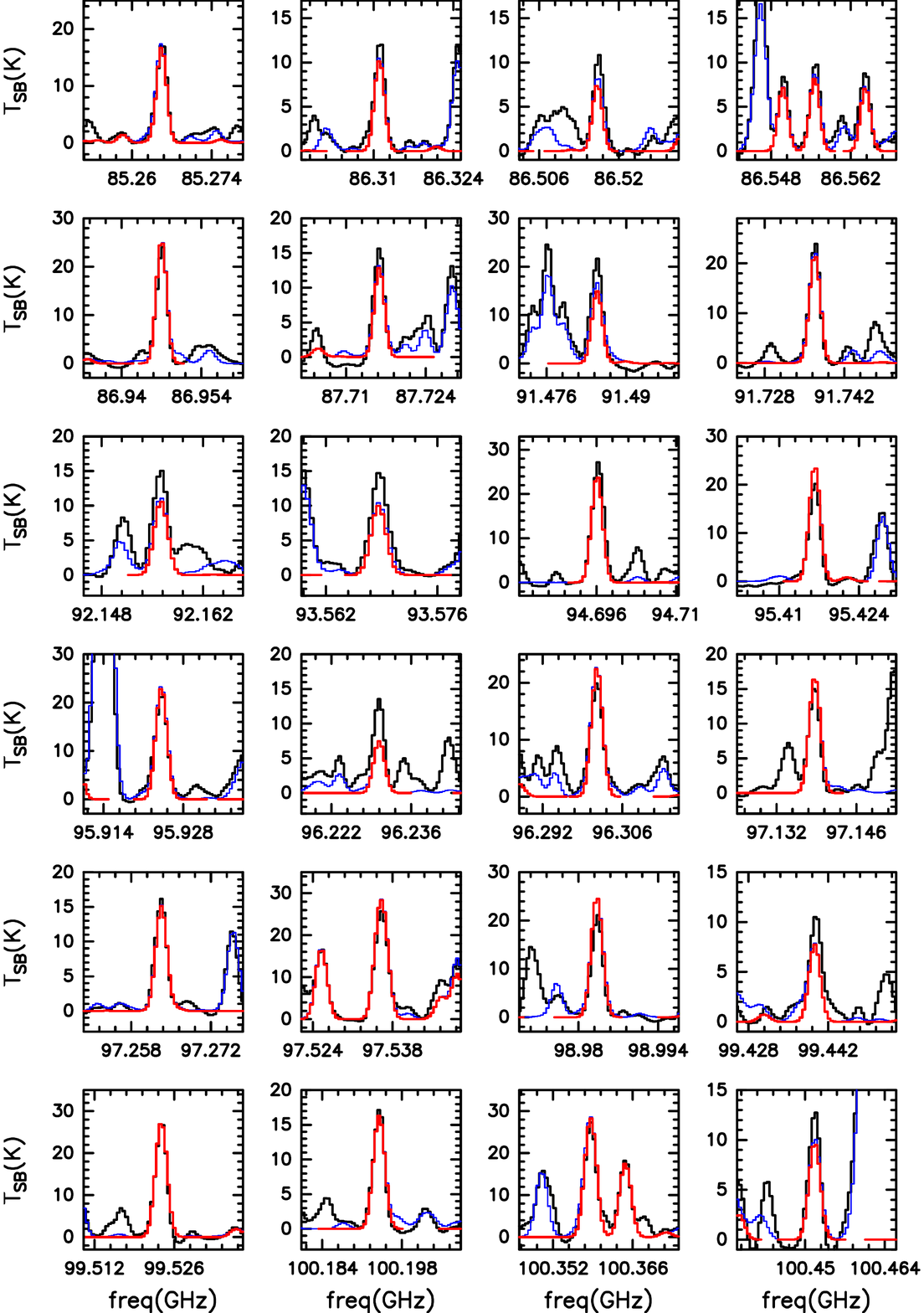}
    \caption{Transitions used to constrain the fit of C$_2$H$_5$OH. In black the
observed spectrum, in red the synthetic spectrum of the best fit for C$_2$H$_5$OH only,
while in blue the spectrum that takes into account all the species identified in the spectrum, including those published in \citet{mininni2020,colziguapos}. $T_{\rm{SB}}$ stands for synthesized beam temperature.
}
    \label{fig:spectrac2h5ohp1}
\end{figure*}
\begin{figure*}
    \centering
    \includegraphics[width=0.8\textwidth, trim={0, 6.5cm, 0, 6.5cm}, clip]{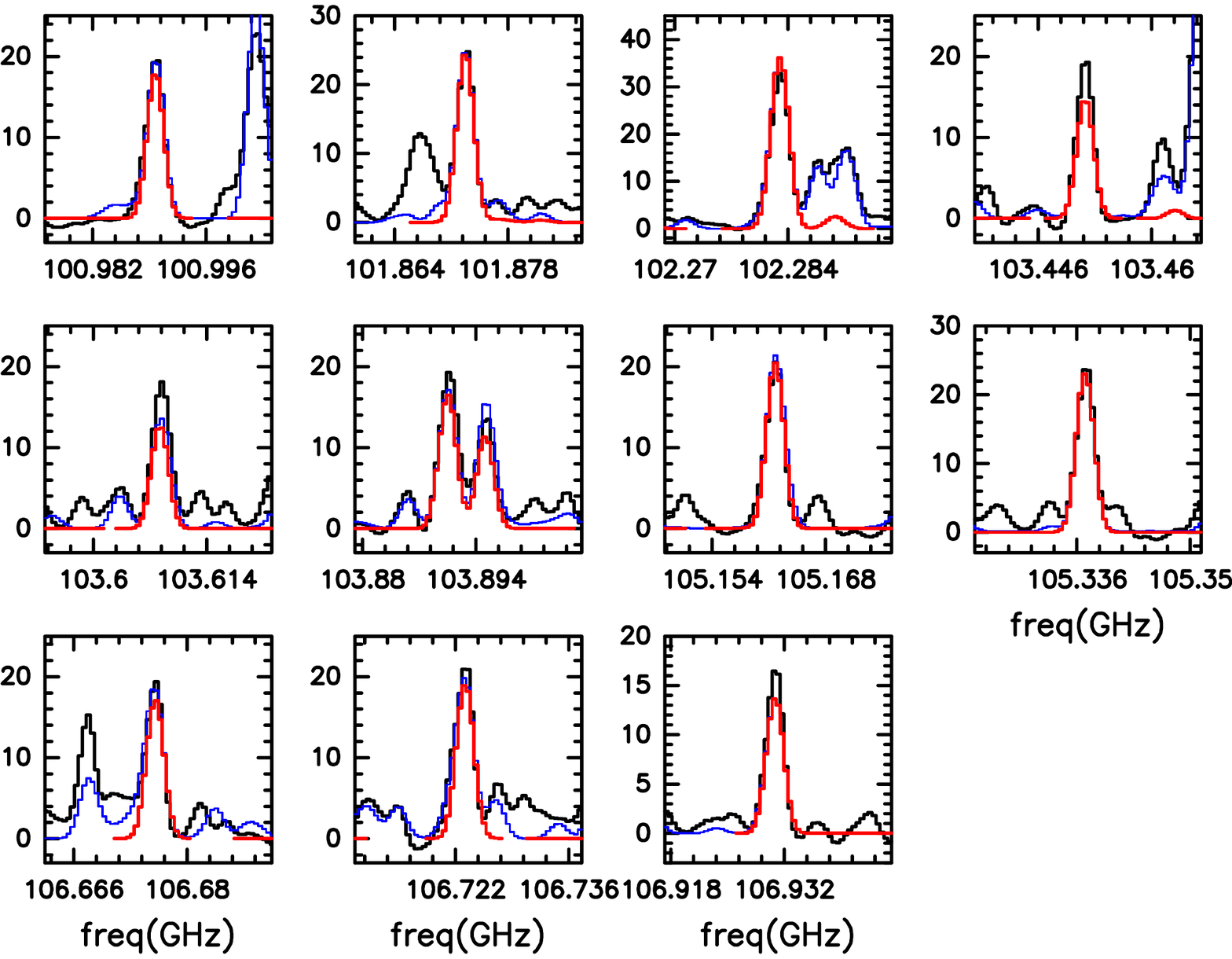}
    \caption{Continue}
    \label{fig:spectrac2h5ohp2}
\end{figure*}

\begin{figure*}
    \centering
    \includegraphics[width=0.8\textwidth, trim={0, 9cm, 0, 9cm}, clip]{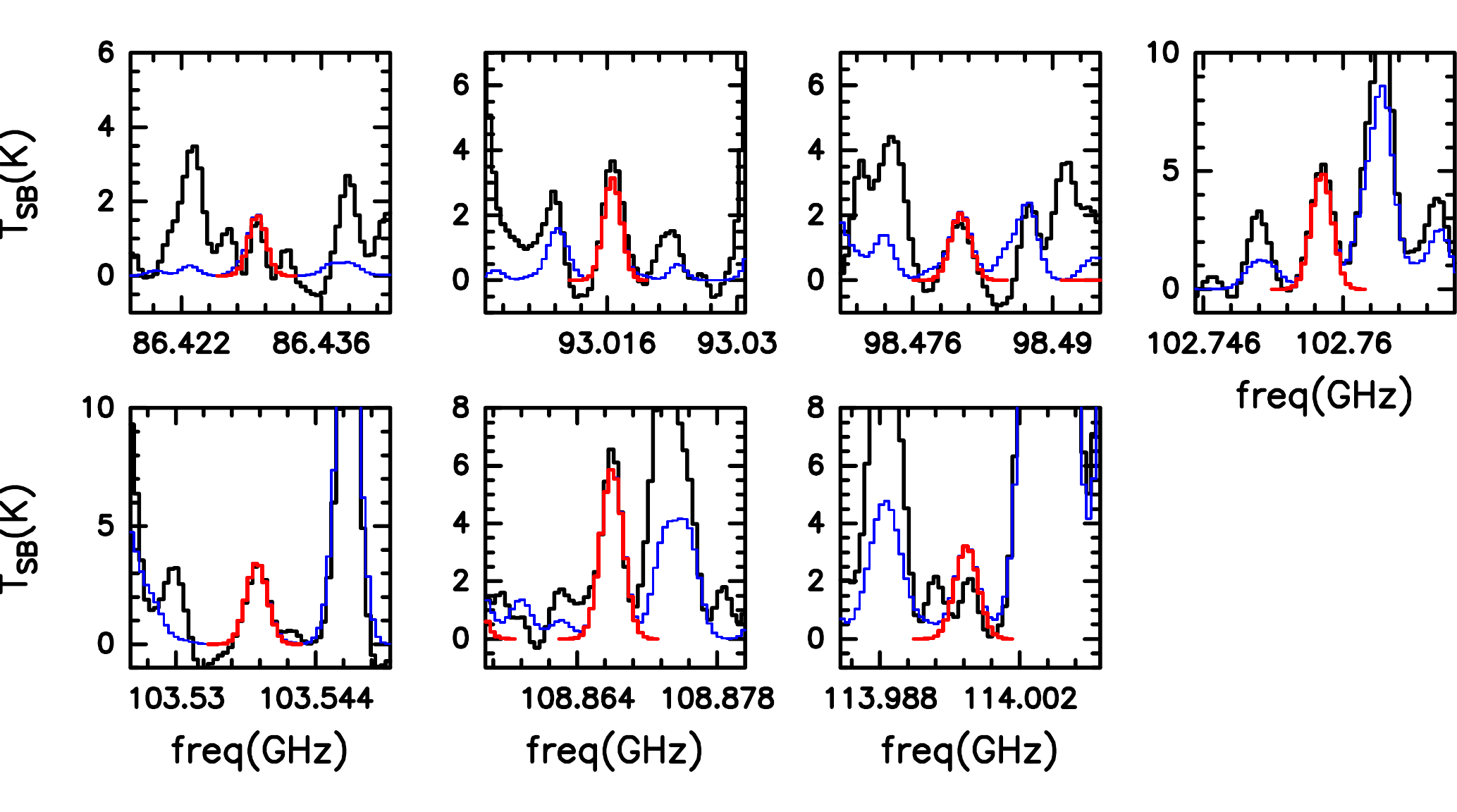}
    \caption{Transitions used to constrain the fit of gGg'-(CH$_2$OH)$_2$. In black the
observed spectrum, in red the synthetic spectrum of the best fit for gGg'-(CH$_2$OH)$_2$ only,
while in blue the spectrum that takes into account all the species identified in the spectrum, including those published in \citet{mininni2020,colziguapos}. $T_{\rm{SB}}$ stands for synthesized beam temperature.
}
    \label{fig:spectra_gGg}
\end{figure*}
\begin{figure*}
    \centering
    \includegraphics[width=0.75\textwidth, trim={0, 2cm, 0, 1.7cm}, clip]{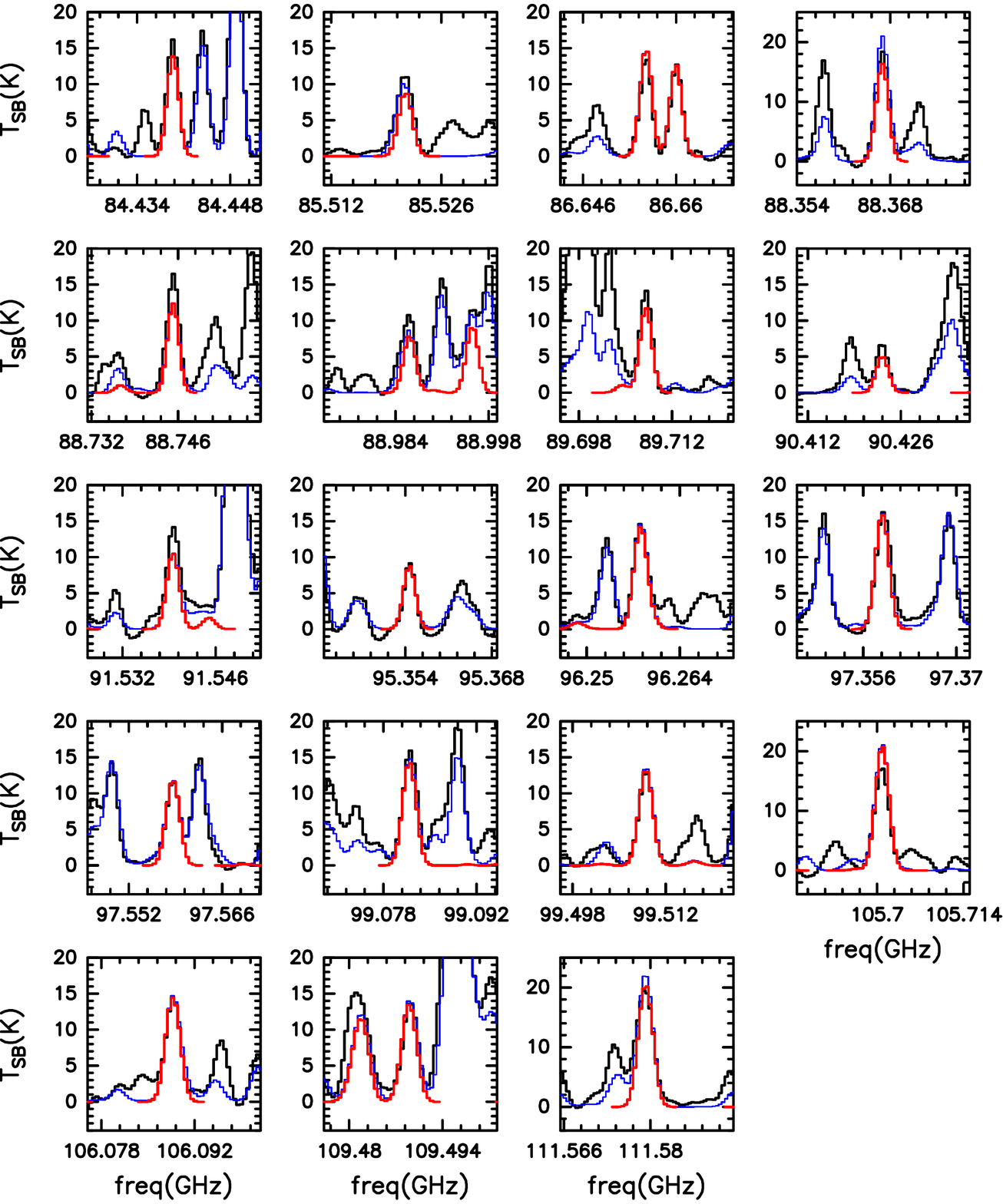}
    \caption{Transitions used to constrain the fit of aGg'-(CH$_2$OH)$_2$. In black the
observed spectrum, in red the synthetic spectrum of the best fit for aGg'-(CH$_2$OH)$_2$ only,
while in blue the spectrum that takes into account all the species identified in the spectrum, including those published in \citet{mininni2020,colziguapos}. $T_{\rm{SB}}$ stands for synthesized beam temperature.
}
    \label{fig:spectra_aGg}
\end{figure*}

\begin{figure*}
    \centering
    \includegraphics[width=0.66\textwidth, trim={0, 11.5cm, 0, 10cm}, clip]{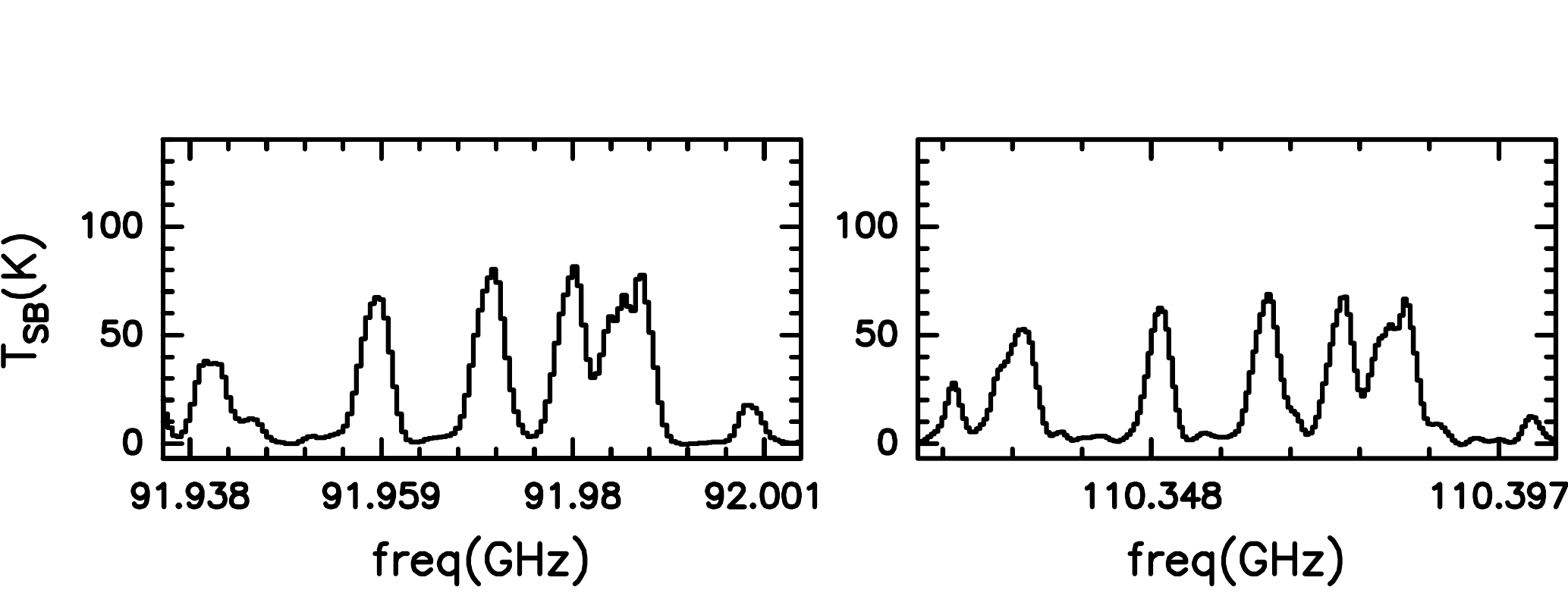}
    \caption{Observed spectrum of CH$_3$CN v=0. $T_{\rm{SB}}$ stands for synthesized beam temperature.
}
    \label{fig:spectrach3cn}
\end{figure*}
\begin{figure*}
    \centering
    \includegraphics[width=0.8\textwidth, trim={0, 9cm, 0, 9cm}, clip]{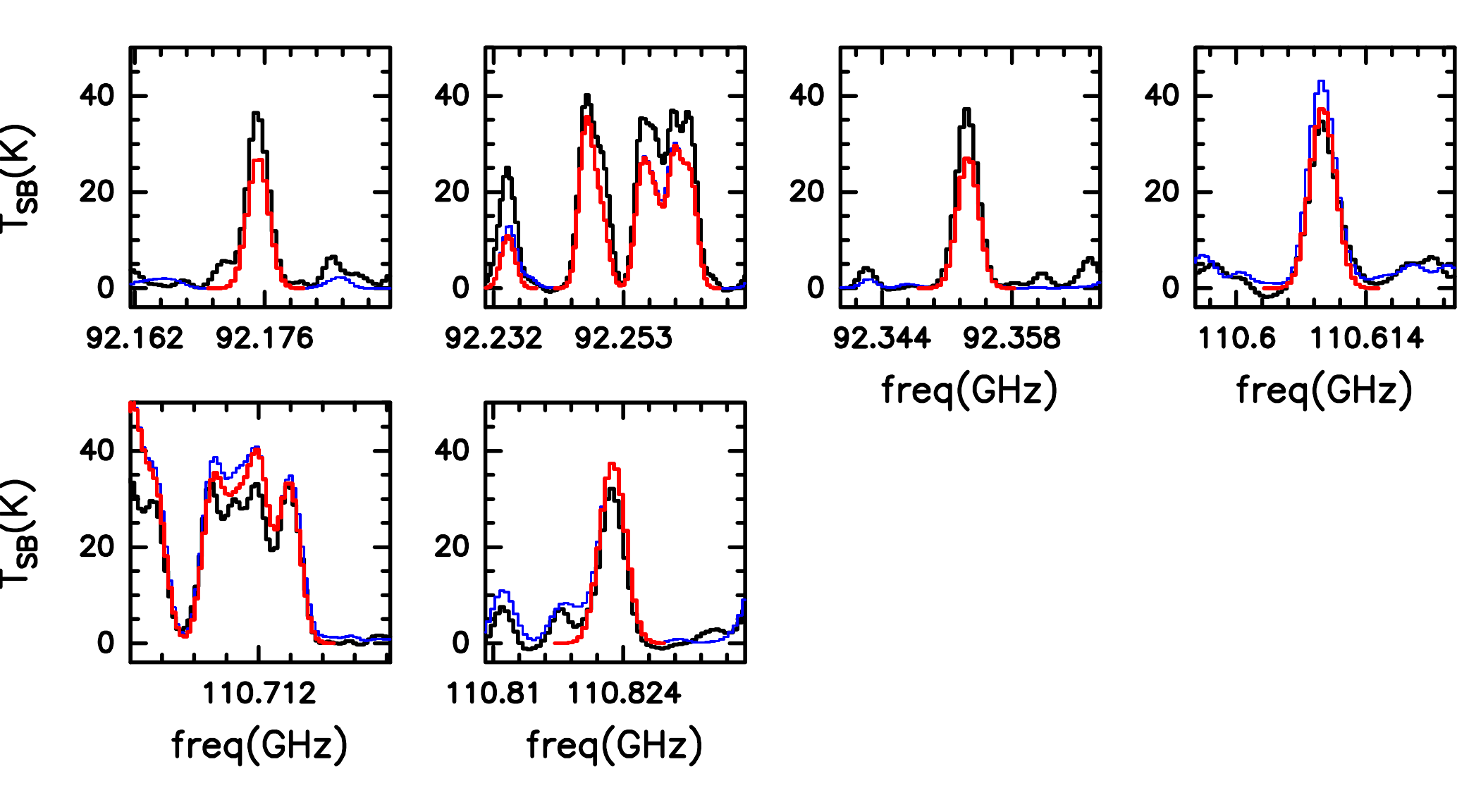}
    \caption{Transitions used to constrain the fit of CH$_3$CN\,v$_8 = 1$. In black the
observed spectrum, in red the synthetic spectrum of the best fit for CH$_3$CN\,v$_8 = 1$ only,
while in blue the spectrum that takes into account all the species identified in the spectrum, including those published in \citet{mininni2020,colziguapos}. $T_{\rm{SB}}$ stands for synthesized beam temperature.
}
    \label{fig:spectrach3cnvib}
\end{figure*}

\begin{figure*}
    \centering
    \includegraphics[width=0.42\textwidth, trim={0, 9cm, 0, 9cm}, clip]{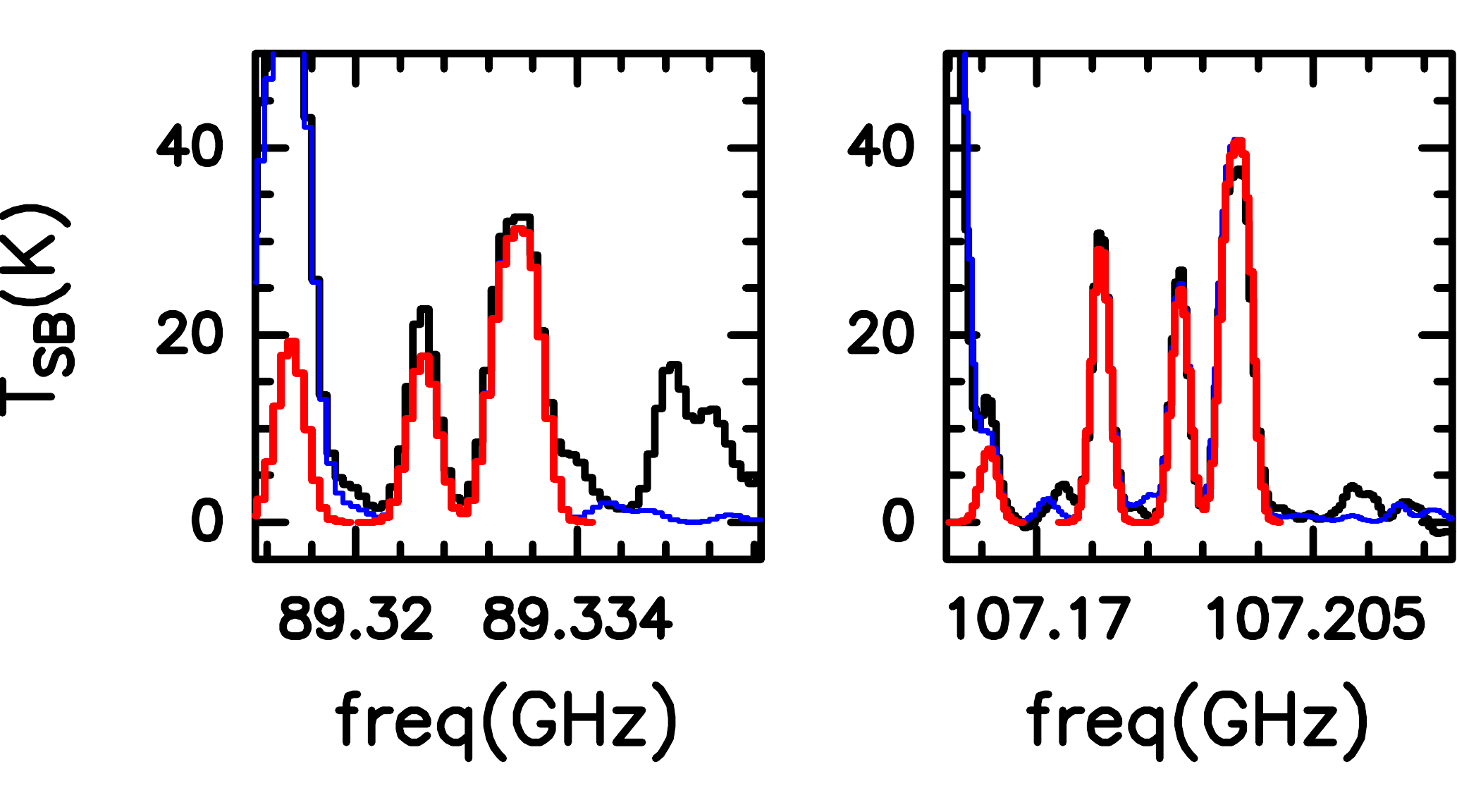}
    \caption{Transitions used to constrain the fit of $^{13}$CH$_3$CN. In black the
observed spectrum, in red the synthetic spectrum of the best fit for $^{13}$CH$_3$CN only,
while in blue the spectrum that takes into account all the species identified in the spectrum, including those published in \citet{mininni2020,colziguapos}. $T_{\rm{SB}}$ stands for synthesized beam temperature.
}
    \label{fig:spectrac13h3cn}
\end{figure*}
\begin{figure*}
    \centering
    \includegraphics[width=0.42\textwidth, trim={0, 9cm, 0, 9cm}, clip]{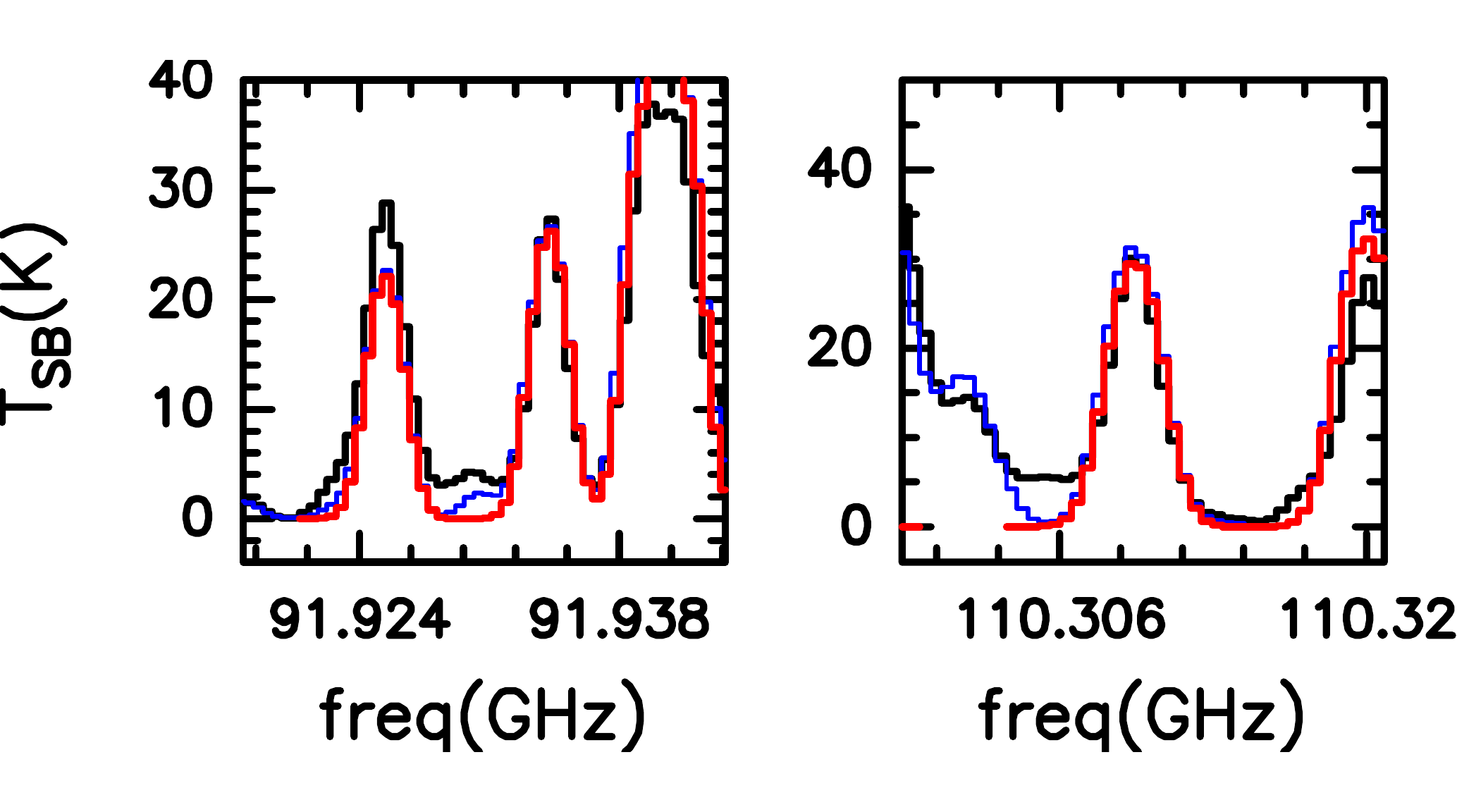}
    \caption{Transitions used to constrain the fit of CH$_3^{13}$CN. In black the
observed spectrum, in red the synthetic spectrum of the best fit for CH$_3^{13}$CN only,
while in blue the spectrum that takes into account all the species identified in the spectrum, including those published in \citet{mininni2020,colziguapos}. $T_{\rm{SB}}$ stands for synthesized beam temperature.
}
    \label{fig:spectrach3c13n}
\end{figure*}
\begin{figure*}
    \centering
    \includegraphics[width=0.8\textwidth, trim={0, 7cm, 0, 7cm}, clip]{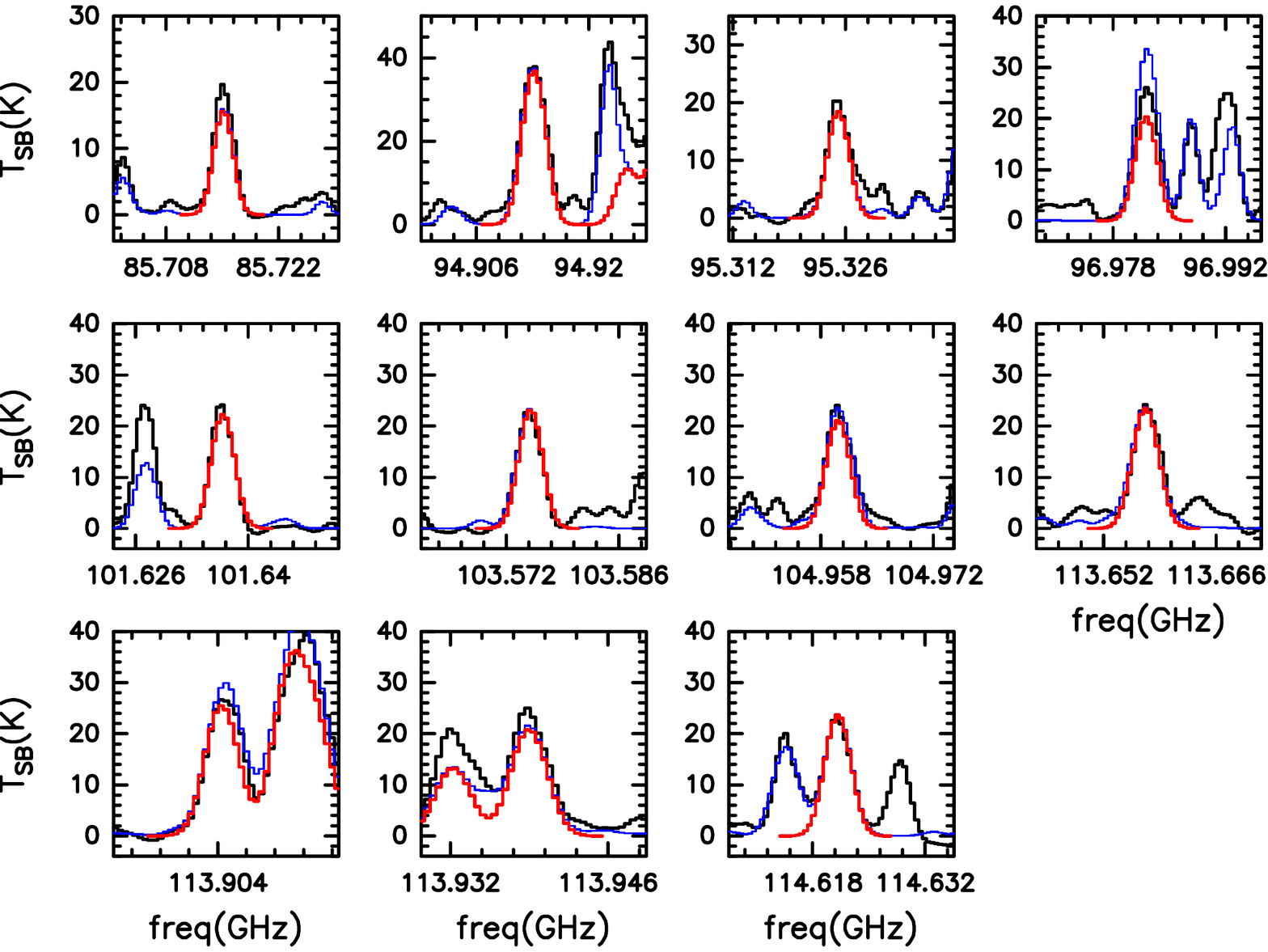}
    \caption{Transitions used to constrain the fit of C$_2$H$_3$CN. In black the
observed spectrum, in red the synthetic spectrum of the best fit for C$_2$H$_3$CN only,
while in blue the spectrum that takes into account all the species identified in the spectrum, including those published in \citet{mininni2020,colziguapos}. $T_{\rm{SB}}$ stands for synthesized beam temperature.
}
    \label{fig:spectrac2h3cn}
\end{figure*}
\begin{figure*}
    \centering
    \includegraphics[width=0.75\textwidth, trim={0, 4.5cm, 0, 4cm}, clip]{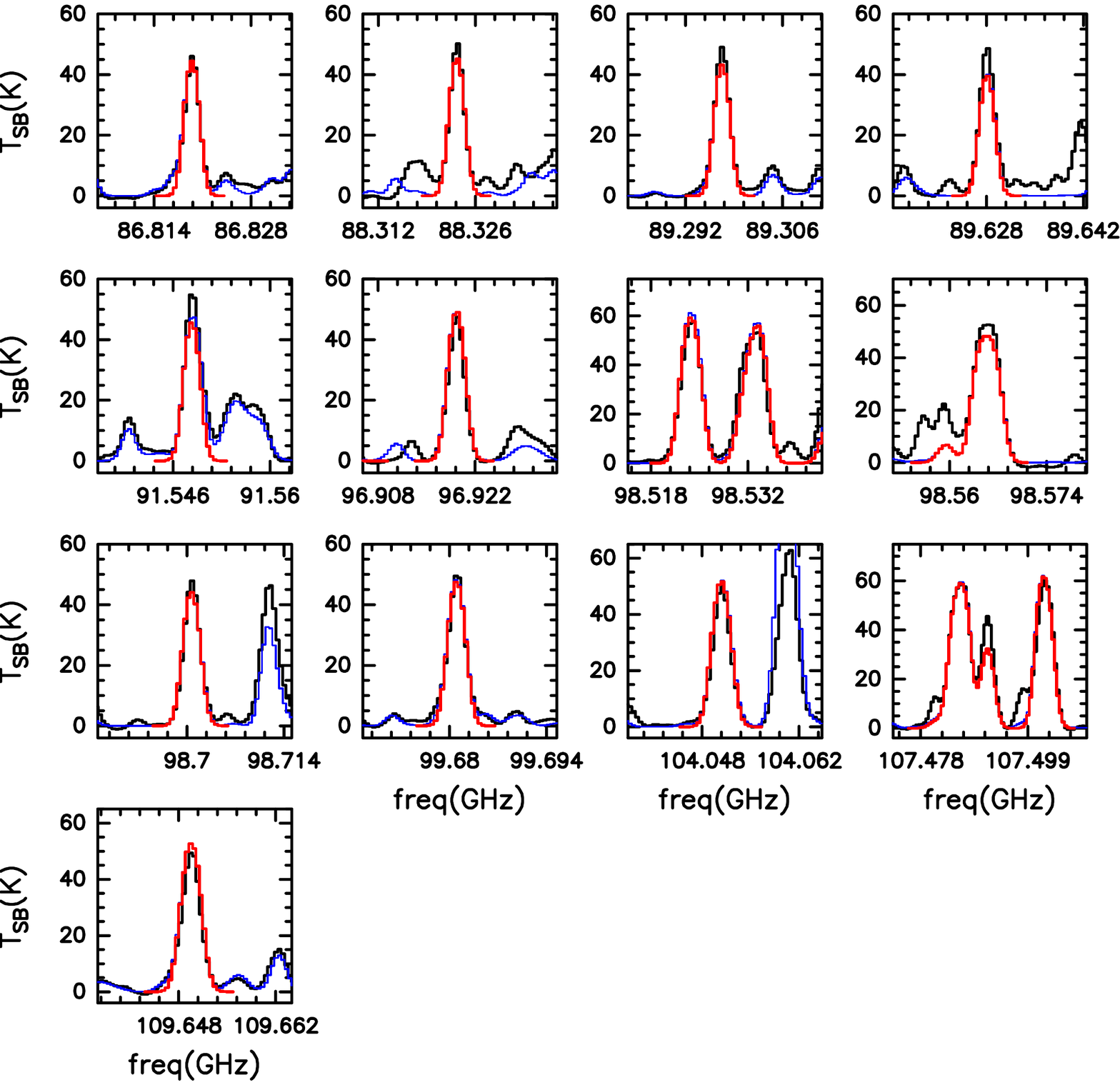}
    \caption{Transitions used to constrain the fit of C$_2$H$_5$CN. In black the
observed spectrum, in red the synthetic spectrum of the best fit for C$_2$H$_5$CN only,
while in blue the spectrum that takes into account all the species identified in the spectrum, including those published in \citet{mininni2020,colziguapos}. $T_{\rm{SB}}$ stands for synthesized beam temperature.
}
    \label{fig:spectrac2h5cn}
\end{figure*}
\begin{figure*}
    \centering
    \includegraphics[width=0.75\textwidth, trim={0, 9.3cm, 0, 9cm}, clip]{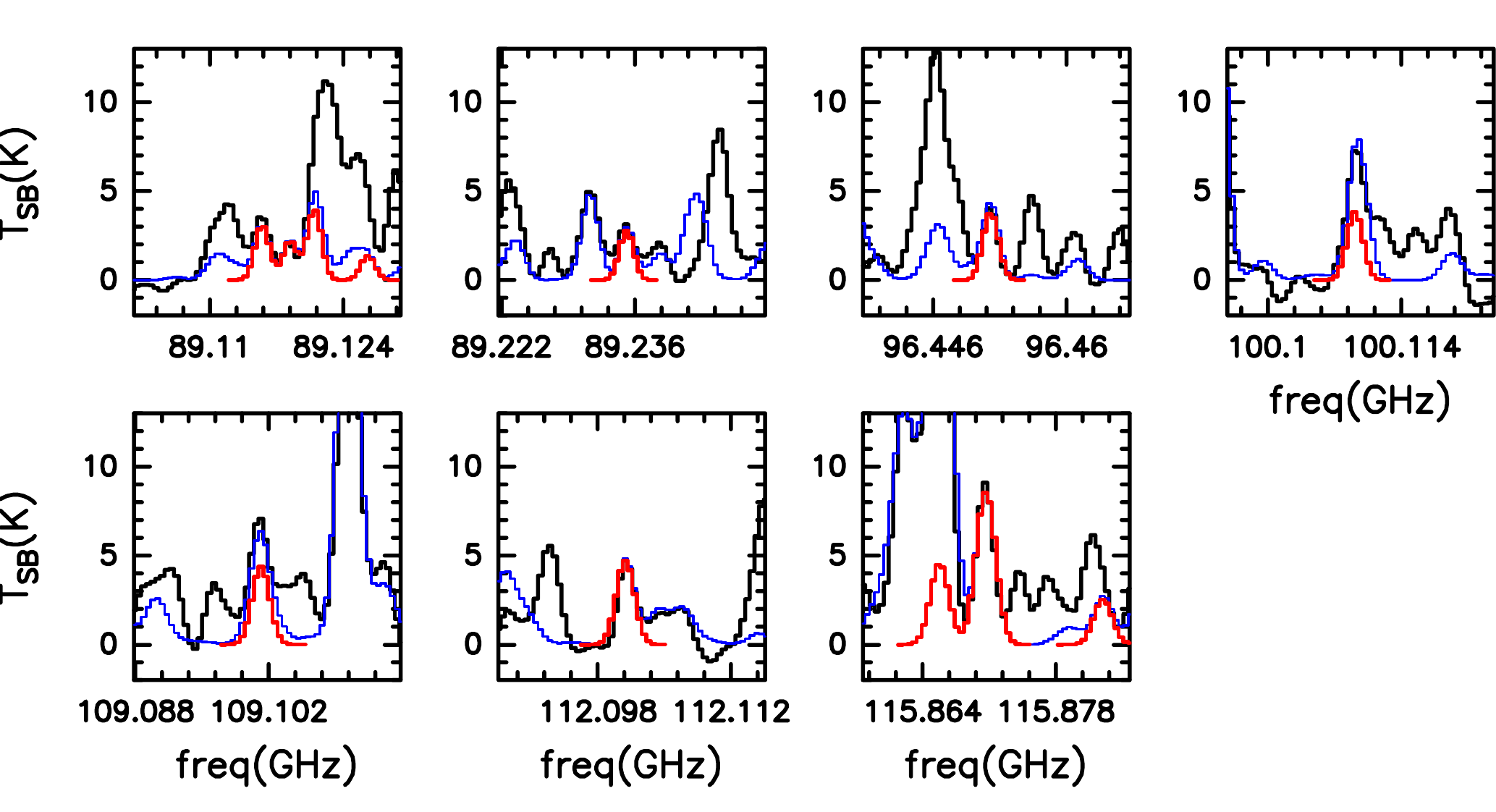}
    \caption{Transitions used to constrain the fit of C$_2$H$_5^{13}$CN. In black the
observed spectrum, in red the synthetic spectrum of the best fit for C$_2$H$_5^{13}$CN only,
while in blue the spectrum that takes into account all the species identified in the spectrum, including those published in \citet{mininni2020,colziguapos}. $T_{\rm{SB}}$ stands for synthesized beam temperature.
}
    \label{fig:spectrac2h5c13n}
\end{figure*}
\clearpage
\onecolumn
\section{Selected transitions for the \\analysis}
{\small
\setlength{\tabcolsep}{3pt}

\renewcommand{\arraystretch}{1.1}

\begin{longtable}[1]{ccccccccccc}
\label{table:ch3ohvib1}\\
\caption{Most unblended transitions of CH$_3$OH v$_t=1$ used to constrain the fit. Database used: CDMS}\\
\hline\hline
Frequency	&log(I)	& $E_{\rm{U}}/k_{\rm{B}}$	&\multicolumn{3}{c}{$Q_{\rm{U}}$}	&\multicolumn{3}{c}{$Q_{\rm{L}}$}	& $\tau _{0}$ & \\
MHz &  &K & $J$ & $K_{\rm{a}}$ & $K_{\rm{c}}$ & $J$ & $K_{\rm{a}}$ & $K_{\rm{c}}$ &  & \\
\hline
\endfirsthead
\caption[]{Continued.} \\
\hline
 Frequency	&log(I)	& $E_{\rm{U}}/k_{\rm{B}}$	&\multicolumn{3}{c}{$Q_{\rm{U}}$}	&\multicolumn{3}{c}{$Q_{\rm{L}}$}	& $\tau _{0}$ & \\
MHz &  &K & $J$ & $K_{\rm{a}}$ & $K_{\rm{c}}$ & $J$ & $K_{\rm{a}}$ & $K_{\rm{c}}$ & & \\
\hline
\endhead
\hline
\endfoot
\endlastfoot
90813.078	&-5.274	& 809.18	&20	&	3	&	18	&	19	&	2	&	18		&0.178& Y	\\ 
93196.672	&-5.514	& 303.22	&1	&	0	&	1	&	2	&	1	&	2		&0.210& Y	\\ 
96492.163	&-5.519	& 298.74	&2	&	1	&	2	&	1	&	1	&	1		&0.201& Y	\\ 
96493.551	&-5.407	& 307.85	&2	&	0	&	2	&	1	&	0	&	1		&0.257& Y	\\ 
99730.940	&-5.137	& 340.46	&6	&	1	&	6	&	5	&	0	&	5		&0.442& Y	\\ 
99772.834	&-5.202	& 903.29	&20	&	3	&	17	&	21	&	4	&	18		&0.167& Y	\\ 
102957.672	&-5.019	& 642.91	&15	&	2	&	14	&	16	&	3	&	14		&0.361& Y	\\ 
105187.877	&-5.434	& 646.65	&7	&	7	&	0	&	6	&	6	&	0		&0.135& Y	\\ 

\hline
\hline
\end{longtable}
{\textbf{Notes:} Frequency: rest frequency of the transitions in units MHz; log(I): logarithm of the line intensity; $\mathrm{E_{\rm{U}}/k_{\rm{B}}}$: energy of the upper state of the transition in units K; $\mathrm{Q_{U}}$: quantum numbers of the upper level; $\mathrm{Q_{L}}$: quantum numbers of the lower level; $\tau_{0} $: optical depth at the center of the line, calculated with the parameters of best fit; flag indicating which transitions have been used to create the mean map.} \vspace{3mm}
}

{\small
\setlength{\tabcolsep}{3pt}

\renewcommand{\arraystretch}{1.1}

\begin{longtable}[2]{ccccccccccc}
\label{table:c13h3oh}\\
\caption{Most unblended transitions of $^{13}$CH$_3$OH used to constrain the fit. Database used: CDMS}\\
\hline\hline
Frequency	&log(I)	& $E_{\rm{U}}/k_{\rm{B}}$	&\multicolumn{3}{c}{$Q_{\rm{U}}$}	&\multicolumn{3}{c}{$Q_{\rm{L}}$}	& $\tau _{0}$ &\\
MHz &  &K & $J$ & $K_{\rm{a}}$ & $K_{\rm{c}}$ & $J$ & $K_{\rm{a}}$ & $K_{\rm{c}}$ & &\\
\hline
\endfirsthead
\caption[]{Continued.} \\
\hline
 Frequency	&log(I)	& $E_{\rm{U}}/k_{\rm{B}}$	&\multicolumn{3}{c}{$Q_{\rm{U}}$}	&\multicolumn{3}{c}{$Q_{\rm{L}}$}	& $\tau _{0}$ &\\
MHz &  &K & $J$ & $K_{\rm{a}}$ & $K_{\rm{c}}$ & $J$ & $K_{\rm{a}}$ & $K_{\rm{c}}$ & &\\
\hline
\endhead
\hline
\endfoot
\endlastfoot
84444.140	&-5.174	&269.32	&1	&	3	&	-3 	&	1	&	2	&	-4	&0.091&Y	\\ 
92588.704	&-4.964	&101.33	&7	&	2	&	&		8	&	1	&		&0.235&Y	\\ 
93441.190	&-5.204	&339.40	&6	&	1	&	&		5	&	0	&		&0.061&Y	\\ 
93619.460	&-5.155	&21.32	&2	&	1	&	&		1	&	1	&		&0.194&Y	\\ 
94256.648	&-5.599	&332.51	&2	&	1	&	&		1	&	1	&		&0.025&	\\ 
94405.163	&-5.135	&12.41	&2	&	-1	&	&		1	&	-1	&		&0.207&	\\ 
94407.129	&-5.002	&6.80	&2	&	0	&	&		1	&	0	&		&0.287&	\\ 
94411.016	&-5.021	&19.93	&2	&	0	&	&		1	&	0	&		&0.263&	\\ 
94420.449	&-5.146	&27.88	&2	&	1	&	&		1	&	1	&		&0.192&Y	\\ 
95208.660	&-5.141	&21.43	&2	&	1	&	&		1	&	1	&		&0.197&Y	\\ 
97219.021	&-5.154	&331.12	&1	&	3	&	-5	&	1	&	4	&	-4	&0.068&	\\ 
105914.401	&-5.413	&232.53	&1	&	3	&	-2	&	1	&	3	&	1	&0.047&Y	\\ 
109164.120	&-5.103	&13.12	&0	&	0	&	&		1	&	-1	&		&0.193&Y	\\ 
112143.546	&-4.614	&28.06	&3	&	1	&	&		4	&	0	&		&0.553&	\\ 
\hline
\hline
\end{longtable}
{\textbf{Notes:} Frequency: rest frequency of the transitions in units MHz; log(I): logarithm of the line intensity; $\mathrm{E_{\rm{U}}/k_{\rm{B}}}$: energy of the upper state of the transition in units K; $\mathrm{Q_{U}}$: quantum numbers of the upper level; $\mathrm{Q_{L}}$: quantum numbers of the lower level; $\tau_{0} $: optical depth at the center of the line. calculated with the parameters of best fit; flag indicating which transitions have been used to create the mean map.} \vspace{3mm}
}
{\small
\setlength{\tabcolsep}{3pt}

\renewcommand{\arraystretch}{1.1}

\begin{longtable}[1]{ccccccccccc}
\label{table:ch3c18oh}\\
\caption{Most unblended transitions of CH$_3^{18}$OH used to constrain the fit. Database used: CDMS}\\
\hline\hline
Frequency	&log(I)	& $E_{\rm{U}}/k_{\rm{B}}$	&\multicolumn{3}{c}{$Q_{\rm{U}}$}	&\multicolumn{3}{c}{$Q_{\rm{L}}$}	& $\tau _{0}$ & \\
MHz &  &K & $J$ & $K_{\rm{a}}$ & $K_{\rm{c}}$ & $J$ & $K_{\rm{a}}$ & $K_{\rm{c}}$ &  & \\
\hline
\endfirsthead
\caption[]{Continued.} \\
\hline
 Frequency	&log(I)	& $E_{\rm{U}}/k_{\rm{B}}$	&\multicolumn{3}{c}{$Q_{\rm{U}}$}	&\multicolumn{3}{c}{$Q_{\rm{L}}$}	& $\tau _{0}$ & \\
MHz &  &K & $J$ & $K_{\rm{a}}$ & $K_{\rm{c}}$ & $J$ & $K_{\rm{a}}$ & $K_{\rm{c}}$ & & \\
\hline
\endhead
\hline
\endfoot
\endlastfoot
84109.417	& -5.000	&  144.29		& 	11	&	1	&	11	&	10	&	2	&	9	&		0.048	& Y \\
84727.078	& -5.179	&  217.58		& 	11	&	4	&	7	&	12	&	3	&	10	&		0.025	&  \\
92723.286	& -5.158	&  7.62		& 	2	&	1	&	2	&	1	&	1	&	1	&		0.048	&   \\
92725.469	& -5.026	&  2.16		& 	2	&	0	&	2	&	1	&	0	&	1	&		0.066	&    \\
92729.399	& -5.045	&  14.66		& 	2	&	0	&	2	&	1	&	0	&	1	&		0.061	&  \\
92733.939	& -5.163	&  22.35		& 	2	&	1	&	1	&	1	&	1	&	0	&		0.045	&  \\
97276.539	& -4.938	&  92.54		& 	7	&	2	&	6	&	8	&	1	&	7	&		0.057	& Y \\
98206.527	& -4.986	&  66.06		& 	6	&	2	&	5	&	7	&	1	&	7	&		0.055	&  \\
100210.505	& -5.011	&  338.39		& 	16	&	3	&	14	&	15	&	4	&	11	&		0.021	&  \\
100604.491	& -5.007	&  338.39		& 	16	&	3	&	13	&	15	&	4	&	12	&		0.021	&  \\
103930.931	& -4.747	&  179.31		& 	12	&	1	&	11	&	11	&	2	&	10	&		0.063	&  \\
104329.446	& -5.162	&  7.62		& 	0	&	0	&	0	&	1	&	1	&	1	&		0.042	&  \\
111209.731	& -4.639	&  21.65		& 	3	&	1	&	3	&	4	&	0	&	4	&		0.126	& Y \\
112948.991	& -4.773	&  52.67		& 	6	&	0	&	6	&	5	&	1	&	4	&		0.082	&  \\
114179.077	& -4.890	&  119.23		& 	8	&	3	&	5	&	9	&	2	&	7	&		0.050	& Y \\
\hline
\hline
\end{longtable}
{\textbf{Notes:} Frequency: rest frequency of the transitions in units MHz; log(I): logarithm of the line intensity; $\mathrm{E_{\rm{U}}/k_{\rm{B}}}$: energy of the upper state of the transition in units K; $\mathrm{Q_{U}}$: quantum numbers of the upper level; $\mathrm{Q_{L}}$: quantum numbers of the lower level; $\tau_{0} $: optical depth at the center of the line, calculated with the parameters of best fit; flag indicating which transitions have been used to create the mean map.} \vspace{3mm}
}

{\small
\setlength{\tabcolsep}{3pt}

\renewcommand{\arraystretch}{1.1}

\begin{longtable}[3]{ccccccccccc}
\label{table:ch3cho}\\
\caption{Most unblended transitions of CH$_3$CHO used to constrain the fit. Database used: JPL}\\
\hline\hline
Frequency	&log(I)	& $E_{\rm{U}}/k_{\rm{B}}$	&\multicolumn{3}{c}{$Q_{\rm{U}}$}	&\multicolumn{3}{c}{$Q_{\rm{L}}$}	& $\tau _{0}$ &\\
MHz &  &K & $J$ & $K_{\rm{a}}$ & $K_{\rm{c}}$ & $J$ & $K_{\rm{a}}$ & $K_{\rm{c}}$ & &\\
\hline
\endfirsthead
\caption[]{Continued.} \\
\hline
 Frequency	&log(I)	& $E_{\rm{U}}/k_{\rm{B}}$	&\multicolumn{3}{c}{$Q_{\rm{U}}$}	&\multicolumn{3}{c}{$Q_{\rm{L}}$}	& $\tau _{0}$& \\
MHz &  &K & $J$ & $K_{\rm{a}}$ & $K_{\rm{c}}$ & $J$ & $K_{\rm{a}}$ & $K_{\rm{c}}$ & &\\
\hline
\endhead
\hline
\endfoot
\endlastfoot
93580.909	&-4.409	&15.76		&5	&	1	&	5	&	4	&	1	&	4			&0.201&	Y	\\ 
93595.235	&-4.409	&15.84		&5	&	1	&	5	&	4	&	1	&	4			&0.201&		\\ 
96343.265	&-4.859	&50.11		&5	&	4	&	2	&	4	&	4	&	1			&0.050&	Y	\\ 
96343.279	&-4.859	&50.11		&5	&	4	&	1	&	4	&	4	&	0			&0.050&	Y	\\ 
96425.614	&-4.454	& 22.40	& 5	&	2	&	4	&	4	&	2	&	3	    	    &0.000&	Y	\\ 
96475.524	&-4.454	&23.05		&5	&	2	&	3	&	4	&	2	&	2			&0.164&	Y	\\ 
96632.663	&-4.450	&22.99		&5	&	2	&	3	&	4	&	2	&	2			&0.166&		\\ 
100127.164	&-4.980	&76.28		&12	&	1	&	11	&	12	&	0	&	12		    &0.028&	Y	\\ 
114940.175	&-4.139	&19.47		&6	&	0	&	6	&	5	&	0	&	5			&0.296&	Y	\\ 
114959.902	&-4.139	&19.37		&6	&	0	&	6	&	5	&	0	&	5			&0.297&	Y	\\ 
115634.704	&-4.440	&55.60		&6	&	4	&	2	&	5	&	4	&	1			&0.104&	Y	\\ 
\hline
\hline
\end{longtable}
{\textbf{Notes:} Frequency: rest frequency of the transitions in units MHz; log(I): logarithm of the line intensity; $\mathrm{E_{\rm{U}}/k_{\rm{B}}}$: energy of the upper state of the transition in units K; $\mathrm{Q_{U}}$: quantum numbers of the upper level; $\mathrm{Q_{L}}$: quantum numbers of the lower level; $\tau_{0} $: optical depth at the center of the line, calculated with the parameters of best fit; flag indicating which transitions have been used to create the mean map.} \vspace{3mm}
}

{\small
\setlength{\tabcolsep}{3pt}

\renewcommand{\arraystretch}{1.1}

\begin{longtable}[4]{ccccccccccc}
\label{table:ch3och3}\\
\caption{Most unblended transitions of CH$_3$OCH$_3$ used to constrain the fit. Database used: CDMS}\\
\hline\hline
Frequency	&log(I)	& $E_{\rm{U}}/k_{\rm{B}}$	&\multicolumn{3}{c}{$Q_{\rm{U}}$}	&\multicolumn{3}{c}{$Q_{\rm{L}}$}	& $\tau _{0}$ & \\
MHz &  &K & $J$ & $K_{\rm{a}}$ & $K_{\rm{c}}$ & $J$ & $K_{\rm{a}}$ & $K_{\rm{c}}$ & & \\
\hline
\endfirsthead
\caption[]{Continued.} \\
\hline
 Frequency	&log(I)	& $E_{\rm{U}}/k_{\rm{B}}$	&\multicolumn{3}{c}{$Q_{\rm{U}}$}	&\multicolumn{3}{c}{$Q_{\rm{L}}$}	& $\tau _{0}$& \\
MHz &  &K & $J$ & $K_{\rm{a}}$ & $K_{\rm{c}}$ & $J$ & $K_{\rm{a}}$ & $K_{\rm{c}}$ & & \\
\hline
\endhead
\hline
\endfoot
\endlastfoot
84631.897	&-6.145	& 11.10	&3	&	2	&	1	&	3	&	1	&	2		&0.083&	Y	\\ 
84632.274	&-6.322	& 11.10	&3	&	2	&	1	&	3	&	1	&	2		&0.055&	Y	\\ 
84634.421	&-5.720	& 11.10	&3	&	2	&	1	&	3	&	1	&	2		&0.223&	Y	\\ 
84636.757	&-5.923	& 11.10	&3	&	2	&	1	&	3	&	1	&	2		&0.139&	Y\\ 
88706.231	&-5.399	& 116.98	&15	&	2	&	13	&	15	&	1	&	14	&0.215	&	Y\\ 
88706.231	&-5.575	& 116.98	&15	&	2	&	13	&	15	&	1	&	14	&0.143	&	Y\\ 
88707.704	&-4.973	& 116.98	&15	&	2	&	13	&	15	&	1	&	14	&0.574	  & Y\\ 
88709.177	&-5.177	& 116.98	&15	&	2	&	13	&	15	&	1	&	14	&0.359	&	Y\\ 
91473.762	&-6.304	& 11.10	&3	&	2	&	2	&	3	&	1	&	3		&0.053&		\\ 
91474.139	&-6.605	& 11.10	&3	&	2	&	2	&	3	&	1	&	3		&0.027&		\\ 
91476.607	&-5.702	& 11.10	&3	&	2	&	2	&	3	&	1	&	3		&0.215&		\\ 
91479.263	&-6.128	& 11.09	&3	&	2	&	2	&	3	&	1	&	3		&0.081&		\\ 
93664.597	&-5.788	& 74.09	&12	&	1	&	11	&	12	&	0	&	12	    &0.112	&	\\ 
93664.597	&-6.089	& 74.09	&12	&	1	&	11	&	12	&	0	&	12	    &0.056	&	\\ 
93666.463	&-5.186	& 74.09	&12	&	1	&	11	&	12	&	0	&	12	    &0.448	&	\\ 
93668.329	&-5.612	& 74.09	&12	&	1	&	11	&	12	&	0	&	12	    &0.168	&	\\ 
93854.438	&-6.152	& 14.73	&4	&	2	&	3	&	4	&	1	&	4		&0.072&	Y	\\ 
93854.560	&-5.976	& 14.73	&4	&	2	&	3	&	4	&	1	&	4		&0.109&	Y	\\ 
93857.113	&-5.550	& 14.73	&4	&	2	&	3	&	4	&	1	&	4   	&0.290	  &  Y    \\ 
93859.727	&-5.754	& 14.73	&4	&	2	&	3	&	4	&	1	&	4		&0.181&	Y	\\ 
95729.780	&-5.526	& 131.94	&16	&	2	&	14	&	16	&	1	&	15	&0.134	&	\\ 
95729.781	&-5.827	& 131.94	&16	&	2	&	14	&	16	&	1	&	15	&0.067	&	\\ 
95731.253	&-4.924	& 131.94	&16	&	2	&	14	&	16	&	1	&	15	&0.539	&	\\ 
95732.726	&-5.350	& 131.94	&16	&	2	&	14	&	16	&	1	&	15	&0.202	&	\\ 
96847.241	&-6.039	& 19.28	&5	&	2	&	4	&	5	&	1	&	5		&0.088&		\\ 
96847.292	&-6.340	& 19.28	&5	&	2	&	4	&	5	&	1	&	5		&0.044&		\\ 
96849.890	&-5.437	& 19.28	&5	&	2	&	4	&	5	&	1	&	5		&0.354&		\\ 
96852.514	&-5.863	& 19.28	&5	&	2	&	4	&	5	&	1	&	5		&0.133&		\\ 
104175.880	&-5.300	& 147.81	&17	&	2	&	15	&	17	&	1	&	16	&0.187	  &  Y	\\ 
104175.880	&-5.476	& 147.81	&17	&	2	&	15	&	17	&	1	&	16	&0.124	  &  Y   \\ 
104177.381	&-4.874	& 147.81	&17	&	2	&	15	&	17	&	1	&	16	&0.498	  &  Y	\\ 
104178.881	&-5.078	& 147.81	&17	&	2	&	15	&	17	&	1	&	16	&0.311	  &  Y    \\ 
104700.568	&-5.866	& 31.09	&7	&	2	&	6	&	7	&	1	&	7		&0.112&		\\ 
104700.581	&-6.167	& 31.09	&7	&	2	&	6	&	7	&	1	&	7		&0.056&		\\ 
104703.262	&-5.264	& 31.09	&7	&	2	&	6	&	7	&	1	&	7		&0.451&		\\ 
104705.949	&-5.690	& 31.09	&7	&	2	&	6	&	7	&	1	&	7		&0.169&		\\ 
106775.679	&-5.641	& 43.48	&9	&	1	&	8	&	8	&	2	&	7		&0.170&		\\ 
106777.372	&-5.436	& 43.48	&9	&	1	&	8	&	8	&	2	&	7		&0.273&		\\ 
106779.061	&-5.862	& 43.48	&9	&	1	&	8	&	8	&	2	&	7		&0.102&	 \\ 
106779.069	&-6.038	& 43.48	&9	&	1	&	8	&	8	&	2	&	7		&0.068&		\\ 
109571.391	&-5.795	& 38.35	&8	&	2	&	7	&	8	&	1	&	8		&0.120&		\\ 
109571.398	&-5.619	& 38.35	&8	&	2	&	7	&	8	&	1	&	8		&0.181&		\\ 
109574.127	&-5.193	& 38.35	&8	&	2	&	7	&	8	&	1	&	8		&0.483&		\\ 
109576.860	&-5.397	& 38.35	&8	&	2	&	7	&	8	&	1	&	8		&0.302&		\\ 
111782.249	&-8.579	& 60.70	&7	&	4	&	4	&	8	&	5	&	3		&0.000&		\\ 
111782.600	&-5.305	& 25.28	&7	&	0	&	7	&	6	&	1	&	6		&0.400&	\\ 
111783.117	&-5.101	& 25.28	&7	&	0	&	7	&	6	&	1	&	6		&0.641&		\\ 
111783.633	&-5.527	& 25.28	&7	&	0	&	7	&	6	&	1	&	6		&0.240&		\\ 
111783.634	&-5.703	& 25.28	&7	&	0	&	7	&	6	&	1	&	6		&0.160&  	\\ 
111812.252	&-5.663	& 169.97	&18	&	3	&	15	&	18	&	2	&	16	&0.064	&	\\ 
111812.253	&-5.362	& 169.97	&18	&	3	&	15	&	18	&	2	&	16	&0.129	&	\\ 
111812.705	&-7.413	& 60.70	&7	&	4	&	4	&	8	&	5	&	3		&0.002&		\\ 
111813.812	&-4.760	& 169.97	&18	&	3	&	15	&	18	&	2	&	16	&0.519    &  	\\ 
111815.372	&-5.186	& 169.97	&18	&	3	&	15	&	18	&	2	&	16	&0.194	  &  	\\ 
111817.549	&-8.579	& 60.70	&7	&	4	&	3	&	8	&	5	&	4		&0.000&		\\ 
113057.591	&-5.190	& 153.24	&17	&	3	&	14	&	17	&	2	&	15	&0.214	&Y	\\ 
113057.593	&-5.366	& 153.24	&17	&	3	&	14	&	17	&	2	&	15	&0.142	&Y	\\ 
113059.352	&-4.764	& 153.24	&17	&	3	&	14	&	17	&	2	&	15	&0.571	&Y \\ 
113061.112	&-4.968	& 153.24	&17	&	3	&	14	&	17	&	2	&	15	&0.356	&Y	\\ 
114003.844	&-5.427	& 164.60	&18	&	2	&	16	&	18	&	1	&	17	&0.113	&	\\ 
114003.844	&-5.728	& 164.60	&18	&	2	&	16	&	18	&	1	&	17	&0.056	&	\\ 
114005.401	&-4.825	& 164.60	&18	&	2	&	16	&	18	&	1	&	17	&0.455	&	\\ 
114006.957	&-5.251	& 164.60	&18	&	2	&	16	&	18	&	1	&	17	&0.170	&	\\ 
115072.305	&-5.731	& 46.51	&9	&	2	&	8	&	9	&	1	&	9		&0.126&Y		\\ 
115072.310	&-6.032	& 46.51	&9	&	2	&	8	&	9	&	1	&	9		&0.063&Y		\\ 
115075.094	&-5.129	& 46.51	&9	&	2	&	8	&	9	&	1	&	9		&0.505&Y  	\\ 
115077.881	&-5.554	& 46.51	&9	&	2	&	8	&	9	&	1	&	9		&0.189&Y		\\ 
115543.994	&-5.760	& 14.63	&5	&	1	&	5	&	4	&	0	&	4		&0.146&Y		\\ 
115543.996	&-6.061	& 14.63	&5	&	1	&	5	&	4	&	0	&	4		&0.073&Y		\\ 
115544.807	&-5.158	& 14.63	&5	&	1	&	5	&	4	&	0	&	4		&0.585&Y		\\ 
115545.619	&-5.584	& 14.63	&5	&	1	&	5	&	4	&	0	&	4		&0.219&Y		\\ 
115710.648	&-5.161	& 225.94	&21	&	3	&	18	&	21	&	2	&	19	&0.135	&Y	\\ 
115710.648	&-5.337	& 225.94	&21	&	3	&	18	&	21	&	2	&	19	&0.090	&Y	\\ 
115711.738	&-4.735	& 225.94	&21	&	3	&	18	&	21	&	2	&	19	&0.361	&Y	\\ 
115712.828	&-4.939	& 225.94	&21	&	3	&	18	&	21	&	2	&	19	&0.226	&Y	\\ 
\hline
\hline
\end{longtable}
{\textbf{Notes:} Frequency: rest frequency of the transitions in units MHz; log(I): logarithm of the line intensity; $\mathrm{E_{\rm{U}}/k_{\rm{B}}}$: energy of the upper state of the transition in units K; $\mathrm{Q_{U}}$: quantum numbers of the upper level; $\mathrm{Q_{L}}$: quantum numbers of the lower level; $\tau_{0} $: optical depth at the center of the line, calculated with the parameters of best fit.} \vspace{3mm}
}

{\small
\setlength{\tabcolsep}{3pt}

\renewcommand{\arraystretch}{1.1}

\begin{longtable}[5]{ccccccccccc}
\label{table:ch3coch3}\\
\caption{Most unblended transitions of CH$_3$COCH$_3$ used to constrain the fit. Database used: JPL}\\
\hline\hline
Frequency	&log(I)	& $E_{\rm{U}}/k_{\rm{B}}$	&\multicolumn{3}{c}{$Q_{\rm{U}}$}	&\multicolumn{3}{c}{$Q_{\rm{L}}$}	& $\tau _{0}$& \\
MHz &  &K & $J$ & $K_{\rm{a}}$ & $K_{\rm{c}}$ & $J$ & $K_{\rm{a}}$ & $K_{\rm{c}}$ & &\\
\hline
\endfirsthead
\caption[]{Continued.} \\
\hline
 Frequency	&log(I)	& $E_{\rm{U}}/k_{\rm{B}}$	&\multicolumn{3}{c}{$Q_{\rm{U}}$}	&\multicolumn{3}{c}{$Q_{\rm{L}}$}	& $\tau _{0}$ &\\
MHz &  &K & $J$ & $K_{\rm{a}}$ & $K_{\rm{c}}$ & $J$ & $K_{\rm{a}}$ & $K_{\rm{c}}$ & &\\
\hline
\endhead
\hline
\endfoot
\endlastfoot
86536.667	& -5.598	&	143.21		&	19 & 10 & 10 & 19 & 9 & 11	&	0.057	& Y  \\
88441.793	& -5.636	&	110.59		&	17 & 7 & 10 & 17 & 6 & 11		&	0.056	&  \\
91634.636	& -5.540	&	17.47		&	8 & 1 & 7 & 7 & 2 & 6			&	0.085	&  \\
91637.465	& -5.540	&	17.47		&	8 & 2 & 7 & 7 & 1 & 6			&	0.085	&  \\
92735.670	& -5.401	&	18.75		&	9 & 0 & 9 & 8 & 1 & 8			&	0.116	&  \\
92735.672	& -5.401	&	18.75		&	9 & 1 & 9 & 8 & 0 & 8			&	0.116	&  \\
92743.361	& -5.605	&	18.64		&	9 & 0 & 9 & 8 & 1 & 8			&	0.072	& Y  \\
92743.364	& -5.827	&	18.64		&	9 & 1 & 9 & 8 & 0 & 8			&	0.043	& Y \\
95669.941	& -5.529	&	156.20		&	20 & 10 & 11 & 20 & 9 & 12	&	0.059	&  \\
98352.377	& -6.273	&	9.36		&	5 & 5 & 1 & 4 & 4 & 1			&	0.015	&  \\
98455.636	& -6.079	&	9.24		&	5 & 5 & 1 & 4 & 4 & 0			&	0.023	& Y  \\
99052.509	& -5.712	&	76.66		&	15 & 4 & 11 & 15 & 3 & 12		&	0.045	&  \\
99052.559	& -5.712	&	76.66		&	15 & 5 & 11 & 15 & 4 & 12		&	0.045	&  \\
99266.432	& -5.854	&	9.21		&	5 & 5 & 0 & 4 & 4 & 1			&	0.039	&  \\
99422.031	& -8.712	&	150.83		&	19 & 12 & 7 & 18 & 15 & 4		&	0.000	& Y  \\
99422.076	& -5.806	&	63.35		&	14 & 3 & 11 & 14 & 2 & 12		&	0.038	&  Y\\
99422.084	& -5.806	&	63.35		&	14 & 4 & 11 & 14 & 3 & 12		&	0.038	&  Y\\
99721.506	& -5.951	&	50.93		&	13 & 2 & 11 & 13 & 1 & 12		&	0.028	&  \\
99721.507	& -5.951	&	50.93		&	13 & 3 & 11 & 13 & 2 & 12		&	0.028	&  \\
99900.091	& -6.153	&	50.83	&	13 & 2 & 11 & 13 & 1 & 12		&	0.017	&  \\
99900.092	& -6.375	&	50.83	&	13 & 3 & 11 & 13 & 2 & 12		&	0.010	&  \\
100350.304	& -5.554	&	19.75		&	8 & 2 & 6 & 7 & 3 & 5			&	0.075	&  \\
101426.663	& -5.819	&	21.97		&	9 & 1 & 8 & 8 & 2 & 7			&	0.040	&Y  \\
101426.759	& -5.995	&	21.97		&	9 & 1 & 8 & 8 & 2 & 7			&	0.026	& Y \\
101427.041	& -6.296	&	21.97		&	9 & 2 & 8 & 8 & 1 & 7			&	0.013	& Y \\
101427.129	& -5.995	&	21.97		&	9 & 2 & 8 & 8 & 1 & 7			&	0.026	& Y \\
101451.058	& -5.393	&	21.87		&	9 & 1 & 8 & 8 & 2 & 7			&	0.107	&  \\
101451.446	& -5.393	&	21.87		&	9 & 2 & 8 & 8 & 1 & 7			&	0.107	&  \\
101475.332	& -5.597	&	21.77		&	9 & 1 & 8 & 8 & 2 & 7			&	0.067	&  \\
101475.733	& -5.819	&	21.77		&	9 & 2 & 8 & 8 & 1 & 7			&	0.040	&  \\
102554.696	& -5.271	&	23.20		&	10 & 0 & 10 & 9 & 1 & 9		&	0.139	&  \\
102554.696	& -5.271	&	23.20		&	10 & 1 & 10 & 9 & 0 & 9		&	0.139	&  \\
102562.281	& -5.697	&	23.09		&	10 & 0 & 10 & 9 & 1 & 9		&	0.052	&  \\
102562.282	& -5.475	&	23.09		&	10 & 1 & 10 & 9 & 0 & 9		&	0.087	&  \\
102901.559	& -5.762	&	16.66		&	7 & 4 & 4 & 6 & 3 & 3			&	0.045	&  \\
106273.673	& -5.505	&	133.47		&	19 & 7 & 12 & 19 & 6 & 13		&	0.059	& Y \\
106274.498	& -5.505	&	133.47		&	19 & 8 & 12 & 19 & 7 & 13		&	0.059	& Y \\
107227.738	& -6.012	&	100.79		&	17 & 5 & 12 & 17 & 4 & 13		&	0.019	& Y \\
107227.778	& -6.489	&	100.79		&	17 & 6 & 12 & 17 & 5 & 13		&	0.006	& Y \\
107227.803	& -6.188	&	100.79		&	17 & 5 & 12 & 17 & 4 & 13		&	0.013	& Y \\
107227.843	& -6.188	&	100.79		&	17 & 6 & 12 & 17 & 5 & 13		&	0.013	& Y \\
107376.810	& -5.585	&	100.73	&	17 & 5 & 12 & 17 & 4 & 13		&	0.053	&  \\
107376.851	& -5.585	&	100.73	&	17 & 6 & 12 & 17 & 5 & 13		&	0.053	&  \\
107797.720	& -5.649	&	85.66		&	16 & 4 & 12 & 16 & 3 & 13		&	0.047	&  \\
107797.728	& -5.649	&	85.66		&	16 & 5 & 12 & 16 & 4 & 13		&	0.047	&  \\
107961.851	& -6.073	&	85.59		&	16 & 4 & 12 & 16 & 3 & 13		&	0.017	&  \\
107961.859	& -5.852	&	85.59		&	16 & 5 & 12 & 16 & 4 & 13		&	0.029	&  \\
107967.774	& -6.171	&	71.55		&	15 & 3 & 12 & 15 & 2 & 13		&	0.014	&  \\
107967.775	& -6.648	&	71.55		&	15 & 4 & 12 & 15 & 3 & 13		&	0.004	&  \\
107967.865	& -6.347	&	71.55		&	15 & 3 & 12 & 15 & 2 & 13		&	0.009	&  \\
107967.866	& -6.347	&	71.55		&	15 & 4 & 12 & 15 & 3 & 13		&	0.009	&  \\
108387.594	& -5.616	&	21.59		&	8 & 3 & 5 & 7 & 4 & 4			&	0.060	&  \\
108434.510	& -5.887	&	58.15		&	14 & 2 & 12 & 14 & 1 & 13		&	0.029	&  \\
108434.511	& -5.887	&	58.15		&	14 & 3 & 12 & 14 & 2 & 13		&	0.029	&  \\
109994.365	& -5.732	&	12.96		&	6 & 5 & 2 & 5 & 4 & 1			&	0.046	& Y \\
110189.150	& -5.401	&	24.56		&	9 & 2 & 7 & 8 & 3 & 6			&	0.096	&  \\
110208.702	& -5.401	&	24.56		&	9 & 3 & 7 & 8 & 2 & 6			&	0.096	&  \\
110490.618	& -5.795	&	21.42		&	8 & 4 & 5 & 7 & 3 & 4			&	0.039	&  \\
110987.131	& -5.876	&	23.14		&	8 & 4 & 4 & 7 & 5 & 3			&	0.032	&  \\
112365.987	& -6.059	&	28.22		&	11 & 1 & 11 & 10 & 0 & 10		&	0.020	&  \\
112365.987	& -5.582	&	28.22		&	11 & 0 & 11 & 10 & 1 & 10		&	0.061	&  \\
112366.032	& -5.758	&	28.22		&	11 & 0 & 11 & 10 & 1 & 10		&	0.041	&  \\
112366.032	& -5.758	&	28.22		&	11 & 1 & 11 & 10 & 0 & 10		&	0.041	&  \\
112381.029	& -5.581	&	28.01		&	11 & 1 & 11 & 10 & 0 & 10		&	0.061	&  \\
112381.029	& -5.359	&	28.01		&	11 & 0 & 11 & 10 & 1 & 10		&	0.102	&  \\
114469.404	& -5.429	&	164.18		&	21 & 8 & 13 & 21 & 7 & 14		&	0.060	&  \\
114469.992	& -5.429	&	164.18		&	21 & 9 & 13 & 21 & 8 & 14		&	0.060	&  \\
115115.835	& -5.451	&	145.61		&	20 & 7 & 13 & 20 & 6 & 14		&	0.060	&  \\
115115.980	& -5.451	&	145.61		&	20 & 8 & 13 & 20 & 7 & 14		&	0.060	&  \\

\hline
\hline
\end{longtable}
{\textbf{Notes:} Frequency: rest frequency of the transitions in units MHz; log(I): logarithm of the line intensity; $\mathrm{E_{\rm{U}}/k_{\rm{B}}}$: energy of the upper state of the transition in units K; $\mathrm{Q_{U}}$: quantum numbers of the upper level; $\mathrm{Q_{L}}$: quantum numbers of the lower level; $\tau_{0} $: optical depth at the center of the line, calculated with the parameters of best fit; flag indicating which transitions have been used to create the mean map.} \vspace{3mm}
}

{\small
\setlength{\tabcolsep}{3pt}

\renewcommand{\arraystretch}{1.1}

\begin{longtable}[6]{ccccccccccc}
\label{table:c2h5oh}\\
\caption{Most unblended transitions of C$_2$H$_5$OH used to constrain the fit. Database used: CDMS}\\
\hline\hline
Frequency	&log(I)	& $E_{\rm{U}}/k_{\rm{B}}$	&\multicolumn{3}{c}{$Q_{\rm{U}}$}	&\multicolumn{3}{c}{$Q_{\rm{L}}$}	& $\tau _{0}$ & \\
MHz &  &K & $J$ & $K_{\rm{a}}$ & $K_{\rm{c}}$ & $J$ & $K_{\rm{a}}$ & $K_{\rm{c}}$ & &\\
\hline
\endfirsthead
\caption[]{Continued.} \\
\hline
 Frequency	&log(I)	& $E_{\rm{U}}/k_{\rm{B}}$	&\multicolumn{3}{c}{$Q_{\rm{U}}$}	&\multicolumn{3}{c}{$Q_{\rm{L}}$}	& $\tau _{0}$ &\\
MHz &  &K & $J$ & $K_{\rm{a}}$ & $K_{\rm{c}}$ & $J$ & $K_{\rm{a}}$ & $K_{\rm{c}}$ & &\\
\hline
\endhead
\hline
\endfoot
\endlastfoot
85265.488	&-5.151	&17.50	&6	&	0	&	6	&	5	&	1	&	5	&0.157 &Y		\\ 
86311.282	&-5.250	&74.28	&5	&	2	&	4	&	4	&	2	&	3	&0.092 &		\\ 
86516.482	&-5.648	&93.82	&5	&	4	&	2	&	4	&	4	&	1	&0.033 &		\\ 
86516.573	&-5.648	&93.82	&5	&	4	&	1	&	4	&	4	&	0	&0.033 &		\\ 
86555.887	&-5.617	&88.94	&5	&	4	&	2	&	4	&	4	&	1	&0.036 &		\\ 
86555.986	&-5.617	&88.94	&5	&	4	&	1	&	4	&	4	&	0	&0.036 &		\\ 
86947.280	&-4.743	&108.18	&15	&	2	&	13	&	15	&	1	&	14	&0.248 &Y	        \\ 
87716.120	&-5.263	&17.62	&5	&	2	&	4	&	5	&	1	&	5	&0.118 &		\\ 
91485.187	&-5.166	&22.65	&6	&	2	&	5	&	6	&	1	&	6	&0.137 &Y		\\ 
91737.135	&-4.784	&115.42	&11	&	0	&	11	&	11	&	1	&	11	&0.207 &	    	\\ 
92156.494	&-4.767	&273.74	&22	&	1	&	21	&	22	&	2	&	21	&0.096 &	        \\ 
93568.917	&-4.749	&292.45	&23	&	1	&	22	&	23	&	2	&	22	&0.089 &		\\ 
94696.389	&-4.625	&158.52	&15	&	0	&	15	&	15	&	1	&	15	&0.232 &		\\ 
95416.567	&-4.599	&171.24	&16	&	0	&	16	&	16	&	1	&	16	&0.229 &		\\ 
95924.494	&-4.582	&184.74	&17	&	0	&	17	&	17	&	1	&	17	&0.222 &Y	\\ 
96230.720	&-4.732	&353.16	&26	&	1	&	25	&	26	&	2	&	25	&0.066 &		\\ 
96301.854	&-4.570	&199.01	&18	&	0	&	18	&	18	&	1	&	18	&0.211 &		\\ 
97139.033	&-4.558	&263.84	&22	&	0	&	22	&	22	&	1	&	22	&0.155 &	\\ 
97263.633	&-4.562	&281.98	&23	&	0	&	23	&	23	&	1	&	23	&0.140 &Y		\\ 
97535.982	&-4.552	&246.48	&21	&	1	&	21	&	21	&	0	&	21	&0.171 &		\\ 
97536.860	&-4.559	&281.99	&23	&	1	&	23	&	23	&	0	&	23	&0.140 &		\\ 
98983.556	&-4.619	&146.69	&14	&	1	&	14	&	14	&	0	&	14	&0.239 &		\\ 
99439.978	&-4.703	&353.24	&26	&	2	&	25	&	26	&	1	&	25	&0.069 &		\\ 
99524.061	&-4.578	&141.48	&17	&	3	&	14	&	17	&	2	&	15	&0.269 &		\\ 
100194.322	&-4.969	&75.04	&6	&	1	&	6	&	5	&	1	&	5	&0.152 &Y		\\ 
100358.937	&-4.588	&126.84	&16	&	3	&	13	&	16	&	2	&	14	&0.280 &		\\ 
100365.042	&-4.925	&79.72	&6	&	1	&	6	&	5	&	1	&	5	&0.164 &		\\ 
100452.033	&-4.687	&312.06	&24	&	2	&	23	&	24	&	1	&	23	&0.087 &		\\ 
100452.089	&-7.897	&385.51	&26	&	5	&	21	&	26	&	5	&	21	&0.000 &		\\ 
100990.113	&-5.008	&35.21	&8	&	2	&	7	&	8	&	1	&	8	&0.169 &		\\ 
101873.097	&-4.548	&173.44	&19	&	3	&	16	&	19	&	2	&	17	&0.239 &		\\ 
102282.773	&-4.676	&274.00	&22	&	2	&	21	&	22	&	1	&	21	&0.106 &	\\ 
102283.550	&-4.597	&113.10	&15	&	3	&	12	&	15	&	2	&	13	&0.289 &		\\ 
103452.151	&-4.993	&79.25	&6	&	2	&	5	&	5	&	2	&	4	&0.136 &	\\ 
103608.576	&-4.674	&256.14	&21	&	2	&	20	&	21	&	1	&	20	&0.116 &		\\ 
103890.656	&-5.197	&93.93	&6	&	4	&	3	&	5	&	4	&	2   &0.078 &		\\ 
103891.101	&-5.197	&93.93	&6	&	4	&	2	&	5	&	4	&	1   &0.078 &		\\ 
103895.500	&-5.088	&90.18	&6	&	3	&	4	&	5	&	3	&	3   &0.103 &		\\ 
105162.141	&-4.789	&98.87	&9	&	1	&	9	&	9	&	0	&	9   &0.194 &Y		\\ 
105337.356	&-4.528	&190.75	&20	&	3	&	17	&	20	&	2	&	18	&0.222 &		\\ 
106676.441	&-4.916	&76.13	&6	&	1	&	5	&	5	&	1	&	4	&0.160 &		\\ 
106723.565	&-4.939	&42.73	&9	&	2	&	8	&	9	&	1	&	9	&0.180 &	\\ 
106931.229	&-4.698	&222.75	&19	&	2	&	18	&	19	&	1	&	18	&0.126 &Y		\\ 
\hline
\hline
\end{longtable}
{\textbf{Notes:} Frequency: rest frequency of the transitions in units MHz; log(I): logarithm of the line intensity; $\mathrm{E_{\rm{U}}/k_{\rm{B}}}$: energy of the upper state of the transition in units K; $\mathrm{Q_{U}}$: quantum numbers of the upper level; $\mathrm{Q_{L}}$: quantum numbers of the lower level; $\tau_{0} $: optical depth at the center of the line, calculated with the parameters of best fit; flag indicating which transitions have been used to create the mean map.} \vspace{3mm}
}

{\small
\setlength{\tabcolsep}{3pt}

\renewcommand{\arraystretch}{1.1}

\begin{longtable}[10]{ccccccccccc}
\label{table:gGg}\\
\caption{Most unblended transitions of gGg'-(CH$_2$OH)$_2$ used to constrain the fit. Database used: CDMS}\\
\hline\hline
Frequency	&log(I)	& $E_{\rm{U}}/k_{\rm{B}}$	&\multicolumn{3}{c}{$Q_{\rm{U}}$}	&\multicolumn{3}{c}{$Q_{\rm{L}}$}	& $\tau _{0}$ & \\
MHz &  &K & $J$ & $K_{\rm{a}}$ & $K_{\rm{c}}$ & $J$ & $K_{\rm{a}}$ & $K_{\rm{c}}$ & &\\
\hline
\endfirsthead
\caption[]{Continued.} \\
\hline
 Frequency	&log(I)	& $E_{\rm{U}}/k_{\rm{B}}$	&\multicolumn{3}{c}{$Q_{\rm{U}}$}	&\multicolumn{3}{c}{$Q_{\rm{L}}$}	& $\tau _{0}$ &\\
MHz &  &K & $J$ & $K_{\rm{a}}$ & $K_{\rm{c}}$ & $J$ & $K_{\rm{a}}$ & $K_{\rm{c}}$ & &\\
\hline
\endhead
\hline
\endfoot
\endlastfoot
86429.874	&	-5.272	&	74.23	&	17	&	3	&	15	&	17	&	2	&	16		& 	0.014	& Y \\
93016.853	&	-5.068	&	21.81	&	9	&	3	&	7	&	8	&	3	&	6		& 	0.027	& Y \\
98481.089	&	-5.024	&	115.43	&	21	&	4	&	18	&	21	&	3	&	19		& 	0.017	& Y \\
102758.214	&	-4.819	&	25.60	&	11	&	1	&	11	&	10	&	1	&	10		& 	0.043	& Y \\
103538.434	&	-4.961	&	29.56	&	10	&	4	&	6	&	9	&	4	&	5		& 	0.030	& Y \\
108867.794	&	-4.739	&	86.56	&	17	&	6	&	12	&	17	&	5	&	13		& 	0.035	& Y \\
108867.930	&	-5.140	&	63.13	&	14	&	6	&	8	&	14	&	5	&	10		& 	0.016	& Y \\
113997.038	&	-4.939	&	34.43	&	11	&	4	&	7	&	10	&	4	&	6		& 	0.028	& Y \\
\hline
\hline
\end{longtable}
{\textbf{Notes:} Frequency: rest frequency of the transitions in units MHz; log(I): logarithm of the line intensity; $\mathrm{E_{\rm{U}}/k_{\rm{B}}}$: energy of the upper state of the transition in units K; $\mathrm{Q_{U}}$: quantum numbers of the upper level; $\mathrm{Q_{L}}$: quantum numbers of the lower level; $\tau_{0} $: optical depth at the center of the line, calculated with the parameters of best fit; flag indicating which transitions have been used to create the mean map.} \vspace{3mm}
}
{\small
\setlength{\tabcolsep}{3pt}

\renewcommand{\arraystretch}{1.1}

\begin{longtable}[10]{ccccccccccc}
\label{table:aGg}\\
\caption{Most unblended transitions of aGg'-(CH$_2$OH)$_2$ used to constrain the fit. Database used: CDMS}\\
\hline\hline
Frequency	&log(I)	& $E_{\rm{U}}$	&\multicolumn{3}{c}{$Q_{\rm{U}}$}	&\multicolumn{3}{c}{$Q_{\rm{L}}$}	& $\tau _{0}$ & \\
MHz &  &K & $J$ & $K_{\rm{a}}$ & $K_{\rm{c}}$ & $J$ & $K_{\rm{a}}$ & $K_{\rm{c}}$ & &\\
\hline
\endfirsthead
\caption[]{Continued.} \\
\hline
 Frequency	&log(I)	& $E_{\rm{U}}$	&\multicolumn{3}{c}{$Q_{\rm{U}}$}	&\multicolumn{3}{c}{$Q_{\rm{L}}$}	& $\tau _{0}$ &\\
MHz &  &K & $J$ & $K_{\rm{a}}$ & $K_{\rm{c}}$ & $J$ & $K_{\rm{a}}$ & $K_{\rm{c}}$ & &\\
\hline
\endhead
\hline
\endfoot
\endlastfoot

84439.522	&	-4.689	&	13.25	& 8	&	0	&	8	&	7	&	0	&	7		& 0.126 & Y \\
85521.565	&	-4.868	&	25.71	& 9	&	4	&	6	&	8	&	4	&	5		& 0.077 &  \\
85522.221	&	-6.560	&	102.24	& 20	&	2	&	18	&	19	&	4	&	15	& 0.001 &  \\
86655.610	&	-4.633	&	18.96	& 9	&	1	&	8	&	8	&	1	&	7		& 0.135 &  \\
86660.118	&	-4.699	&	22.37	& 9	&	3	&	6	&	8	&	3	&	5		& 0.114 & Y \\
88366.963	&	-4.572	&	21.45	& 10	&	1	&	10	&	9	&	1	&	9	& 0.151 & Y \\
88745.339	&	-4.702	&	21.40	& 10	&	0	&	10	&	9	&	0	&	9	& 0.111 &  \\
88986.213	&	-4.907	&	21.47	& 8	&	4	&	5	&	7	&	4	&	4		& 0.069 &  \\
89708.194	&	-4.728	&	18.09	& 8	&	3	&	5	&	7	&	3	&	4		& 0.106 &  \\
90423.320	&	-5.094	&	24.01	& 10	&	1	&	9	&	9	&	2	&	8	& 0.043 &  \\
91539.725	&	-4.776	&	15.88	& 8	&	2	&	6	&	7	&	2	&	5		& 0.094 &  \\
95354.813	&	-5.007	&	53.52	& 10	&	8	&	3	&	9	&	8	&	2	& 0.034 & Y \\
95354.813	&	-5.116	&	53.52	& 10	&	8	&	2	&	9	&	8	&	1	& 0.043 & Y \\
96258.221	&	-4.602	&	23.412	& 10	&	1	&	9	&	9	&	1	&	8	& 0.128 &  \\
96259.944	&	-5.614	&	106.12	& 20	&	4	&	17	&	20	&	3	&	18	& 0.008 &  \\
97359.027	&	-4.547	&	19.37	& 9	&	2	&	8	&	8	&	2	&	7		& 0.146 & Y \\
97558.756	&	-4.669	&	26.83	& 10	&	3	&	7	&	9	&	3	&	6	& 0.106 &  \\
99082.055	&	-4.804	&	35.13	& 9	&	6	&	4	&	8	&	6	&	3		& 0.057 & Y \\
99082.055	&	-4.913	&	35.13	& 9	&	6	&	3	&	8	&	6	&	2		& 0.073 & Y \\
99509.149	&	-4.608	&	24.80	& 10	&	2	&	8	&	9	&	2	&	7	& 0.121 &  \\
105701.002	&	-4.364	&	28.32	& 11	&	1	&	10	&	10	&	1	&	9	& 0.197 & Y \\
106088.917	&	-4.510	&	39.38	& 11	&	5	&	6	&	10	&	5	&	5	& 0.132 & Y \\
109489.186	&	-4.540	&	34.16	& 10	&	5	&	5	&	9	&	5	&	4	& 0.123 &  \\
111579.436	&	-4.355	&	25.66	& 11	&	1	&	11	&	10	&	1	&	10	& 0.192 &  \\
\hline
\hline
\end{longtable}
{\textbf{Notes:} Frequency: rest frequency of the transitions in units MHz; log(I): logarithm of the line intensity; $\mathrm{E_{U}}$: energy of the upper state of the transition in units K; $\mathrm{Q_{U}}$: quantum numbers of the upper level; $\mathrm{Q_{L}}$: quantum numbers of the lower level; $\tau_{0} $: optical depth at the center of the line, calculated with the parameters of best fit; flag indicating which transitions have been used to create the mean map.} \vspace{3mm}
}

{\small
\setlength{\tabcolsep}{3pt}

\renewcommand{\arraystretch}{1.1}

\begin{longtable}[7]{ccccccccccc}
\label{table:ch3cnvib}\\
\caption{Most unblended transitions of CH$_3$CN\,v$_8=1$ used to constrain the fit. Database used: CDMS}\\
\hline\hline
Frequency	&log(I)	& $E_{\rm{U}}/k_{\rm{B}}$	&\multicolumn{3}{c}{$Q_{\rm{U}}$}	&\multicolumn{3}{c}{$Q_{\rm{L}}$}	& $\tau _{0}$& \\
MHz &  &K & $J$ & $K_{\rm{a}}$ & $K_{\rm{c}}$ & $J$ & $K_{\rm{a}}$ & $K_{\rm{c}}$ & & \\
\hline
\endfirsthead
\caption[]{Continued.} \\
\hline
 Frequency	&log(I)	& $E_{\rm{U}}/k_{\rm{B}}$	&\multicolumn{3}{c}{$Q_{\rm{U}}$}	&\multicolumn{3}{c}{$Q_{\rm{L}}$}	& $\tau _{0}$ &\\
MHz &  &K & $J$ & $K_{\rm{a}}$ & $K_{\rm{c}}$ & $J$ & $K_{\rm{a}}$ & $K_{\rm{c}}$ & &\\
\hline
\endhead
\hline
\endfoot
\endlastfoot
92175.520	&-4.029	& 532.85		&5	&	1	&	3	&	4	&	-1	&	3		&0.150&Y		\\ 
92247.240	&-4.176	& 594.26		&5	&	-2	&	2	&	4	&	2	&	2		&0.096&Y		\\ 
92247.240	&-4.176	& 594.26		&5	&	2	&	2	&	4	&	-2	&	2		&0.096&Y		\\ 
92249.653	&-4.552	& 600.11		&5	&	-4	&	3	&	4	&	4	&	3		&0.040&Y		\\ 
92249.653	&-4.552	& 600.11		&5	&	4	&	3	&	4	&	-4	&	3		&0.040&Y		\\ 
92256.288	&-4.067	& 559.50		&5	&	1	&	2	&	4	&	1	&	2		&0.131&		\\ 
92258.412	&-4.249	& 563.40		&5	&	3	&	3	&	4	&	3	&	3		&0.086&		\\ 
92261.440	&-4.020	& 539.03		&5	&	0	&	2	&	4	&	0	&	2		&0.152&	\\ 
92263.992	&-4.098	& 540.98		&5	&	2	&	3	&	4	&	2	&	3		&0.126&	\\ 
92353.516	&-4.028	& 532.87		&5	&	-1	&	3	&	4	&	1	&	3		&0.150&Y	\\ 
110609.594	&-3.793	& 538.16		&6	&	-1	&	3	&	5	&	1	&	3		&0.214&Y		\\ 
110706.340	&-3.831	& 564.82		&6	&	1	&	2	&	5	&	1	&	2		&0.187&		\\ 
110709.354	&-3.950	& 568.72		&6	&	3	&	3	&	5	&	3	&	3		&0.141&		\\ 
110712.220	&-3.789	& 544.35		&6	&	0	&	2	&	5	&	0	&	2		&0.213&Y		\\ 
110716.278	&-3.843	& 540.98		&6	&	2	&	3	&	5	&	2	&	3		&0.188&		\\ 
110823.126	&-3.792	& 532.87		&6	&	1	&	3	&	5	&	-1	&	3		&0.214&Y		\\ 
\hline
\hline
\end{longtable}
{\textbf{Notes:} Frequency: rest frequency of the transitions in units MHz; log(I): logarithm of the line intensity; $\mathrm{E_{\rm{U}}/k_{\rm{B}}}$: energy of the upper state of the transition in units K; $\mathrm{Q_{U}}$: quantum numbers of the upper level; $\mathrm{Q_{L}}$: quantum numbers of the lower level; $\tau_{0} $: optical depth at the center of the line, calculated with the parameters of best fit; flag indicating which transitions have been used to create the mean map.} \vspace{3mm}
}

{\small
\setlength{\tabcolsep}{3pt}

\renewcommand{\arraystretch}{1.1}

\begin{longtable}[8]{ccccccccc}
\label{table:c13h3cn}\\
\caption{Most unblended transitions of $^{13}$CH$_3$CN used to constrain the fit. Database used: CDMS}\\
\hline\hline
 Frequency	&log(I)	& $E_{\rm{U}}/k_{\rm{B}}$	&\multicolumn{2}{c}{$Q_{\rm{U}}$}	&\multicolumn{2}{c}{$Q_{\rm{L}}$}	& $\tau _{0}$& \\
MHz &  &K & $J$ & $K$ &  $J$ & $K$ & & \\
\hline
\endfirsthead
\caption[]{Continued.} \\
\hline
 Frequency	&log(I)	& $E_{\rm{U}}/k_{\rm{B}}$	&\multicolumn{2}{c}{$Q_{\rm{U}}$}	&\multicolumn{2}{c}{$Q_{\rm{L}}$}	& $\tau _{0}$ &\\
MHz &  &K & $J$ & $K$ &  $J$ & $K$ & & \\
\hline
\endhead
\hline
\endfoot
\endlastfoot
89324.555	&-3.238	&41.54		&5	&	2	&	4	&	2		&0.179&Y		\\ 
89329.600	&-3.149	&15.75		&5	&	1	&	4	&	1		&0.248&Y		\\ 
89331.282	&-3.121	&8.58		&5	&	0	&	4	&	0		&0.276&Y		\\ 
107178.424	&-2.808	&77.36		&6	&	3	&	5	&	3		&0.320&Y		\\ 
107188.500	&-2.983	&41.54		&6	&	2	&	5	&	2		&0.261&Y		\\ 
107194.550	&-2.913	&20.04		&6	&	1	&	5	&	1		&0.347&Y		\\ 
107196.570	&-2.890	&12.87		&6	&	0	&	5	&	0		&0.381&Y		\\ 
\hline
\hline
\end{longtable}
{\textbf{Notes:} Frequency: rest frequency of the transitions in units MHz; log(I): logarithm of the line intensity; $\mathrm{E_{\rm{U}}/k_{\rm{B}}}$: energy of the upper state of the transition in units K; $\mathrm{Q_{U}}$: quantum numbers of the upper level; $\mathrm{Q_{L}}$: quantum numbers of the lower level; $\tau_{0} $: optical depth at the center of the line. calculated with the parameters of best fit; flag indicating which transitions have been used to create the mean map.} \vspace{3mm} 
}

{\small
\setlength{\tabcolsep}{3pt}

\renewcommand{\arraystretch}{1.1}

\begin{longtable}[9]{ccccccccc}
\label{table:ch3c13n}\\
\caption{Most unblended transitions of CH$_3^{13}$CN used to constrain the fit. Database used: CDMS}\\
\hline\hline
Frequency	&log(I)	& $E_{\rm{U}}/k_{\rm{B}}$	&\multicolumn{2}{c}{$Q_{\rm{U}}$}	&\multicolumn{2}{c}{$Q_{\rm{L}}$}	& $\tau _{0}$ &\\
MHz &  &K & $J$ & $K$ & $J$ & $K$ & &\\
\hline
\endfirsthead
\caption[]{Continued.} \\
\hline
 Frequency	&log(I)	& $E_{\rm{U}}/k_{\rm{B}}$	&\multicolumn{2}{c}{$Q_{\rm{U}}$}	&\multicolumn{2}{c}{$Q_{\rm{L}}$}	& $\tau _{0}$& \\
MHz &  &K & $J$ & $K$ & $J$ & $K$ & &\\
\hline
\endhead
\hline
\endfoot
\endlastfoot
91925.704	&-3.070	&77.62		&5	&3	&	4	&	3		&0.578&Y		\\ 
91934.531	&-3.201	&41.86		&5	&2	&	4	&	2		&0.736&Y		\\ 
110309.800	&-2.771	&82.92		&6	&3	&	5	&	3		&0.891&Y	\\ 
\hline
\hline
\end{longtable}
{\textbf{Notes:} Frequency: rest frequency of the transitions in units MHz; log(I): logarithm of the line intensity; $\mathrm{E_{\rm{U}}/k_{\rm{B}}}$: energy of the upper state of the transition in units K; $\mathrm{Q_{U}}$: quantum numbers of the upper level; $\mathrm{Q_{L}}$: quantum numbers of the lower level; $\tau_{0} $: optical depth at the center of the line, calculated with the parameters of best fit; flag indicating which transitions have been used to create the mean map.} \vspace{3mm}
}

{\small
\setlength{\tabcolsep}{3pt}

\renewcommand{\arraystretch}{1.1}

\begin{longtable}[10]{ccccccccccc}
\label{table:c2h3cn}\\
\caption{Most unblended transitions of C$_2$H$_3$CN used to constrain the fit. Database used: CDMS}\\
\hline\hline
Frequency	&log(I)	& $E_{\rm{U}}/k_{\rm{B}}$	&\multicolumn{3}{c}{$Q_{\rm{U}}$}	&\multicolumn{3}{c}{$Q_{\rm{L}}$}	& $\tau _{0}$ & \\
MHz &  &K & $J$ & $K_{\rm{a}}$ & $K_{\rm{c}}$ & $J$ & $K_{\rm{a}}$ & $K_{\rm{c}}$ & &\\
\hline
\endfirsthead
\caption[]{Continued.} \\
\hline
 Frequency	&log(I)	& $E_{\rm{U}}/k_{\rm{B}}$	&\multicolumn{3}{c}{$Q_{\rm{U}}$}	&\multicolumn{3}{c}{$Q_{\rm{L}}$}	& $\tau _{0}$ &\\
MHz &  &K & $J$ & $K_{\rm{a}}$ & $K_{\rm{c}}$ & $J$ & $K_{\rm{a}}$ & $K_{\rm{c}}$ & &\\
\hline
\endhead
\hline
\endfoot
\endlastfoot
85715.423	&-3.937	& 29.22	&9	&	2	&	7	&	8	&	2	&	6		&0.170&Y		\\ 
94913.115	&-3.900	& 59.72	&10	&	4	&	7	&	9	&	4	&	6		&0.138&	Y	\\ 
94913.227	&-3.900	& 59.72	&10	&	4	&	6	&	9	&	4	&	5		&0.138&	Y	\\ 
94913.958	&-3.978	& 79.18	&10	&	5	&	6	&	9	&	5	&	5		&0.102&	Y	\\ 
94913.959	&-3.978	& 79.18	&10	&	5	&	5	&	9	&	5	&	4		&0.102&	Y	\\ 
95325.476	&-3.801	& 33.80	&10	&	2	&	8	&	9	&	2	&	7		&0.203&		\\ 
96982.442	&-3.764	& 27.81	&10	&	1	&	9	&	9	&	1	&	8		&0.226&	Y\\ 
101637.231	&-3.686	& 31.49	&11	&	1	&	11	&	10	&	1	&	10	    &0.252	&	Y\\ 
103575.395	&-3.664	& 30.00	&11	&	0	&	11	&	10	&	0	&	10  	&0.263	&	Y\\ 
104960.538	&-3.680	& 38.84	&11	&	2	&	9	&	10	&	2	&	8		&0.237&	Y	\\ 
113657.635	&-3.578	& 44.19	&12	&	2	&	11	&	11	&	2	&	10	    &0.268	&	Y\\ 
113904.798	&-3.712	& 89.66	&12	&	5	&	7	&	11	&	5	&	6		&0.147&	Y	\\ 
113904.798	&-3.712	& 89.66	&12	&	5	&	8	&	11	&	5	&	7		&0.147&	Y	\\ 
113939.397	&-3.607	& 55.06	&12	&	3	&	10	&	11	&	3	&	9		&0.233&		\\ 
114621.567	&-3.571	& 44.35 &12	&	2	&	10	&	11	&	2	&	9		&0.270&		\\ 
\hline
\hline
\end{longtable}
{\textbf{Notes:} Frequency: rest frequency of the transitions in units MHz; log(I): logarithm of the line intensity; $\mathrm{E_{\rm{U}}/k_{\rm{B}}}$: energy of the upper state of the transition in units K; $\mathrm{Q_{U}}$: quantum numbers of the upper level; $\mathrm{Q_{L}}$: quantum numbers of the lower level; $\tau_{0} $: optical depth at the center of the line, calculated with the parameters of best fit; flag indicating which transitions have been used to create the mean map.} \vspace{3mm}
}

{\small
\setlength{\tabcolsep}{3pt}

\renewcommand{\arraystretch}{1.1}

\begin{longtable}[11]{ccccccccccc}
\label{table:c2h5cn}\\
\caption{Most unblended transitions of C$_2$H$_5$CN used to constrain the fit. Database used: CDMS}\\
\hline\hline
Frequency	&log(I)	& $E_{\rm{U}}/k_{\rm{B}}$	&\multicolumn{3}{c}{$Q_{\rm{U}}$}	&\multicolumn{3}{c}{$Q_{\rm{L}}$}	& $\tau _{0}$ &\\
MHz &  &K & $J$ & $K_{\rm{a}}$ & $K_{\rm{c}}$ & $J$ & $K_{\rm{a}}$ & $K_{\rm{c}}$ & &\\
\hline
\endfirsthead
\caption[]{Continued.} \\
\hline
 Frequency	&log(I)	& $E_{\rm{U}}/k_{\rm{B}}$	&\multicolumn{3}{c}{$Q_{\rm{U}}$}	&\multicolumn{3}{c}{$Q_{\rm{L}}$}	& $\tau _{0}$ &\\
MHz &  &K & $J$ & $K_{\rm{a}}$ & $K_{\rm{c}}$ & $J$ & $K_{\rm{a}}$ & $K_{\rm{c}}$ & &\\
\hline
\endhead
\hline
\endfoot
\endlastfoot
86819.845	&-3.738	& 24.09		&10	&	1	&	10	&	9	&	1	&	9		&0.822&		\\ 
88323.735	&-3.719	& 23.48		&10	&	0	&	10	&	9	&	0	&	9		&0.850&Y	\\ 
89297.660	&-3.732	& 28.08		&10	&	2	&	9	&	9	&	2	&	8		&0.782&	Y	\\ 
89628.485	&-3.760	& 33.70		&10	&	3	&	8	&	9	&	3	&	7		&0.695&		\\ 
91549.112	&-3.694	& 25.35		&10	&	1	&	9	&	9	&	1	&	8	    &0.855&	\\ 
96919.762	&-3.603	& 28.13		&11	&	0	&	11	&	10	&	0	&	10	    &0.972	&	Y\\ 
98523.872	&-3.799	& 68.47		&11	&	6	&	5	&	10	&	6	&	4		&0.428&	Y\\ 
98523.872	&-3.799	& 68.47		&11	&	6	&	6	&	10	&	6	&	5		&0.428&	Y\\ 
98524.672	&-3.892	& 82.92		&11	&	7	&	5	&	10	&	7	&	4		&0.304&	Y	\\ 
98524.672	&-3.892	& 82.92		&11	&	7	&	4	&	10	&	7	&	3		&0.304&	Y	\\ 
98532.084	&-4.018	& 99.59		&11	&	8	&	4	&	10	&	8	&	3		&0.197&	Y	\\ 
98532.084	&-4.018	& 99.59		&11	&	8	&	3	&	10	&	8	&	2		&0.197&	Y	\\ 
98533.987	&-3.729	& 56.23		&11	&	5	&	6	&	10	&	5	&	5		&0.560&	Y	\\ 
98533.987	&-3.729	& 56.23		&11	&	5	&	7	&	10	&	5	&	6		&0.560&	Y	\\ 
98564.827	&-3.675	& 46.22		&11	&	4	&	8	&	10	&	4	&	7		&0.692&	Y	\\ 
98566.792	&-3.675	& 46.22		&11	&	4	&	7	&	10	&	4	&	6		&0.692&	Y	\\ 
98701.101	&-3.635	& 38.44		&11	&	3	&	8	&	10	&	3	&	7		&0.811&	Y	\\ 
99681.461	&-3.599	& 33.04		&11	&	2	&	9	&	10	&	2	&	8		&0.914&		\\ 
104051.276	&-3.514	& 33.67		&12	&	1	&	12	&	11	&	1	&	11	    &1.062	&	\\ 
107485.160	&-3.741	& 88.09		&12	&	7	&	6	&	11	&	7	&	5		&0.378&		\\ 
107485.160	&-3.741	& 88.09		&12	&	7	&	5	&	11	&	7	&	4		&0.378&		\\ 
107486.949	&-3.665	& 73.63		&12	&	6	&	6	&	11	&	6	&	5		&0.512&		\\ 
107486.949	&-3.665	& 73.63		&12	&	6	&	7	&	11	&	6	&	6		&0.512&		\\ 
107491.574	&-3.840	& 104.76		&12	&	8	&	5	&	11	&	8	&	4	&0.260&		\\ 
107491.574	&-3.840	& 104.76		&12	&	8	&	4	&	11	&	8	&	3	&0.260&		\\ 
107502.432	&-3.605	& 61.40		&12	&	5	&	7	&	11	&	5	&	6		&0.654&	Y	\\ 
107502.432	&-3.605	& 61.40		&12	&	5	&	8	&	11	&	5	&	7		&0.654&	Y	\\ 
107503.686	&-3.971	& 123.64		&12	&	9	&	4	&	11	&	9	&	3	&0.163&	Y	\\ 
107503.686	&-3.971	& 123.64		&12	&	9	&	3	&	11	&	9	&	2	&0.163&	Y	\\ 
109650.295	&-3.470	& 35.45		&12	&	1	&	11	&	11	&	1	&	10	    &1.097	&	\\ 
\hline
\hline
\end{longtable}
{\textbf{Notes:} Frequency: rest frequency of the transitions in units MHz; log(I): logarithm of the line intensity; $\mathrm{E_{\rm{U}}/k_{\rm{B}}}$: energy of the upper state of the transition in units K; $\mathrm{Q_{U}}$: quantum numbers of the upper level; $\mathrm{Q_{L}}$: quantum numbers of the lower level; $\tau_{0} $: optical depth at the center of the line, calculated with the parameters of best fit; flag indicating which transitions have been used to create the mean map.} \vspace{3mm}
}

{\small
\setlength{\tabcolsep}{3pt}

\renewcommand{\arraystretch}{1.1}

\begin{longtable}[11]{ccccccccccc}
\label{table:c2h5c13n}\\
\caption{Most unblended transitions of C$_2$H$_5^{13}$CN used to constrain the fit. Database used: CDMS}\\
\hline\hline
Frequency	&log(I)	& $E_{\rm{U}}/k_{\rm{B}}$	&\multicolumn{3}{c}{$Q_{\rm{U}}$}	&\multicolumn{3}{c}{$Q_{\rm{L}}$}	& $\tau _{0}$ & \\
MHz &  &K & $J$ & $K_{\rm{a}}$ & $K_{\rm{c}}$ & $J$ & $K_{\rm{a}}$ & $K_{\rm{c}}$ & &\\
\hline
\endfirsthead
\caption[]{Continued.} \\
\hline
 Frequency	&log(I)	& $E_{\rm{U}}/k_{\rm{B}}$	&\multicolumn{3}{c}{$Q_{\rm{U}}$}	&\multicolumn{3}{c}{$Q_{\rm{L}}$}	& $\tau _{0}$ &\\
MHz &  &K & $J$ & $K_{\rm{a}}$ & $K_{\rm{c}}$ & $J$ & $K_{\rm{a}}$ & $K_{\rm{c}}$ & &\\
\hline
\endhead
\hline
\endfoot
\endlastfoot
89116.016	&-3.972	& 63.61	&10	&	6	&	5	&	9	&	6	&	4		&0.012&	Y	\\ 
89116.016	&-3.972	& 63.61	&10	&	6	&	4	&	9	&	6	&	3		&0.012&	Y	\\ 
89235.646	&-3.774	& 33.58	&10	&	3	&	7	&	9	&	3	&	6		&0.022&	Y	\\ 
96452.595	&-3.617	& 27.99	&11	&	0	&	11	&	10	&	0	&	10	    &0.031	&	Y\\ 
100109.732	&-3.591	& 30.04	&11	&	1	&	10	&	10	&	1	&	9		&0.031&	Y	\\ 
109101.845	&-3.485	& 35.28	&12	&	1	&	11	&	11	&	1	&	10	    &0.036	&	Y\\ 
112101.543	&-3.431	& 38.90	&13	&	1	&	13	&	12	&	1	&	12	    &0.039	&	Y\\ 
115871.121	&-3.561	& 79.02	&13	&	6	&	8	&	12	&	6	&	7		&0.023&	Y	\\ 
115871.122	&-3.561	& 79.02	&13	&	6	&	7	&	12	&	6	&	6		&0.023&	Y\\ 
115871.741	&-3.709	& 110.14	&13	&	8	&	6	&	12	&	8	&	5	&0.014&	Y\\ 
115871.741	&-3.709	& 110.14	&13	&	8	&	5	&	12	&	8	&	4	&0.014&	Y\\ 
\hline
\hline
\end{longtable}
{\textbf{Notes:} Frequency: rest frequency of the transitions in units MHz; log(I): logarithm of the line intensity; $\mathrm{E_{\rm{U}}/k_{\rm{B}}}$: energy of the upper state of the transition in units K; $\mathrm{Q_{U}}$: quantum numbers of the upper level; $\mathrm{Q_{L}}$: quantum numbers of the lower level; $\tau_{0} $: optical depth at the center of the line, calculated with the parameters of best fit.} \vspace{3mm}
}

\section{Moment maps of low-energy transitions}
\begin{figure*}[h!]
    \centering
    \includegraphics[width=18.5cm]{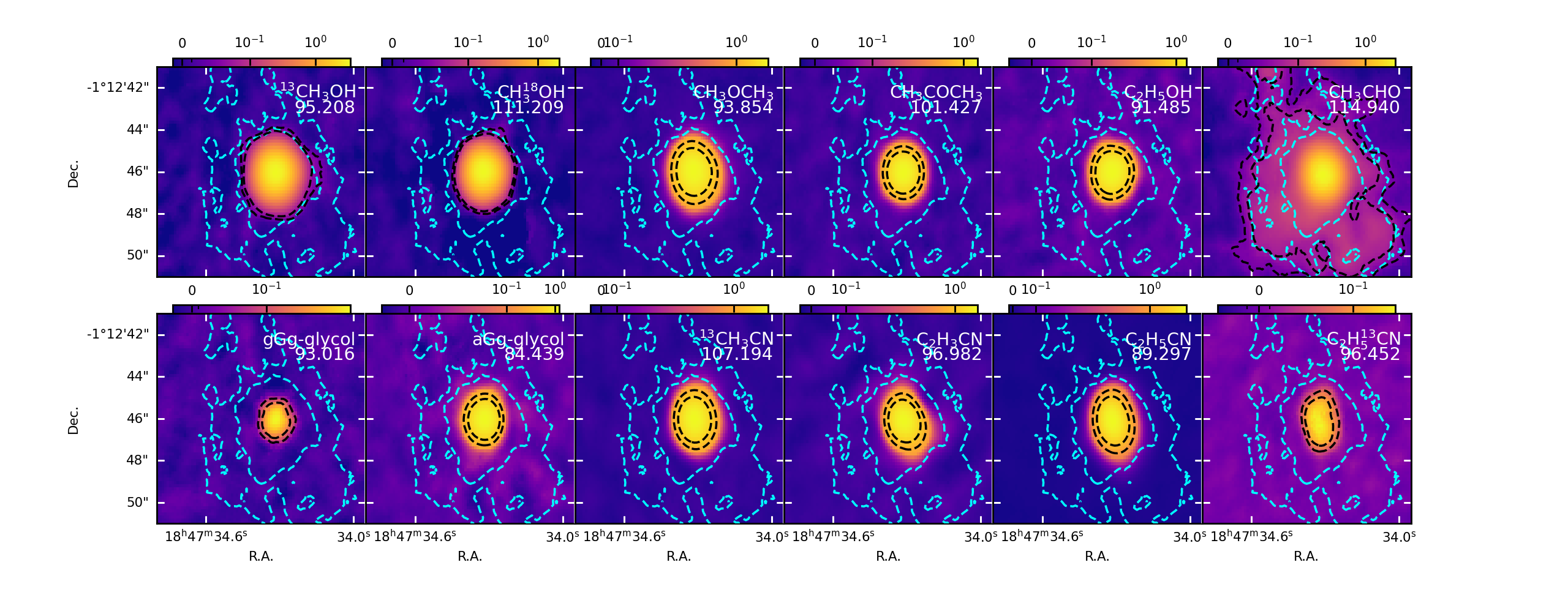}
    \caption{Moment-0 maps of low-energy transitions for the molecular species analyzed in this paper. We do not include CH$_3$OH vt=1, CH$_3$CN v8=1, and CH$_3^{13}$CN since there were not available transitions with $E_{\rm{U}}/\kappa_{\rm{B}}<30\,\rm{K}$. The moment-0 maps are not normalised to the peak intensity (i.e. to 1, as done for the mean maps shown in Fig. 1). The units are Jy/beam\,km\,s$^{-1}$. The black-dashed contours are the 5\textit{rms} and 10\textit{rms} contour of the color-scale image, while the cyan-dashed contours are the  5\textit{rms} and 10\textit{rms} contours of the mean map of CH$_3$CHO shown in Fig. 1.}
    \label{fig:lowenergy}
\end{figure*}


\end{appendix}
\end{document}